\documentclass{tcibook}
\usepackage{fancyhea}
\usepackage{amssymb}
\usepackage{work}
\usepackage{bm}       
\usepackage{graphicx}
\usepackage{multirow}
\usepackage{lineno}
\usepackage{threeparttable}
\usepackage[normalem]{ulem}
\usepackage{color}
\usepackage[multiple]{footmisc}
\usepackage[colorlinks = true,
            linkcolor = blue,
            urlcolor  = blue,
            citecolor = blue,
            anchorcolor = blue,
            hyperfootnotes=false]{hyperref}
\usepackage[numbers,square,sort&compress]{natbib}

\def\beq{\begin{equation}}
\def\eeq#1{\label{#1}\end{equation}}
\def\eeqn{\end{equation}}

\newenvironment{Eqnarray}%
   {\arraycolsep 0.14em\begin{eqnarray}}{\end{eqnarray}}
\def\beqa{\begin{Eqnarray}}
\def\eeqa#1{\label{#1}\end{Eqnarray}}
\def\eeqan{\end{Eqnarray}}

\let\bar=\overbar

\def\lsim{\mathrel{\raise.3ex\hbox{$<$\kern-.75em\lower1ex\hbox{$\sim$}}}}
\def\gsim{\mathrel{\raise.3ex\hbox{$>$\kern-.75em\lower1ex\hbox{$\sim$}}}}

\def\del{\partial}
\def\Dslash{\not{\hbox{\kern-4pt $D$}}}
\def\dslash{\not{\hbox{\kern-2pt $\del$}}}
\def\pslash{\not{\hbox{\kern-2pt $p$}}}
\def\ETmiss{\not{\hbox{\kern-4pt $E$}}_T}

\def\Dlr{\mathrel{\raise1.5ex\hbox{$\leftrightarrow$\kern-1em\lower1.5ex\hbox{$D$}}}}

\def\MSB{{\bar{M \kern -2pt S}}}
\def\msb{{\bar{\scriptsize M \kern -1pt S}}}

\def\drb{{\bar{\scriptsize D \kern -1pt R}}}

\makeatletter
\@ifpackageloaded{hyperref}{%
  \renewcommand\footref[1]{%
    \begingroup 
    \unrestored@protected@xdef\@thefnmark{%
      \ref*{#1}%
    }%
    \endgroup 
    \ifHy@hyperfootnotes 
       \expandafter\@firstoftwo 
    \else 
       \expandafter\@secondoftwo 
    \fi 
    {\hyperref[#1]{\strut\H@@footnotemark}}{\@footnotemark}%
  }%
}{}%

\usepackage{endnotes}
\let\footnote=\endnote

\begin{document}


\pagenumbering{roman}

\parindent=0pt
\parskip=8pt
\setlength{\evensidemargin}{0pt}
\setlength{\oddsidemargin}{0pt}
\setlength{\marginparsep}{0.0in}
\setlength{\marginparwidth}{0.0in}
\marginparpush=0pt


\pagenumbering{arabic}

\renewcommand{\chapname}{chap:intro_}
\renewcommand{\chapterdir}{.}
\renewcommand{\arraystretch}{1.25}
\addtolength{\arraycolsep}{-3pt}

\thispagestyle{empty}
\begin{centering}

{\Huge\bf The Future of US Particle Physics}

{\Large \bf Report of the 2021 Snowmass Community Study}

\vskip 0.4in

{\Huge\bf Chapter 5: Cosmic Frontier}

\begin{figure}[h!]
\begin{center}
\includegraphics[width=\linewidth]{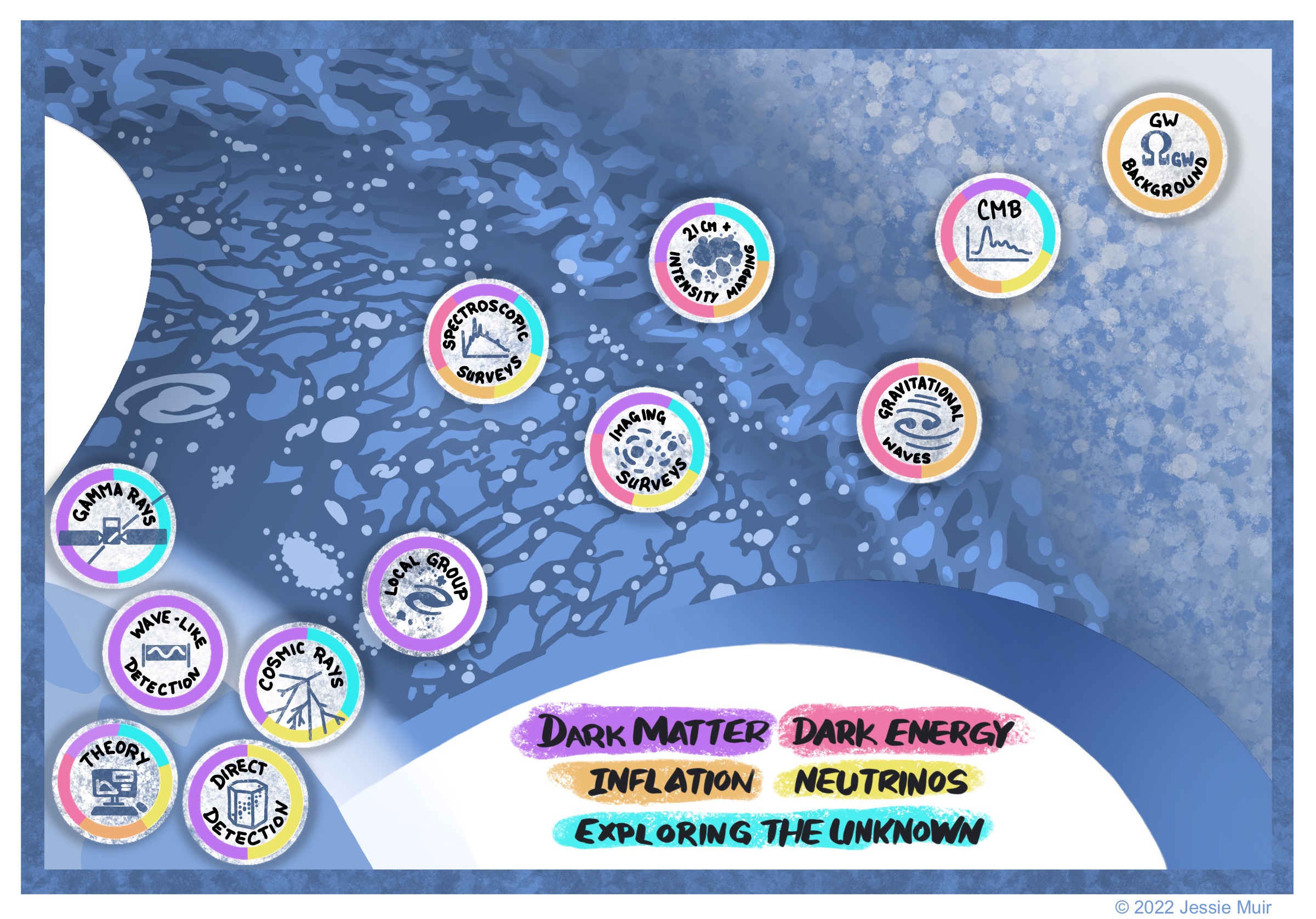}
\end{center}
\end{figure}

{\bf Frontier Conveners:} 
Aaron~S.~Chou\footnote{\label{fnal}Fermi National Accelerator Laboratory, Batavia, IL, 60510, USA},
Marcelle~Soares-Santos\footnote{\label{umich}Department of Physics, University of Michigan, Ann Arbor, MI, 48109, USA},
Tim~M.P.~Tait\footnote{\label{ucirvine}Department of Physics and Astronomy, University of California, Irvine, CA, 92697, USA}
\\
{\bf Topical Group Conveners:} 
Rana~X.~Adhikari\footnote{\label{caltech}Division of Physics, Math, and Astronomy, California Institute of Technology,	Pasadena,	CA,	91125,	USA},
Luis~A.~Anchordoqui\footnote{\label{cuny}Department of Physics and Astronomy, 	Lehman College, CUNY,	Bronx,	NY,	10468,	USA},
James~Annis\footref{fnal},
Clarence~L.~Chang\footnote{\label{anl}High Energy Physics Division,	Argonne National Laboratory,	Lemont,	IL,	60439,	USA}$^,$\footnote{\label{uchicago}Department of Astronomy \& Astrophysics,	University of Chicago,	Chicago,	IL,	60637,	USA}$^,$\footnote{\label{kicp}Kavili Institute for Cosmological Physics,	University of Chicago,	Chicago,	IL,	60637,	USA},
Jodi~Cooley\footnote{\label{snolab}SNOLAB,	Lively,	Ontario,	P3Y 1N2,	Canada}$^,$\footnote{\label{smu}Department of Physics,	Southern Methodist University,	Dallas,	TX,	75275,	USA},
Alex~Drlica-Wagner\footref{fnal}\footref{uchicago}\footref{kicp},
Ke~Fang\footnote{\label{uwisc}Department of Physics, University of Wisconsin-Madison, Madison, WI, 53705, USA},
Brenna~Flaugher\footref{fnal},
Joerg~Jaeckel\footnote{Institute for Theoretical Physics,	Heidelberg University,	Heidelberg,		69120,	Germany},
W.~Hugh~Lippincott\footnote{\label{ucsb}Department of Physics, University of California, Santa Barbara, CA, 93106, USA},
Vivian~Miranda\footnote{\label{stonybrook}Department of Physics and Astronomy, Stony Brook University, Stony Brook, NY, 11794, USA},
Laura Newburgh\footnote{Department of Physics, Yale University, New Haven, CT, USA},
Jeffrey~A.~Newman\footnote{Department of Physics and Astronomy and PITT PACC, University of Pittsburgh, Pittsburgh, PA, 15260, USA},
Chanda~Prescod-Weinstein\footnote{Department of Physics and Astronomy, University of New Hampshire, Durham, NH, 03824, USA},
Gray~Rybka\footnote{Department of Physics, University of Washington, Seattle, WA, 98195, USA},
B.~S. ~Sathyaprakash\footnote{Department of Astronomy and Astrophysics, The Pennsylvania State University, University Park, PA, 16802, USA},
David~J.~Schlegel\footnote{Physics Division, Lawrence Berkeley National Lab, Berkeley, CA, 94720, USA},
Deirdre~M.~Shoemaker\footnote{Department of Physics, University of Texas at Austin, Austin, TX, 78712, USA}
Tracy~R.~Slatyer\footnote{Center for Theoretical Physics, Massachusetts Institute of Technology, Cambridge, MA, 02139, USA},
An\v{z}e~Slosar\footnote{Department of Physics, Brookhaven National Laboratory, Upton, NY, 11973, USA},
Kirsten~Tollefson\footnote{Department of Physics and Astronomy, Michigan State University, East Lansing, MI, 48824, USA},
Lindley~Winslow\footnote{Laboratory for Nuclear Science, Massachusetts Institute of Technology, Cambridge, MA, 02139, USA},
Hai-Bo~Yu\footnote{\label{ucr}Department of Physics and Astronomy, University of California, Riverside , CA, 92521, USA},
Tien-Tien~Yu\footnote{Department of Physics,	University of Oregon,	Eugene,	OR,	97403,	USA}$^,$\footnote{Institute for Fundamental Science	University of Oregon,	Eugene,	OR,	97403,	USA}
\\
{\bf Liaisons:} 
Kristi~Engel\footnote{Department of Physics, University of Maryland, College Park, MD, USA},
Susan~Gardner\footnote{Department of Physics and Astronomy,	University of Kentucky,	Lexington,	KY,	40506, USA},
Tiffany~R.~Lewis\footnote{Astroparticle Physics Laboratory,	NASA Goddard Space Flight Center,	Greenbelt,	MD,	20771,	USA},
Bibhushan~Shakya\footnote{Deutsches Elektronen-Synchrotron DESY, Hamburg, 22607,	Germany},
Phillip~Tanedo\footref{ucr}

\renewcommand\enoteformat{%
  \flushleft
  \parskip -2ex 
  \leftskip=1.8em
  \makebox[0pt][r]{\textsuperscript\theenmark\enspace %
\rule{0pt}{\dimexpr\ht\strutbox}}%
}
\renewcommand{\notesname}{}
\def\enotesize{\footnotesize}
\vspace{-2em}
\theendnotes

\thispagestyle{empty}

\end{centering}

\newpage

\begin{center}
{\bf \large Abstract}

\begin{minipage}{0.8\linewidth}
We present the Snowmass 2021 strategy to delve deep, search wide, and aim high for new discoveries at the Cosmic Frontier of particle physics. The Cosmic Frontier is the bedrock of High Energy Physics in the twenty-first century, providing evidence for Beyond the Standard Model physics that motivates much of the Snowmass 2021 program across all frontiers. The scientific scope of the Cosmic Frontier encompasses four of the five science drivers of the field: dark matter, dark energy and cosmic acceleration, neutrinos, and exploring the unknown in search of new particles, new fields, and new principles of Nature. Covering a time frame of 20 years, our plan features a portfolio of small, medium, and large projects which are ready to produce a continuous stream of groundbreaking science results within this decade along with pathfinders for the next generation of projects to come in the following decade. In this report, we articulate the fundamental questions to be addressed for each science driver and identify key measurements required to achieve well-defined scientific thresholds for discovery. We describe the ecosystem of experiments designed to obtain those measurements as well as the associated developments in theory and technology. We plan to delve deep in sensitivity and search wide across many orders of magnitude in mass to discover the particle nature of dark matter. Moreover, we aim high, through billions of light years of cosmic history, to discover the time-evolution of dark energy, make the first experimental observations of the physics of inflation, and search for new physics at the highest energy scales. 
\end{minipage}

\vfill

\begin{minipage}{0.8\linewidth}
{\bf About Snowmass:} 
Snowmass is the U.S. High Energy Physics (HEP) Community Planning Exercise organized by the Division of Particles and Fields (DPF) of the American Physical Society (APS).  Through the Snowmass process, the community comes together to formulate a 10-year scientific vision for the field within a 20-year global context. The Particle Physics Project Prioritization Panel (P5)  takes the scientific input from Snowmass and develops a strategic plan that will inform decisions by the funding agencies. The Cosmic Frontier is one of the major Snowmass sub-communities, focused on cosmic probes of fundamental physics. Cosmic Frontier key topics include: dark matter, dark energy, and inflation. During the Snowmass 2021 process period, from April 2020 to July 2022,  the Cosmic Frontier community produced white papers, participated in town-hall meetings and workshops, and gathered input more broadly from all 10 Snowmass frontiers and several related fields. As a result, seven topical reports were produced. The community consensus is consolidated in this frontier-level report, led by the three Cosmic Frontier conveners and co-authored by the topical group conveners and liaisons.

\vspace{1em}
{\bf Please Note:} 
The information in this report spans a broad and complex research program with experiments of different scales and time frames which are envisioned as a whole package designed to achieve the full range of cosmic frontier science goals. We explicitly avoid prioritization of science drivers and of components of the experimental program. To indicate this, we have deliberately chosen to vary the order in which topics are presented throughout the report.  The length devoted to each topic represents a measure of how complicated it is to explain and is not indicative of its relative importance. 
Moreover, there is a non-trivial question about whether to structure 
the report around science topics or 
experimental/observational technique or facility.  We have chosen the former.
As a result, we describe specific projects in detail only under one of the science topics even though they can and do cover multiple ones.  

\end{minipage}
\end{center}

\thispagestyle{empty}

\setcounter{tocdepth}{2}
\begingroup
\let\cleardoublepage\clearpage
\tableofcontents
\endgroup
\thispagestyle{empty}

\setcounter{chapter}{4} 

\begingroup
\let\cleardoublepage\clearpage
\chapter{Cosmic Frontier}
\vskip 0.2in
\endgroup

\setcounter{page}{1} 

\section{Executive Summary}
\label{sec:summary}

The Cosmic Frontier realizes the High Energy Physics (HEP) community's goal of understanding the fundamental physics that governs the Universe and its constituents through a rich scientific research program designed to {\it delve deep}, \textit{search wide}, and \textit{aim high} towards discovery. Cosmic observations are exclusively responsible for our knowledge of the need to extend the Standard Model to describe dark matter, dark energy, and cosmic inflation. As a result, the Cosmic Frontier is uniquely positioned to shed light on these mysteries.  These efforts have led to tremendous advances in recent decades, including discoveries that the Universe appears to have experienced a period of early exponential expansion with quantum fluctuations seeding later large-scale structures, that the growth of these cosmic structures is influenced by the dynamics of dark matter and dark energy, and that the Universe is again expanding at an accelerating rate today.  The Cosmic Frontier research portfolio presents opportunities for large, medium, and small projects that will leverage new developments in both technology and theory to produce a continuous stream of high-impact science over the coming decades.

Cosmic surveys are the primary means by which we can study the origin, structure, composition, and evolution of our Universe.  Future surveys will explore its development from the earliest moments and potentially shed light on the connection between gravity and quantum mechanics; these surveys \textit{aim high} to study the Universe under extreme conditions that are impossible to achieve in terrestrial laboratories.  Proposed experiments can simultaneously provide precision measurements of dark energy, determine the energy scale and dynamics of cosmic inflation, search for new light relics and other Beyond the Standard Model physics, and probe the nature of dark matter. 

New technologies, as well as advances in theory and simulations, are providing exciting opportunities to advance our understanding of the fundamental physics that governs the Universe.  Current and near-future large wide-area optical/near-infrared imaging and spectroscopic surveys, proposed cosmic microwave background and spectroscopic survey projects of unprecedented power, and pathfinders for future gravitational wave and line-intensity mapping experiments will greatly extend our scientific reach.
New messengers such as gravitational waves provide unobscured views of the Universe at its earliest moments, and could reveal new physics through dynamical processes such as early phase transitions. Substantial investment in the next generation of cosmic surveys should be a priority to advance the broad range of profound science that only they can address.

Understanding the identity, nature, and origin of {dark matter} is one of the grandest challenges in physics and spans all Snowmass frontiers. Well-studied theoretical models provide a compelling scientific case to make broad and rapid inroads into unexplored dark matter parameter space via a {\it search wide, delve deep} strategy.  With new experiments that will come online in this decade, the HEP community will search wide to efficiently probe broad, logarithmic ranges of parameter space, including hidden sectors and axion-like particles, deploying experiments that exploit new cosmic probes and cutting edge technologies.  Concurrently, the community will delve deep to comprehensively explore the high-priority science targets of weakly interacting massive particles and the QCD axion. Given the importance of dark matter and the broad range of opportunities, aggregate investment in a coordinated dark matter program that spans a wide variety of direct, indirect, and cosmic probe experiments at multiple scales should be a priority with overall funding comparable to that of other large experimental programs, though distributed over a number of complementary efforts.  

The HEP community has identified potentially transformative opportunities to address fundamental physics questions via Cosmic Frontier programs. We aspire to {\bf Aim High}, {\bf Search Wide}, and {\bf Delve Deep}:

\begin{itemize}

\item Complete the CMB-S4 cosmic microwave background experiment and build a large spectroscopic facility (Spec-S5) to study physics including inflation, dark energy, light relics, modifications to general relativity, and dark matter.

\item Pursue a broad program investigating the full landscape of dark matter candidates, including implementation of the existing Dark Matter New Initiatives (DMNI) portfolio and development of future DMNI-like programs focused on small projects; investment in new quantum technologies; and engagement of the HEP community in the development and execution of cosmic and indirect searches for dark matter, to take full advantage of the unique opportunities provided by cosmological and astrophysical probes (e.g., Rubin LSST and AugerPrime).

\item Scale up mature technologies for weakly-interacting massive particle (WIMP) direct detection, fully exploring the parameter space down to the neutrino fog, and support high-energy gamma-ray telescopes (e.g., SWGO and CTA) to probe thermal WIMPs up to tens-of-TeV mass scales. Move ahead with new, construction-ready DMNI experiments with the capacity to probe the QCD axion over most of its viable mass range. 

\end{itemize}

Fully realizing the scientific potential of this exciting experimental program will require support for a range of non-project research activities, including theory studies and simulations, funding for analysis including cross-experiment joint efforts, and the acquisition of key complementary datasets (e.g., follow-up observations to enable supernova and gravitational wave source cosmology; spectroscopy for photometric redshift training and calibration; and measurements of dark matter density distribution and astrophysical backgrounds to inform direct and indirect dark matter searches).
To take maximum advantage of the remarkable datasets produced by the cosmic frontier projects, the science collaborations analyzing these data will need  funding for both scientific infrastructure and research activities.



\section{Search Wide, Aim High, Delve Deep: the Cosmic Frontier Strategy for Discovery}
\label{sec:overview}

The Cosmic Frontier (CF) comprises a broad set of activities aimed at understanding the fundamental physics which governs
the evolution of the Universe and its constituents.  
For Snowmass 2021, the CF is organized into  topical  groups, each of which produced a report summarizing the open questions
and scientific opportunities in its area:
\begin{itemize}
\item CF1 : particle-like dark matter \cite{Cooley:2022ufh};
\item CF2 : wave-like dark matter \cite{Jaeckel:2022kwg};
\item CF3 : cosmic probes of dark matter \cite{Drlica-Wagner:2022lbd};
\item CF4 : dark energy and cosmic acceleration in the modern Universe \cite{Annis:2022xgg};
\item CF5 : dark energy and cosmic acceleration at and before cosmic dawn \cite{Chang:2022lrw};
\item CF6 : new facilities and complementarity in probes of dark energy and cosmic acceleration \cite{Flaugher:2022rob}; and 
\item CF7 : cosmic probes of fundamental physics \cite{Adhikari:2022sve}.
\end{itemize}
Taken together, the topical working group reports present a broad program of inquiry that offers unique opportunities to understand the physics of dark matter, dark energy, and inflation, and to search for physics Beyond the Standard Model (BSM) in a way that is highly complementary to the other Snowmass frontiers.

The Cosmic Frontier engages with all five of the 2014 P5 science drivers, with particular
connections to: 
(i) identify the new physics of dark matter; (ii) understand cosmic acceleration: dark energy and inflation;
and (iii) explore the unknown: new particles, interactions, and physical principles.
Currently, the 
incontrovertible experimental evidence for
physics Beyond the Standard Model of particle physics consists of:
\begin{itemize}
\item dark matter;
\item dark energy;
\item cosmic inflation;
\item the baryon asymmetry of the Universe; and
\item neutrino masses and mixing.
\end{itemize}
With the exception of neutrino masses, all of these are the direct result of observational data from the Cosmic Frontier.  
Moreover, the techniques of the Cosmic Frontier remain the only currently established way to access several of these BSM physics topics.
The successful road map laid out by P5 in 2014 has resulted in a remarkable period
of investigation into fundamental physics through CF projects, which have made great progress exploring the parameter space of dark matter models and pinning down the nature of dark energy.  There is strong motivation to build and expand upon that vision to complete exploration of the physics of dark matter, dark energy and cosmic inflation, as well as to search for new physics that is too massive, too weakly-interacting, and/or requires too high energies to access terrestrially.  A discovery in any of these areas would be transformational for particle physics, and would suggest new priorities for HEP research across all the Snowmass frontiers in the future.

\subsection{Fundamental Questions}

CF projects engage with fundamental questions that are at the core of the quest to extend/replace the Standard Model and are well-aligned with the 2014 P5 science drivers .. A non-exhaustive list of the central questions includes:
\begin{itemize}
\item What is the nature of dark matter?  Is it a single particle or a set of multiple components?  Does it belong to a dark sector?
\item How does dark matter interact  with the Standard Model?  Does the Higgs serve as a portal between the two?  Are there new bosonic mediators?
\item Is the dark matter itself bosonic and does it exhibit macroscopic wave-like phenomena like oscillatory forces?
\item What are the dynamics within the dark sector?  Does dark matter have important self-interactions?
\item How was the dark matter produced in the Early Universe?
\item Does the nature of dark matter reveal further secrets informing physics Beyond the Standard Model or cosmology?
\item What is the nature of dark energy?  Is the present acceleration due to a new energy density component or does it demand a change in general relativity?
\item If dark energy is an energy density, is it a cosmological constant, or a dynamical quantity changing with time?
\item When did dark energy become important in the history of the Universe?
\item Why are there two eras of acceleration in the history of the Universe?
\item How is the inflationary paradigm realized in nature?
\item What is the energy scale of inflation?
\item Does inflation have dynamics that manifest themselves as an observable imprint on the primordial distribution of matter fluctuations?
\item Did BSM degrees of freedom influence the thermal history of the Universe?
\item What is the scale of neutrino masses?  Can cosmological observations sufficiently constrain it to distinguish between normal and inverted mass hierarchies?
\item Can gravitational waves reveal new dynamics or early cosmological phase transitions?
\item How can cosmic sources of high energy particles reveal their participation in new interactions or new physics?
\end{itemize}

\subsection{State of the Cosmic Frontier} 

Since the previous Snowmass in 2013, the Cosmic Frontier has made great progress towards realizing the science goals prioritized by P5 in 2014.
Searches for WIMP dark matter scattering with nuclei
have advanced to Generation-2, and searches for axions have begun to probe
the parameter space relevant to explain the strong CP problem.  These searches `delve deep' to study some of the high priority
accessible targets in dark matter parameter space.  At the same time, new techniques such as those sensitive to electron scattering are able to probe dark matter to sub-MeV masses and `search wide', 
providing coverage for a wide parameter space consistent with dark matter production
in the early Universe.

The current U.S. priority on quantum information science has produced fruitful collaborations between HEP and researchers in other fields, as ultra-weakly interacting dark matter is a natural science target for quantum sensing.  Cross-disciplinary collaborations with the atomic, molecular, and optical (AMO) and condensed matter fields have also produced novel dark matter search techniques to provide broad coverage of dark matter parameter space.

Indirect searches for dark matter using the Galactic center,  Milky Way dwarf galaxies, and our local Galactic halo as observational targets have produced broad constraints on dark matter mass and self-annihilation cross section.  However, despite a strong endorsement from the previous P5, indirect searches for dark matter with gamma-ray observatories have not been prioritized and the U.S. is ceding leadership in this area to other countries.

Investments in optical cosmological surveys are now coming to fruition --  results from the Dark Energy Survey (DES) and BOSS/eBOSS surveys have cemented $\Lambda$CDM as a description of the expansion of the Universe, 
the Dark Energy Spectroscopic Instrument (DESI) is obtaining data at a phenomenal rate, and construction of the Vera C. Rubin Observatory is nearing completion.  These experiments 
`aim high', offering unprecedented opportunity to understand the nature of cosmic acceleration/dark energy as well as providing
unique probes of the properties of dark matter such as self-interactions. In addition to these powerful ground-based, U.S. led surveys, there are also space missions and other large ground-based cosmology projects being carried out by other communities world-wide. Ensuring a return on the investment that was made in these Stage IV facilities by the U.S. community requires that the science collaborations analyzing these datasets have robust funding for both scientific infrastructure and research activities.

Major advances have also been made in cosmic microwave background (CMB) science including tight constraints on the physics of the early Universe that broadly support inflation, and rule out once plausible theories; constraints on BSM particles (light relativistic species); and tight constraints on $\Lambda$CDM.
The U.S. CMB community has coalesced around a single, large next-generation project, CMB-S4, which will observe the microwave sky with unprecedented statistical precision, targeting the physics of cosmic inflation, searches for BSM dark radiation, and measuring neutrino masses.

The detection of gravitational waves from merging binary neutron stars and black holes by the LIGO and Virgo Collaborations was a watershed moment for cosmic science.  This new observational tool will be important for many areas of particle astrophysics and cosmology including measurements of cosmic acceleration at high redshift, searches for dark matter, and detection of relic gravitational wave radiation from early Universe phase transitions.

\subsection{Science Opportunities} 
We identify science opportunities across the broad themes of dark matter detection, discovering the physics of inflation, precision cosmology for understanding both dark matter and dark energy, and the use of gravitational waves and high energy cosmic particles for these HEP science goals.

\begin{itemize}

\item \textbf{CMB probes of inflationary cosmology}: Completion of CMB-S4 will confirm or rule out large classes of models of cosmic inflation, pinpoint or limit the energy scale of inflation, and probe the dynamics of this ultra-high-energy sector.  

\item \textbf{CMB probes of dark radiation / light thermal relics}: CMB-S4 will discover or rule out new light relics during and after the QCD phase transition era, and uniquely probe a large parameter space for light particles Beyond the Standard Model of particle physics.

\item \textbf{CMB probes of neutrino physics}: CMB-S4 will constrain the neutrino mass scale, including potential sensitivity to the normal/inverted mass hierarchy, and explore other neutrino properties Beyond the Standard Model of particle physics.  Further advances in CMB physics will eventually be enabled by planning and executing CMB-S5.

\item \textbf{Precision cosmology with existing telescopes}: Cosmic surveys including DESI and the Rubin Observatory Legacy Survey of Space and Time (LSST) will provide strongly-constraining data sets for precision cosmology.  One percent measurements of the dark energy density to $z \sim 1.5$ will provide crucial tests of dynamical dark energy models.  Measurements of the overall matter power spectrum will enable tests for new dynamics in the early Universe and for precision constraints on the total mass of neutrinos.  By probing both the growth rate of density perturbations and the expansion history of the Universe, these experiments will provide important tests of whether cosmic acceleration is in fact due to a dark energy-like component of the Universe or if it instead requires a non-Einsteinian theory of gravity on cosmic scales.  Increased statistics for Milky Way dwarf galaxies and studies of smaller dark halos will provide information on dark matter masses and self-interactions and also provide new targets for indirect detection searches. For these surveys to reach their full potential, funding for innovative science analyses -- including simulations and cross-survey measurements -- will be needed. New data from other facilities will be needed as a complement to unlock the full constraining power of LSST, including follow-up observations of strong gravitational lenses, supernovae, and gravitational wave standard sirens, as well as measurements of spectroscopic redshifts for deep training samples of objects to enable precision photometric redshift measurements.

\item \textbf{Stage V spectroscopic probes of cosmic acceleration and inflationary cosmology:}
The development of a powerful new Stage V spectroscopic facility (Spec-S5) will allow precision  measurements of the cosmic expansion history to be extended to the limits of the modern Universe ($z \lesssim 5$), simultaneously providing new constraints on the inflationary epoch of cosmic acceleration.  Accessing high redshifts greatly expands the cosmic volume surveyed, enabling structures on the largest scales where inflationary signatures are expected to manifest to be explored. 
A Stage V spectroscopic facility also enables a search for new light relics and other Beyond the Standard Model particles including measurement of the sum of the masses of neutrinos, and probes of the physics of dark matter via a variety of precision studies of nearby galaxies and of stars in the Milky Way halo.

\item \textbf{New pathfinder experiments for Stage VI precision cosmology}: Pathfinder experiments will enable progress on powerful new probes of cosmology such as line intensity mapping methods in 21\,cm and millimiter wave molecular lines. Those new probes can extend matter power spectrum studies to still higher redshifts and greater volumes than Stage V spectroscopy can reach. For example, LuSEE-Night is a mission to the far side of the Moon being carried out in partnership between DOE and NASA as a pathfinder for a future Stage VI neutral hydrogen 21-cm experiment. 

\item \textbf{Direct detection of WIMP dark matter}: The next generation of direct dark matter experiments will be able to probe the WIMP paradigm down to and into the ``neutrino fog" over several orders of dark matter mass, using a combination of scaling up existing technologies and exploring new detector concepts.  

\item \textbf{Direct detection of axion dark matter}: A portfolio of axion dark matter search experiments enabled by new quantum sensing technologies will ``delve deep" in searches for the ultraweak QCD axion signal in most of its predicted band.  The Dark Matter New Initiatives (DMNI) has identified promising small projects to explore wide swaths of the parameter space.

\item \textbf{Cosmic probes of dark matter}: Cosmological and astrophysical observations, coupled with numerical simulations and theory, will widely probe and constrain the fundamental nature of dark matter over an expansive range of parameter space, some of which is inaccessible to terrestrial and laboratory searches. In the coming decade, Rubin LSST and DESI have the potential to revolutionize our understanding of microscopic properties of dark matter, such as mass, self-interactions, interactions with radiation, and quantum wave features.  On a longer time horizon, CMB-S4 and a planned stage V spectroscopy project will have access to rich dark matter physics as well. The inclusion of dark matter physics in the research programs of these experiments (together with necessary investments in theory and simulation) alongside studies of dark energy and inflation should be strongly supported. 

\item \textbf{Indirect detection of dark matter via high energy astrophysics}: New $> 100$~GeV gamma ray observatories including CTA will study dark matter via its annihilations or decays, with sensitivity to the thermal freeze-out benchmark annihilation rate up to masses approaching the 100 TeV scale, where model-agnostic limits on the early Universe cross section come into play.  Planned or proposed space missions will set stringent new bounds in the keV-GeV range, and new cosmic-ray detectors will perform the first dedicated searches for low-energy antinuclei as a background-free discovery channel. Planned and ongoing upgrades to existing observatories including IceCube and AugerPrime will provide unique probes of ultra-heavy dark matter. 

\item \textbf{Multi-pronged detection of low mass dark matter in dark portal scenarios}: A variety of new techniques including low-threshold microcalorimetry will be deployed to search for sub-GeV portal dark matter.  Concurrently, accelerator-based experiments will search for dark matter production and for new bosons which serve as the messengers between the dark and visible sectors.  The DMNI has identified promising small projects to advance these searches into motivated regions of parameter space.

\item \textbf{Direct detection of ultralow mass dark matter using cross-disciplinary technologies}: Engagement with the fields of quantum information science and atomic, molecular, and optical physics has brought new ideas and technologies to bear on searching for ultra-low-mass wave-like dark matter, which produce periodic perturbations detectable via ultra-sensitive instruments such as atomic clocks and interferometers.

\item \textbf{New experimental windows on fundamental physics}:  Pathfinder experiments including multi-messenger cosmology with gravitational waves, new techniques for precision measurements in astrometry, spectroscopy and photometry, and new sky surveys from space at radio frequencies below Earth ionospheric cut-off have potential to unlock hitherto unused probes of fundamental physics through cosmological measurements.

\item \textbf{Ultra-high-energy cosmic rays as a probe of new physics up to and beyond the GUT scale}: The new level of precision achieved by AugerPrime, IceCube, and next-generation experiments (GCOS, GRAND, POEMMA, and IceCube-Gen2 with its surface array) will provide opportunities to study non-perturbative QCD and physics Beyond the Standard Model using natural accelerators well beyond the reach of human-made ones. The  already-established anomalous muon production in extensive air showers (which has eluded explanation in models tuned to LHC data) guarantees that discoveries will be made.  

\item \textbf{Gravitational waves as a new probe of particle physics and cosmology}: Gravitational waves have emerged as a new and unique probe of dark sector physics, especially if the dark sector is only minimally coupled to the Standard Model via gravity.  In addition to searching for relic gravitational waves from early Universe phase transitions (including leptogenesis models or the QCD axion) and measuring the equation of state of quark matter at high density in neutron stars, current and future gravitational wave observatories promise a powerful new means of performing precision measurements of cosmic acceleration through a new class of standard distance indicators based on mergers of neutron stars and black holes. These large interferometers also have a large effective collection area, enabling searches for ultra-heavy dark matter.

\item \textbf{High-energy neutrinos as a new venue to study neutrino interaction and mixing}: The cosmic neutrinos detected by IceCube provide a new opportunity to learn neutrino physics at TeV-PeV energies. The upcoming IceCube Upgrade and future GeV cosmic neutrino experiments will measure the flavor composition and tau neutrino appearance with high precision. Future $>$100 PeV neutrino experiments may uncover cosmogenic neutrinos and probe BSM features at extreme energies.

\end{itemize}

\subsection{The Experimental Program}

The CF strategy detailed in this report encompasses a science-rich program with small, medium, and large experiments which are planned on both near-term and long-term timescales. The community consensus is that strong support for this entire program is essential to assure that HEP shares in and benefits from the upcoming era of discovery.  Groundbreaking precision measurements anticipated in the next two decades, as well as even more powerful experiments to follow, will reach key science discovery thresholds which should be comparable in their impact to the discovery of the Higgs boson. An overview of the CF program is provided in this section. Details about the projects and activities mentioned here are provided in later sections of this report. 

\begin{figure}[th!]
    \centering
    \includegraphics[width=\linewidth]{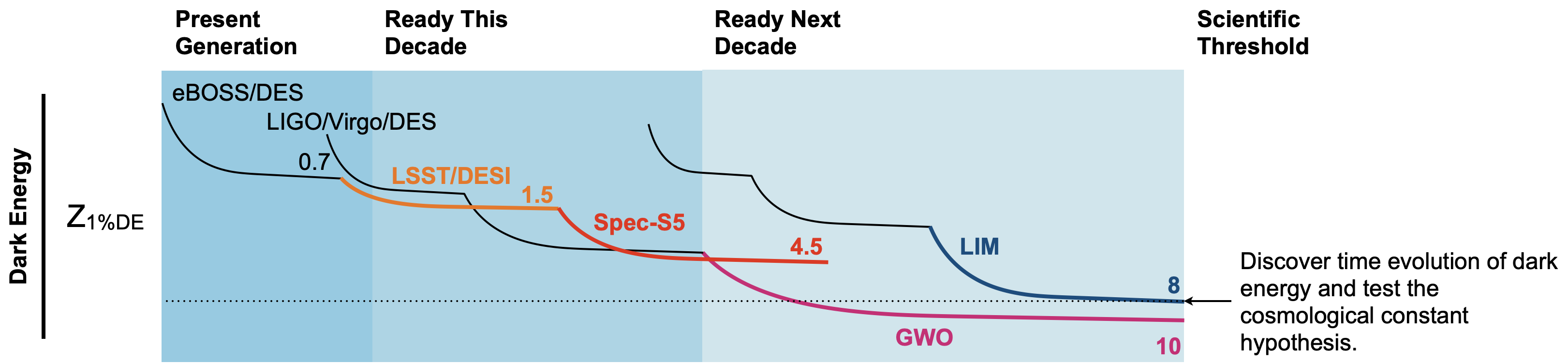}
    \includegraphics[width=\linewidth,trim={0 0 0 80},clip]{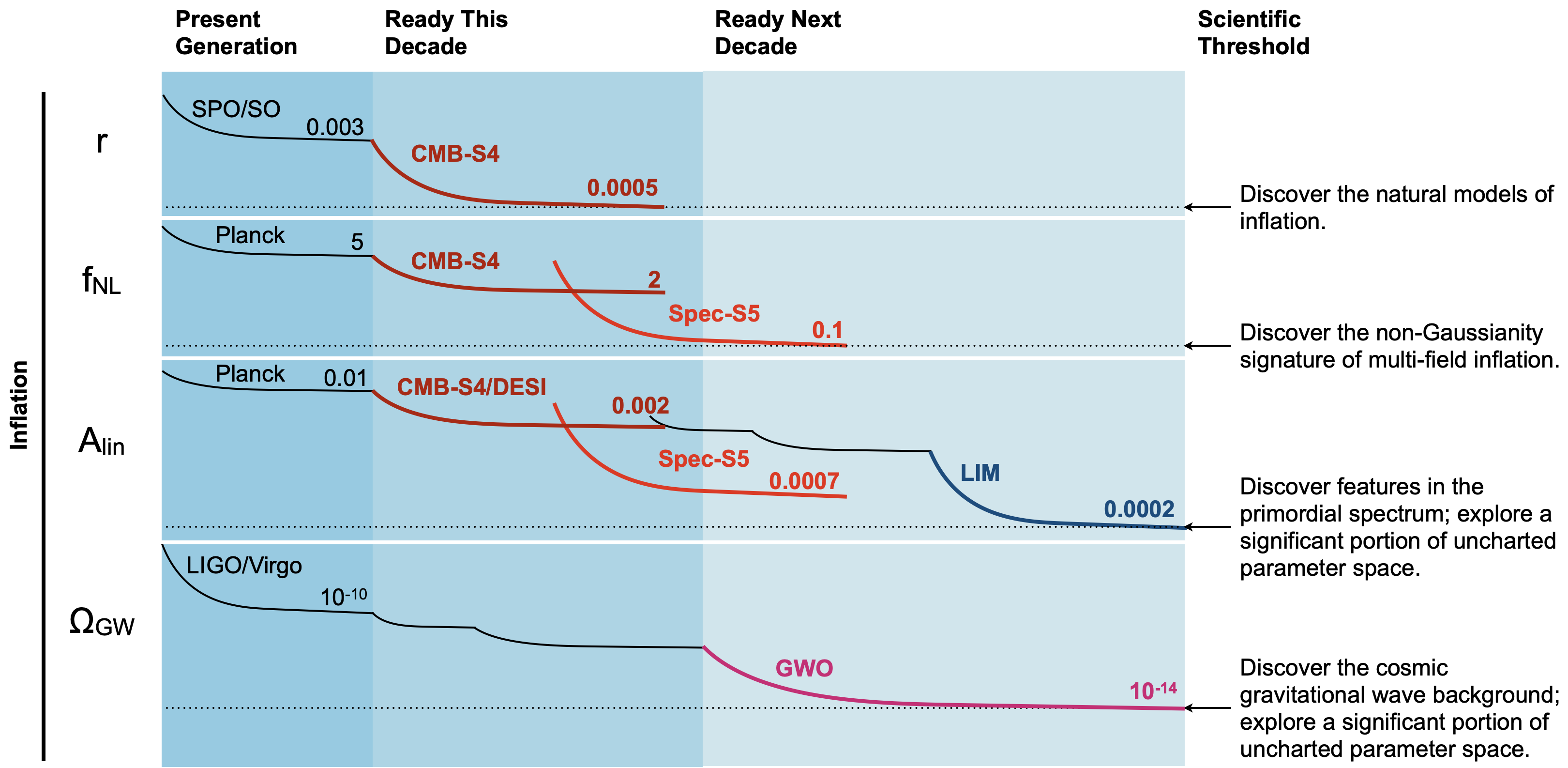}
    \includegraphics[width=\linewidth,trim={0 0 0 80},clip]{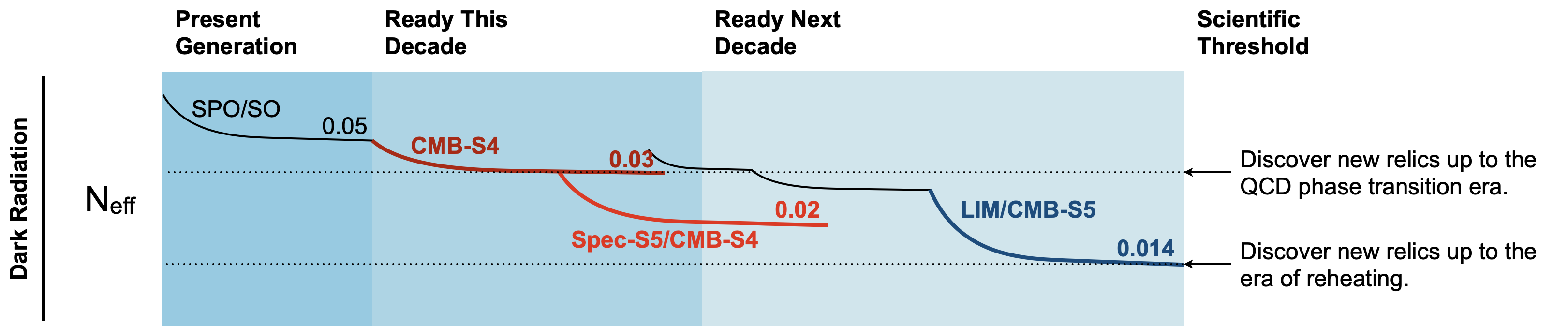}
    \includegraphics[width=\linewidth,trim={0 0 0 80},clip]{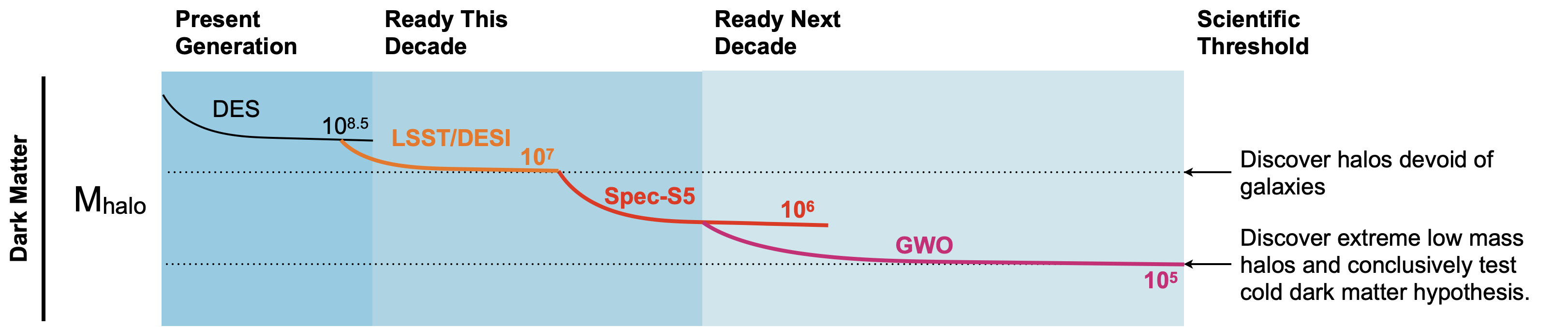}
    \caption{The landscape of cosmic survey projects and their discovery potential. For each major science topic, we identified a key measurement threshold for discovery. Bold lines show medium and large projects ready to produce science in this decade or the next.  In particular, CMB-S4 is a large project   currently with CD-0 and progressing to CD-1. Labeled thin lines represent the results from current surveys and unlabelled thin lines are pathfinders for the next generation of projects. Note that the  labels Spec-S5, GWO, LIM, and CMB-S5 do not represent specific projects. Instead, they represent categories of future experiments for which multiple proposed projects may currently exist in the community: a Stage V Spectroscopic Facility, next-generation gravitational wave observatory, 21cm/mm-wave line intensity mapping, and next-generation CMB, respectively. Adapted from the CF5 report \cite{Chang:2022lrw}.}
    \label{fig:CosmicSurveyProjects}
\end{figure}

\textbf{Cosmic surveys:} Figure~\ref{fig:CosmicSurveyProjects}
summarizes the cosmic survey projects and their discovery potential across multiple science areas of our program. 
For each fundamental question, we have identified a key measurement with a clear threshold for discovery. Together, the planned projects should deliver results spanning large swaths of cosmic frontier science, including studies of cosmic acceleration/dark energy, inflation, searches for indications of new particles (e.g., dark radiation), and exploration of dark matter properties. Cosmic surveys generally provide datasets that can be used for many purposes, addressing multiple HEP science drivers simultaneously. The figure shows a rough timeline for when each major project could come into play and gradually reach its full discovery power.   The large CMB-S4 project and the next large Stage V Spectroscopic Facility (Spec-S5) should play a key role in advances over the next decade and beyond. The figure also includes potential future surveys that will be enabled by near-term R\&D and small pathfinder initiatives; these projects would employ new technologies and techniques such as 21-cm and mm-wave Line Intensity Mapping (LIM), Gravitational Wave Observatories (GWO), and future broadband CMB imaging (CMB-S5). Note that  Spec-S5, GWO, LIM, and CMB-S5 are categories of future experiments for which multiple proposed projects may currently exist in the community (see the CF4--6 reports \cite{Annis:2022xgg,Chang:2022lrw,Flaugher:2022rob} and reference therein for details).  This planned sequence of experiments ensures that CF will have a continuous stream of data to support scientific breakthroughs without gaps during the construction of the largest facilities. 

In the Cosmic Frontier survey ecosystem  each type of program brings unique strengths to the portfolio, including galaxy shape measurements that can only come from imaging, precise and robust redshift measurements that require spectroscopy, and multi-messenger data that exploits new gravitational wave detection capabilities; by also spanning a range in wavelength  (e.g., Optical/IR vs. mm-wave vs. 21-cm), information from different redshift ranges can be accessed. While each survey will deliver groundbreaking results on its own, no single experiment can meet all of the discovery thresholds without complementary information from other parts of the program. The surveys complementary information provide both systematic rigor and complete coverage of the scientific goals. This diverse portfolio of multiple experiments is, thus, required to achieve the scientific reach of the program. 

As a result, the CF survey program is designed to leverage the strengths of multiple experiments simultaneously. One example of this complementarity is that to go beyond discovery of the energy scale of  inflation (which CMB-S4 could accomplish on its own) to conclusively probing its physics via precision measurements of the $f_{\rm NL}$ parameter, we would require measurements from a Stage V Spectroscopic Facility (Spec-S5). 
Experiments using well-developed techniques, like CMB-S4 and Spec-S5, are ready to be immediately implemented, driving forward our understanding of inflation and dark energy in this decade. 
Other science goals such as studying the time evolution of dark energy,  conclusively testing the cold dark matter hypothesis, searching for features in the primordial power spectrum, and discovering the cosmic gravitational wave background will require the next-generation of surveys for which pathfinders and R\&D work should be supported in this decade. These and many other science opportunities realized by the CF's survey ecosystem have been discussed in detail in the topical group reports \cite{Drlica-Wagner:2022lbd,Annis:2022xgg,Chang:2022lrw,Flaugher:2022rob}. In any funding scenario to be considered by the next P5, the Snowmass community consensus is that support for the CF cosmic surveys program should be at a level that enables discovery across the entire scope of the science program. 

Note that in Figure \ref{fig:CosmicSurveyProjects} the scientific thresholds for dark matter and dark energy are parametrized by the smallest dark matter halo mass ($M_{\mathrm{halo}}$) and the highest redshift ($z_{\mathrm{1\%DE}}$) out to which the contribution of dark energy to the  total mass-energy density of the Universe can be detected with percent-level precision, respectively. The discovery threshold for features in the primordial spectrum due to the physics of inflation is 
given in terms of the parameter $A_{\mathrm{lin}}$. The primordial gravitational wave background discovery threshold is represented by the parameter $\Omega_{\mathrm{GW}}$. For the inflation science parameters corresponding to the energy scale ($r$) and the non-Gaussianity features ($f_{\mathrm{NL}}$),  we have  $r<0.001$  as the threshold that would rule out models that naturally explain the spectral index $n_s$ within Planck characteristic scales while $f_{\mathrm{NL}}>1$ is the discovery threshold for models of simple single field inflation. 
These, in turn, translate into sensitivity thresholds for experiments: $\sigma(r) = 0.0005$ and 
$\sigma(f_\mathrm{NL}) = 0.1$, which are the values shown in Figure \ref{fig:CosmicSurveyProjects}. The rationale for new radiation in the early Universe is similar: the threshold for light thermal relics out to 100 GeV is $\Delta N_{\mathrm{eff}}=0.06$ and for exclusion of all thermal light relics it is $\Delta N_{\mathrm{eff}}=0.027$. These correspond to sensitivities of $\sigma(\Delta N_{\mathrm{eff}})=0.03$ and $\sigma(\Delta N_{\mathrm{eff}})=0.013$, respectively.

\begin{figure}[th!]
\begin{center}
\includegraphics[width=0.85\linewidth]{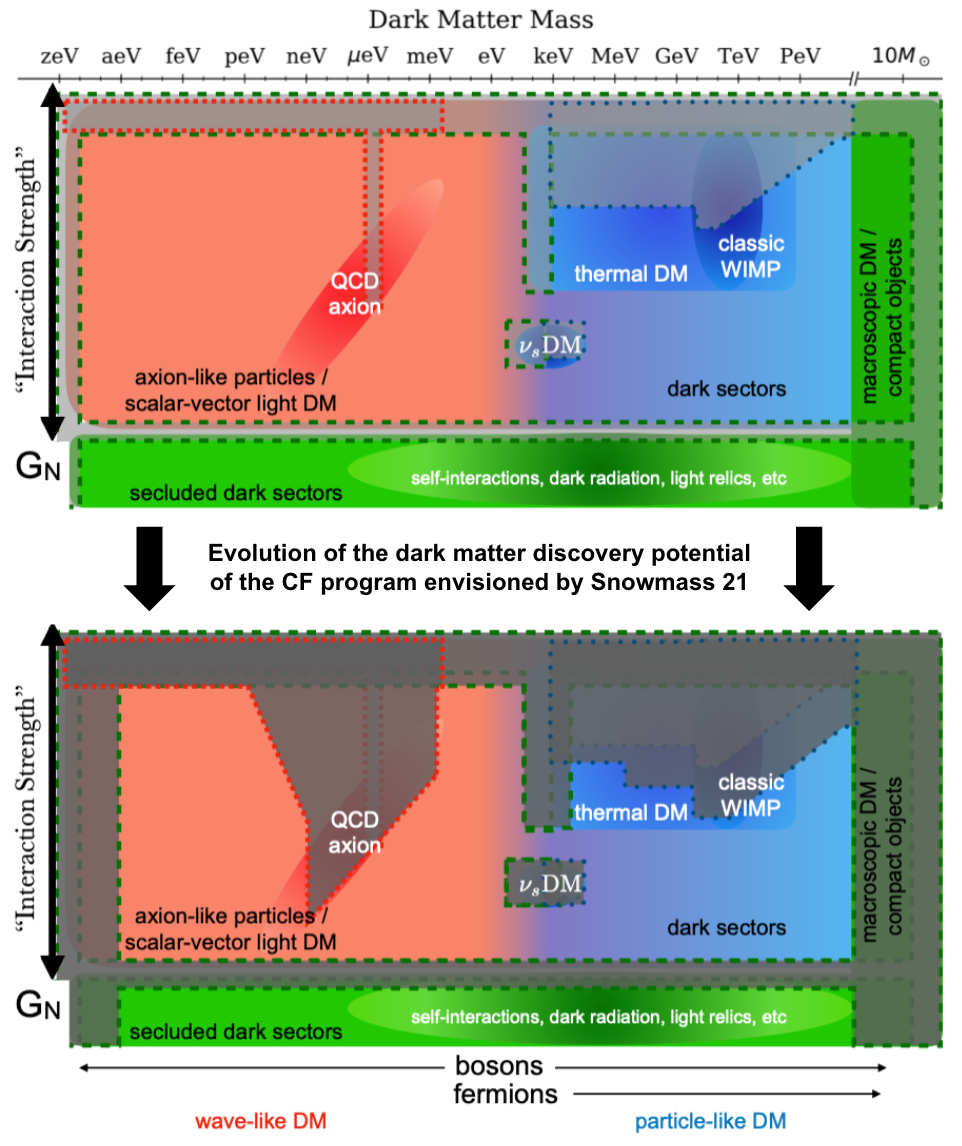}
\caption{Cartoon (not to scale) of the gains in sensitivity enabled by the search strategy outlined in this report (from light to dark gray shaded regions).  Regions where dark matter typically manifests as a wave are shaded in red, whereas regions where it manifests as individual particles are blue.  Broad, logarithmic gains in coverage are attainable through a range of newly developed techniques, which will require an ambitious and significantly expanded program of mid- and small-scale experiments as well as novel cosmic probes using existing and planned survey instruments (shaded green).  See Section~\ref{sec:dm} for details.}
\label{fig:dm_now_and_future}
\end{center}
\end{figure}

\textbf{Searching for dark matter:} Figure~\ref{fig:dm_now_and_future} shows two snapshots of the key dark matter discovery parameter space, which has greatly expanded since the last Snowmass.  They illustrate the broad coverage spanning many decades of dark matter parameters which will be achieved by the ambitious experimental program envisioned in this report, comprising both a broad portfolio of small/medium-scale direct and indirect detection experiments optimized for each decade in dark matter mass, as well as an expanded cosmic probes program to identify the properties of dark matter via cosmic surveys, which is highly synergistic with the other science these surveys will undertake. The top panel shows current experimental limits via the grey shaded regions, while the bottom panel shows the projected sensitivity over the same parameter space after completion of our program.  In many cases, utilizing new detector technologies will require cross-disciplinary collaborations to access the expertise, technology, and facilities of neighboring fields of study such as AMO, condensed matter physics, quantum information science, and gravitational physics.  Similarly, for indirect probes of dark matter scattering, annihilation, or decay in galactic halos, HEP provides a critical role in developing instrumentation for astroparticle observatories.  Understanding the astrophysical backgrounds to these searches will require engagement and collaboration with the broader non-HEP astronomy and astrophysics community.

Accomplishing the full scientific scope of the overall cosmic frontier program will require that robust support for ongoing, new, and future projects be well-matched by funding mechanisms for research groups to do science with data from those projects. This includes research support for  teams within experimental collaborations and beyond (e.g., theory, computing, simulation developments as well as detector R\&D studies that may not be part of an existing experiment in our portfolio today, but that will enable the next generation of experiments).  Even for ongoing experiments, a strong feature of the CF program is that coordinated analyses of data from multiple projects can yield results of greater impact than the sum of their parts. Thus, new mechanisms to support such cross-project initiatives will be an important component of the CF strategy for discovery.


\section{Dark Energy}
\label{sec:de}

The accelerating expansion of the modern universe over the last 5 billion years of cosmic history represents one of the most formidable scientific problems of our time. Its discovery was awarded the 2011 Nobel Prize in Physics.  Similarly to what has happened after the discovery of the Higgs boson by our colleagues in the energy frontier, this discovery has launched our field into a new era of precision measurement studies aiming to understand the physics of cosmic acceleration --- the explanation for which must clearly lie Beyond the Standard Model. 

Developing an understanding of the cause of cosmic acceleration in the modern universe should revolutionize our understanding of fundamental physics. Either we will discover a new form of energy whose source lies Beyond the Standard Model of particle physics -- some form of dark energy; or we will motivate a new theory of gravity beyond Einstein’s general relativity. If dark energy is the solution, it then accounts for two-thirds of the total present mass-energy density of the universe and permeates the empty space between galaxies. It hinders the growth of density perturbations such as those which can form clusters of galaxies, and makes distant astrophysical sources appear dimmer than expected by increasing their distances. Detailed measurements of both the expansion history of the universe and of the rate at which cosmic structures grow are central to understanding the nature of cosmic acceleration; in combination, they allow us to distinguish explanations associated with dark energy from those which require modifications to the theory of gravity.

Cosmology research in the twenty-first century is dominated by large cosmic survey experiments carried out by world-wide collaborations comprising hundreds of members from dozens of institutions. These large teams have developed specialized equipment and novel analysis tools which enable us to carry out observations of unprecedented scope and precision. For example, the Dark Energy Survey (DES) collaboration uses a 570-megapixel camera, the Dark Energy Camera (DECam), installed on a 4-meter telescope in Chile; the Dark Energy Spectroscopic Instrument (DESI) collaboration uses a system of 5000 robotic fiber positioners on a 4-meter telescope in Kitt Peak, Arizona; and the Rubin LSST Dark Energy Science Collaboration (DESC) will use a 3.2 gigapixel camera on a 8.5-meter telescope at the Vera Rubin Observatory in Chile.

As a result of this robust series of cosmic survey experiments, the community has advanced in leaps and bounds towards our goal of understanding cosmic acceleration via precision measurements of key cosmological parameters. 
These key parameters include: 
\begin{itemize}
    \item {\bf The equation of state parameter of dark energy ($w$).} --- If $w$ is exactly equal to $-1$ and does not change throughout the cosmic history, then dark energy acts as a cosmological constant -- the simplest, yet non-trivial, dark energy model. One of the key questions that our community wants to shed light upon is whether or not dark energy is consistent with this model.  
    Determining the dark energy equation of state with percent-level precision over a wide redshift range is key for addressing many fundamental questions related to dark energy and cosmic acceleration. Over the past twenty years of CF research cosmic surveys have made huge progress in this area. We now can measure $w$ well at $z<0.7$; observations so far are consistent with a cosmological constant model, but the time evolution of the cosmic equation of state is only poorly constrained so far.  
    Our next leap will be to study the signatures of dark energy to the limits of the modern universe ($z \lesssim 5$), which will provide novel insights into  the earliest times when it began to become significant.      
    \item {\bf The rate of expansion of the universe today ($H_0$).} --- This parameter is related to dark energy because it provides the low-redshift anchor to measurements of the cosmic expansion history. As a result, the value of $H_0$ has an impact on many constraints on $w$. $H_0$ has been  determined with good precision by experiments relying on data from two vastly different eras of the universe. Early-universe measurements (using for example CMB observations) would be expected to agree with values derived from galaxies, supernovae and other observables in the dark energy-dominated modern universe. A discrepancy between the two regimes would indicate either that the model is incomplete or that there are systematic issues in the distance determinations used to anchor sets of $H_0$ measurements.  
    At this point, measurements of $H_0$ with precision of 1-2\% have been made with a variety of techniques; the tension between determinations tied to the low-redshift versus high-redshift distance scales is starting to exceed the 5-sigma level. Several new measurements that will help uncover the origins of this discrepancy over the next two decades are described in this report.   
    \item {\bf The amplitude of fluctuations in the density of matter in the universe ($S_8$).} --- Precision measurements of the amplitude of density fluctuations in the universe and their growth over time are highly complementary to measurements of the cosmic expansion history (which should directly depend on the dark energy equation of state). For instance, if general relativity provides the correct description of gravity on large scales, the growth rate of density perturbations can be predicted directly from the expansion history; however, if cosmic acceleration occurs due to departures from Einsteinian gravity, this relationship is broken. 
    Early and late-universe measurements of the amplitude of the matter power spectrum, as measured by the $S_8$ parameter, are currently in tension. The significance of this tension is lower than for $H_0$, but the ongoing CF experimental program is rapidly improving the precision of measurements in this area, and percent-level determinations of $S_8$ in both the modern and CMB eras will soon be possible. 
\end{itemize}

\subsection{The Next Stages of Dark Energy Experiments} 

The dark energy component of the CF program has long been planned to proceed via a series of cosmic survey {\it Stages}, each greater in scope and precision measurement capability than its predecessor, as defined in the Dark Energy Task Force (DETF) report \cite{Albrecht:2006um}. The current state-of-the-art results are coming from DETF Stage III surveys (e.g., BOSS, eBOSS and DES); DETF Stage IV experiments are now underway (e.g., DESI) or soon to begin (LSST). These new experiments should provide percent-level constraints on $w$ (for models where its value does not evolve with time), driven primarily by measurements at $z<1.5$. 

Since the last Snowmass report, CMB experiments have become a significant part of the CF program. The next-generation project, CMB-S4, was ranked highly by the previous Snowmass/P5 process and is currently at the CD-0 stage. The data obtained by CMB experiments plays a vital role in exploring the fundamental HEP questions of cosmic acceleration in both the modern universe (dark energy) and the early universe (inflation). 

Looking forward, within the next two decades, our community seeks to build CMB-S4 
as well as to develop new Stage V and Stage VI projects. This suite of experiments will make precision measurements of $w$, $H_0$, and $S_8$ and other cosmological parameters with data sets that probe not only the current era in which dark energy is firmly dominant, but also detailed measurements going well beyond the transition between the dark matter- and dark energy- dominated eras and new tests of consistency with early universe data. 

In this report, we categorize the next stages of dark energy and cosmic acceleration experiments as follows:
\begin{itemize}
    \item Stage IV -- In this category we include those experiments which meet the DETF Stage IV definition  (such as DESI and Rubin LSST), as well as fourth-generation CMB experiments. Both classes of experiment complement each other, as the science of early and late-time cosmic acceleration are strongly intertwined. The Stage IV CMB project CMB-S4 is currently under development and should play a key role in cosmic frontier science. Stage IV cosmic survey experiments will provide precision constraints on dark energy, predominantly over the redshift range $z \sim 1.5$ (vs. $z \sim 0.7$ for Stage III). Novel analysis techniques that can take advantage of the leap in constraining power provided by Stage IV experiments, both when they are considered on their own and especially when analyzed in combination, should reduce systematic uncertainties and establish the true significance of the emerging tensions in  $H_0$ and $S_8$. 
    
    \item Stage V -- In this category we include those experiments that  would represent a substantial improvement  ($3-4\times$ or better) in capabilities or cosmological constraining power over the equivalent Stage IV experiment.
    These projects will take the current precision cosmology program to the next level, providing percent-level constraints on the contribution of dark energy to the total density of the universe at redshifts up to $z \sim 5$; enabling new probes of the physics of inflation, with the potential to rule out broad swaths of models and to measure inflationary energy scales; and exploring in depth possible explanations for the origins of the current $H_0$ and $S_8$ tensions. 
    
    \item Stage VI -- This category encompasses a future generation of dark energy experiments which can further increase constraining power over Stage V and begin to test cosmology in the `dark ages' of the Universe at $5 \lesssim z \lesssim 10$.  Such projects will require smaller pathfinder efforts now to develop key technologies and methods, which should enable them to become ready to reach project status $\sim 10$ years from now. These ambitious experiments will incorporate novel technologies and take advantage of new opportunities, driving the field to its next leap in dark energy precision studies. Higher-redshift experiments should be able to explore even larger scales than Stage V, expanding the sensitivity to inflationary signatures, as well as 
    providing sensitive tests for the presence of early dark energy. 
\end{itemize}

\subsection{A Stage V Spectroscopic Facility (Spec-S5)}

When Vera Rubin LSST, DESI, and CMB-S4 data are in-hand,
the greatest near-term opportunity to revolutionize our understanding of cosmic acceleration both in the modern universe and the inflationary epoch will be provided by a new Stage V Spectroscopic Facility (Spec-S5). Spec-S5 would feature a large telescope aperture, wide field of view, and high multiplexing. The technology for such a facility is at this point well-developed, building substantially on the legacy of DESI and Rubin Observatory.  A Spec-S5 can simultaneously provide a dense sample of galaxies at lower redshifts to provide robust measurements of the growth of structure at small scales, as well as a sample at redshifts $2<z<5$ to measure cosmic structure at the largest scales, spanning a sufficient volume to probe primordial non-Gaussianity from inflation and to search for features in the inflationary power spectrum on a broad range of scales, while also testing dark energy models in poorly-explored regimes,  determining the total neutrino mass, and strongly constraining the effective number of light relics. A Spec-S5 would also be able to probe the nature of dark matter using the kinematics of stars in the Milky Way halo and measurements of the matter power spectrum at small scales. 

\begin{figure}[t]
    \centering
    \includegraphics[width=0.9\linewidth]{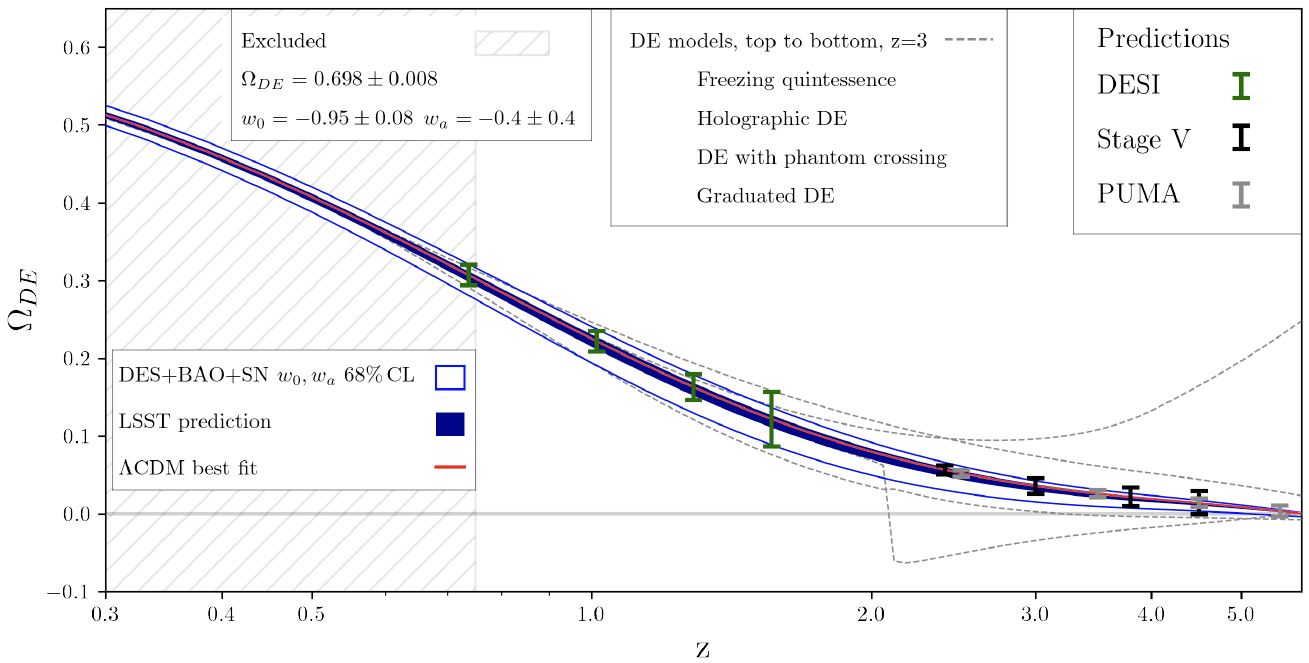}
    \caption{Projected sensitivity to the fractional dark energy density relative to closure density as a function of redshift.  Solid lines indicate extrapolations to high redshift of current conventional models fitted to data at low redshift, $0.1 < z < 0.75$.  Dashed lines indicate predictions of new models designed to address various tensions in existing datasets.  Data points with error bars indicate how the how the $2\%$-level sensitivity that should be obtained in ongoing and future high redshift surveys can directly constrain and discriminate between these models. From the CF4 report~\cite{Annis:2022xgg}.}
    \label{fig:de_models_projections}
\end{figure}

\begin{figure}[t]
    \centering
    \includegraphics[width=0.55\linewidth]{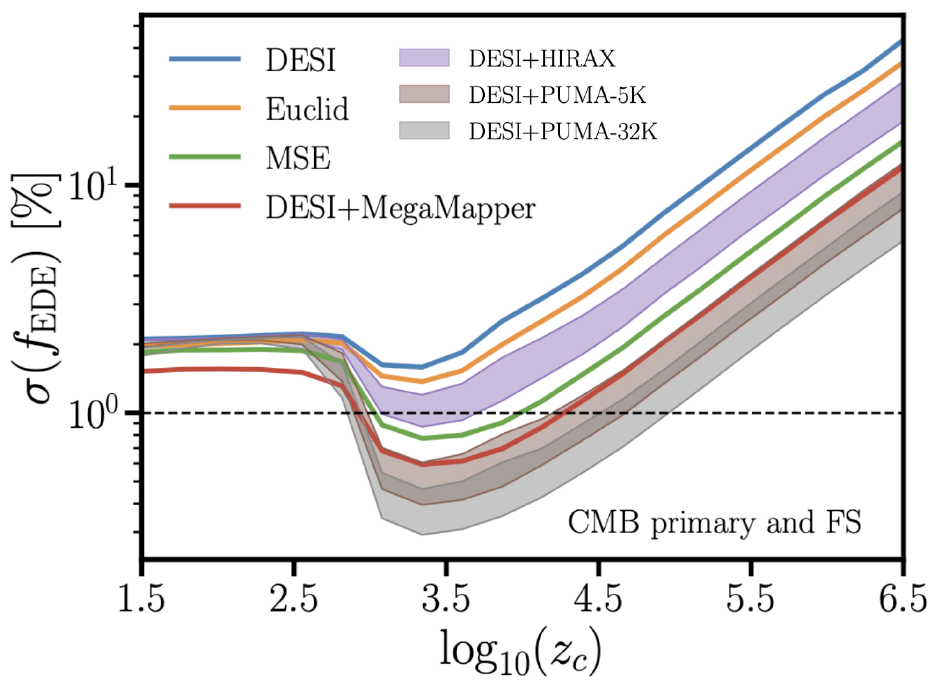}
    \caption{Projected sensitivity of current and next-generation high redshift cosmic surveys to the fractional dark energy density relative to closure density near the CMB recombination epoch.  The signal of early cosmic acceleration is an alteration to the growth of structure as measured as measured via the matter power spectrum of the large scale structure.  In particular, these probes will test models of early dark energy designed to explain the ``Hubble tension," a discrepancy between high redshift CMB measurements and low redshift optical survey measurements of the Hubble expansion rate which has increased in statistical significance.  From the CF4 report~\cite{Annis:2022xgg}.}
    \label{fig:de_early_dark_energy}
\end{figure}

Multiple options for facilities that would qualify as a Spec-S5 have been proposed \cite{Ferraro:2022cmj,Schlegel:2022vrv,DESI:2022lza,Sailer:2021yzm}. These projects, whose concepts are well advanced, are described in the CF4 report \cite{Annis:2022xgg}, which also outlines the key factors to be considered when evaluating candidate facilities. The key requirement is that a Spec-S5 should represent a significant advance over what is possible with Stage IV facilities, should enable transformational progress on all of the Stage V science goals, and open up new discovery space, including:
\begin{itemize}
    \item Measuring the contribution of dark energy to the total mass-energy density in the universe with uncertainties below 2\% of $\Omega_{\rm total}$ out to $z = 5$, redshifts at which dark energy becomes dynamically negligible in $w_0/w_a$ models, while simultaneously constraining modified gravity models over this range; cf. \autoref{fig:de_models_projections}.  Spec-S5 measurements will also provide sensitive tests for early dark energy (at $500 < z < 50000$), as illustrated in \autoref{fig:de_early_dark_energy}.
    \item Exploring the physical nature of the tensions in current measurements of $H_0$ (e.g., via BAO-based measurements spanning an unprecedented range of redshifts) and $S_8$ (with dense low-$z$ samples providing many cross-checks for systematics by enabling measurements using many separate galaxy populations as well as multiple lensing methods). 
    \item Testing for signatures of non-Gaussianity from inflation at a sufficient precision to be able to rule out all non fine-tuned multi-field models, while simultaneously searching for inflationary primordial features that signal the breaking of scale invariance and constraining contributions from dark radiation (sub-eV particles thermalized in the Universe). The Spec-S5 studies of inflation and dark radiation are points of synergy with CMB-S4 and will help the community to build a unified picture of the physics underlying both the early and late-time eras of cosmic acceleration.
\end{itemize}

A major strength of Spec-S5, in any of its proposed implementations, will be its ability to advance our understanding on multiple science fronts simultaneously, while also enhancing the science outcomes from other experiments in the CF portfolio. 
Cross-correlation measurements that combine Spec-S5 data with other surveys can be used to improve photometric redshifts for LSST as well as to unlock additional cosmological information (generally with reduced systematic uncertainties).  A Spec-S5 could also play an important role in obtaining training redshifts for LSST photo-z’s, improving cosmological constraints from this key dataset.  Construction of the instrument for a Spec-S5 facility in collaboration with other partners could build upon HEP strengths and experience with DESI. Realizing Spec-S5 in any of the funding scenarios that P5 might consider would play a key role in advancing the scientific goals of the CF community.

\subsection{Gravitational Waves}

The emergence of gravitational wave observatories (GWO) sensitive enough to detect sources at cosmic distances has revolutionized humanity's view of the universe. The HEP community, and the Cosmic Frontier sub-community in particular has been quick in realizing the implications of this new development and has been working towards incorporating GWO projects in our portfolio. Gravitational-wave standard sirens (merging compact object binary systems) allow measurement of the luminosity distance of the source and, together with redshift measurements, can be used to measure $H_0$ via the distance-redshift relation. Measurement of the Hubble parameter using standard sirens does not require a cosmic distance ladder and is model-independent: the absolute luminosity distance is directly calibrated by the theory of general relativity. 
Aiming to develop standard sirens into fully-fledged probes of dark energy, the DES collaboration launched a search and discovery program for the electromagnetic (EM) counterparts of events detected by the current GWO projects LIGO/Virgo. DES participated in the first multi-messenger discovery, of the  binary neutron star merger GW170817, and contributed to the first standard siren-based measurements of $H_0$. Approximately fifty additional multi-messenger binary neutron star observations would be needed to reach the required precision to weigh in the Hubble tension debate~\cite{Abdalla:2022yfr}.
The community is currently planning upgrades to the existing GWO network as well as a next-generation GWO network. One proposed next-generation project led by the US with participation from international partners is known as Cosmic Explorer. The CF community plans to use standard sirens from a Cosmic Explorer-like GWO as a powerful sample of independent distance indicators going all the way up to $z=10$. Combined with deep optical-to-near-infrared Stage IV and Stage V EM observatories, we can reach the required precision for Stage VI dark energy science. 
The potential impact of adding GWO data to our dark energy science program is seen in the top panel of Figure \ref{fig:CosmicSurveyProjects} based on forecasts performed by the community \cite{Abdalla:2022yfr,Kim:2022iud,Annis:2022xgg,Chang:2022lrw}.

Gravitational waves can probe dark energy and cosmic acceleration 
throughout the entire history of the universe with an observable that is novel and largely independent from the traditional observables employed thus far in the field. The next-generation GWO network will also have access to the binary black hole population when the universe was still in its infancy, to the equation of state of matter at neutron star cores at supranuclear densities and quark deconfinement phase transitions in hot merger remnants, and the ability to measure the properties of dark energy and dark matter, to stochastic gravitational-waves from early-universe phase transitions, and to the highly warped space-time in the strong-field and high-velocity limit of general relativity. 

The CF community plans to incorporate GWO into its portfolio of tools for discovery with a long term strategic vision. We will pursue EM counterparts of events detected by the growing GWO network while launching new pathfinder R\&D efforts to enable the HEP community to participate in the next-generation GWO project in a leading role.  The sensitivity goals for new detectors,  roughly 10 times better than the planned LIGO upgrade, require significantly larger facilities and a number of technological upgrades. Both are challenging requirements that the HEP community is well-equipped to meet. 

This is likely a once-in-a-century opportunity for the HEP community to make new breakthroughs in an entirely new class of experiments; we can take advantage of this new opportunity to advance on our scientific drivers at a much faster pace than previously anticipated. 

\subsection{Experimental Program \& Facilities: Cross-Experiment Strategic Plan}

In light of the opportunities for new breakthroughs in dark energy and cosmic acceleration over the next decades, the CF community has converged on a strategic plan with both near-term and longer-term action items, built upon a set of highly complementary projects/facilities and research initiatives. The top panel in Figure \ref{fig:CosmicSurveyProjects} summarizes how cosmic surveys will advance over Stages IV-VI using a key statistic: the  maximum redshift where we can measure the contribution of dark energy to the total mass-energy density of the universe with percent-level precision. We set the threshold for discovery at $z=8$, which will allow us to rigorously test the cosmological constant hypothesis far beyond the dark energy-dominated era of the universe.  Stage III has achieved this at $z<0.7$, and Stage IV should reach this goal for $z \lesssim 1.5$.  In contrast Stage V and VI experiments should extend our reach to  $z \sim 5$ and $z \lesssim$ 8-10, respectively. These measurements will allow us to definitively test whether simple cosmological constant or time-varying $w_0/w_a$ models are sufficient descriptions of cosmic acceleration, or if instead more exotic explanations with more complicated time evolution (such as those invoked to explain existing tensions, such as the models illustrated in \autoref{fig:de_models_projections}) will be needed.   In either scenario, Stages V and VI will lead to a breakthrough comparable to, if not surpassing, the discovery of dark energy itself. 

Figures \ref{fig:de_models_projections} and \ref{fig:de_early_dark_energy} show projections for the sensitivity of example proposed Stage V and Stage VI surveys relative to current Stage IV surveys for two key parameters: the energy density of dark energy ($\Omega_{\mathrm{DE}}$), and the contribution of early dark energy to the overall cosmological model ($f_{\mathrm{EDE}}$). The program devised by the CF community will enable us to test a wide variety of dark energy models that have been proposed to explain the observed tensions in $H_0$ and $S_8$. 

We summarize here the expected progression of projects and the key areas of support needed to address the questions of cosmic acceleration in both the modern and inflationary eras as described in the CF4~\cite{Annis:2022xgg}, CF5~\cite{Chang:2022lrw}, and CF6~\cite{Flaugher:2022rob} reports (noting that there is no significance to the ordering of items):

\textbf{Near-future efforts:} In the near term, it will be important to complete the Stage IV projects and fund the science efforts needed to take advantage of these powerful new facilities, including obtaining complementary data which will make them more powerful. Key needs to ensure the success of this program are to:
\begin{itemize}
    \item Build and operate CMB-S4.
    \item Begin operation of Rubin LSST.
    \item Continue operation of DESI (via a new DESI-II program) to constrain dark energy in new domains and as a step towards a Stage V spectroscopic facility (Spec-S5).
    \item Fund science analyses of the Stage IV datasets.
    \item Establish funding mechanisms for small and medium scale projects to follow up transients discovered by LSST and standard sirens discovered by gravitational wave observatory (GWO) facilities, as well as to enable photometric redshift training spectroscopy for LSST.
\end{itemize}

\textbf{Longer-term efforts:} In order to enable longer-term progress, it will be vital to establish a flagship Spec-S5 program, as well as to explore new opportunities which can enable other Stage V and even Stage VI experiments.  Key needs for this progress are to:
\begin{itemize}
    \item Build and operate Spec-S5.
    \item Support R\&D and pathfinder studies for a next-generation GWO.
    \item Support R\&D and pathfinder studies for a next-generation CMB experiment (at the Stage V or VI level).
    \item Support R\&D and small projects to develop technologies and methods that can enable a future surveys (e.g., LIM and CMB-S5) and new precision probes of cosmology (e.g., redshift drift measurements).
    \item Support operations of Rubin Observatory beyond the initially planned 10 year LSST survey; the best future use of this facility should be evaluated later this decade after the first LSST analyses have been done.
\end{itemize}

\textbf{Taking advantage of complementary experiments:} The experiments in our program will probe dark energy physics in a variety of different ways, enabling cross-checks for and control of systematic uncertainties to obtain robust and rigorous results. For example, standard siren measurements based on sources from gravitational wave observatories will be subject to very different systematics from observables such as galaxy clustering or supernovae that have been commonly employed in precision cosmology studies to date. Furthermore, different experiments provide complementary information about the universe that yields more powerful constraints on cosmology when analyzed in combination.  

However, such combined analyses present more challenges (particularly organizationally) than those which only involve one science collaboration.  Key needs to ensure the success of multi-experiment analyses are to:
\begin{itemize}
    \item Create funding streams and support for cross-survey analyses.
    \item Develop and support coordination between large facilities for optimized design, timely execution, and joint analyses. 
    \item Create data archive centers for long term storage where data from cosmological surveys is replicated, continuously available, and easily accessible with software and computing resources for developing joint constraints on dark energy, inflation and dark matter.
    \item Create and support development of a diverse set of simulated data sets that could be used in joint analyses.  Support should include common access to data and super-computing resources.
\end{itemize}

\begin{figure}[th!]
    \centering
    \includegraphics[width=\linewidth]{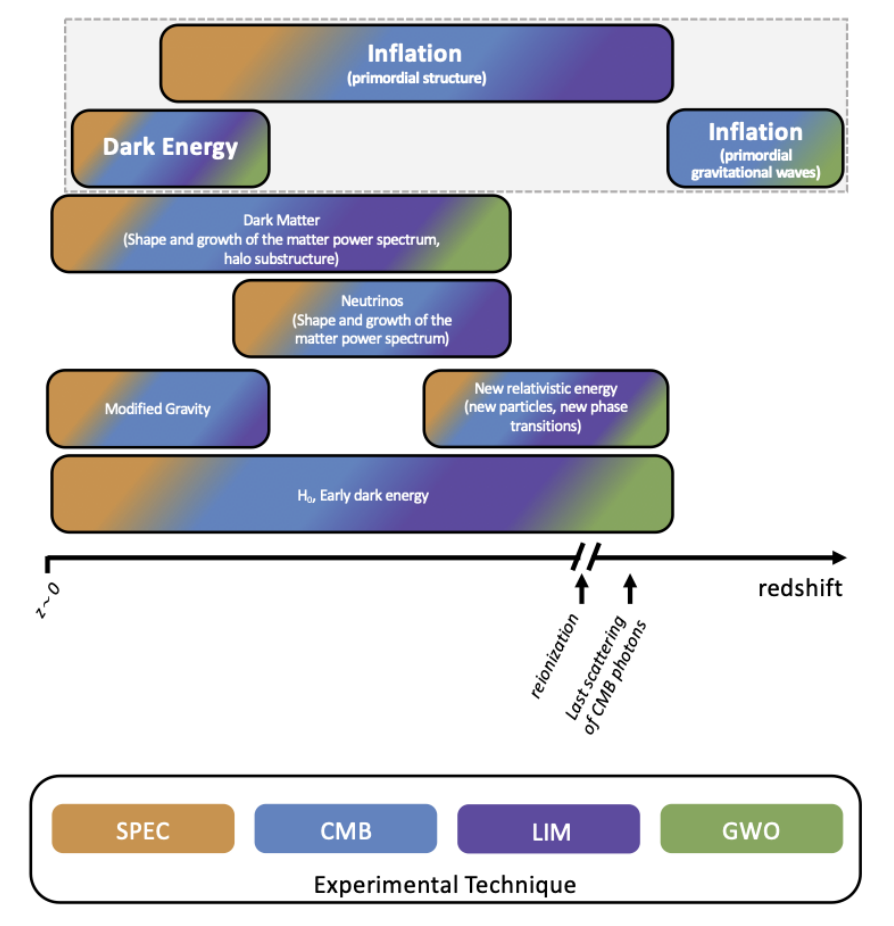}
    \caption{A high-level summary of the key scientific opportunities. The horizontal
extent of each box corresponds to the redshift-range of the tracer, while the coloring
indicates the experimental technique used to measure the signal. The dashed grey box
emphasizes dark energy and inflationary probes. From the CF6 report \cite{Flaugher:2022rob}.}
    \label{fig:allsurveyslandscape}
\end{figure}

The CF strategy for medium and large cosmic survey facilities is designed to take advantage of the interconnected nature of the science questions we aim to approach. The same facilities which can provide compelling tests of the nature of cosmic acceleration can also probe the physics of inflation, neutrino masses, dark matter, and dark radiation, as described in other sections of this report.  
This is illustrated in Figure \ref{fig:allsurveyslandscape}, taken from the CF6 report \cite{Flaugher:2022rob}, which depicts the different areas of cosmic frontier science addressed by wide-field spectroscopic facilities (such as Spec-S5), CMB experiments (such as CMB-S4), line intensity mapping (which may play a key role in Stage VI), and gravitational wave observatories. 

By incorporating both near-term and long-term projects at a range of scales, this experimental program should ensure that high-impact science results are continuously produced, avoiding lengthy gaps during project construction. In sum, the ecosystem of cosmic surveys proposed by the community should efficiently address many of the greatest scientific questions of our time. Accomplishing this program in its entirety will ensure that the HEP community will continue its path of leadership in studies of the accelerating universe.

\section{Inflation}
\label{sec:inflation}
There is strong evidence that the universe went through an extended period of accelerated expansion during its earliest moments.  The observed homogeneity and isotropy of the universe can be explained by 60 e-foldings of expansion of a much smaller, causally connected volume.  Furthermore, the fluctuations in the hot plasma that ultimately become the large scale structure (LSS) of the matter distribution are elegantly explained as originating from quantum fluctuations sourced throughout inflation and stretched to enormous distances by the exponential expansion.  Also, during this epoch, the universe effectively conducted the most extreme high energy physics experiment imaginable, and remnants of the dynamics of primordial ultra-high-energy collisions may remain imprinted as patterns in the density fluctuations of the resulting matter distribution.

\begin{figure}[t]
    \centering
    \includegraphics[width=0.6\linewidth]{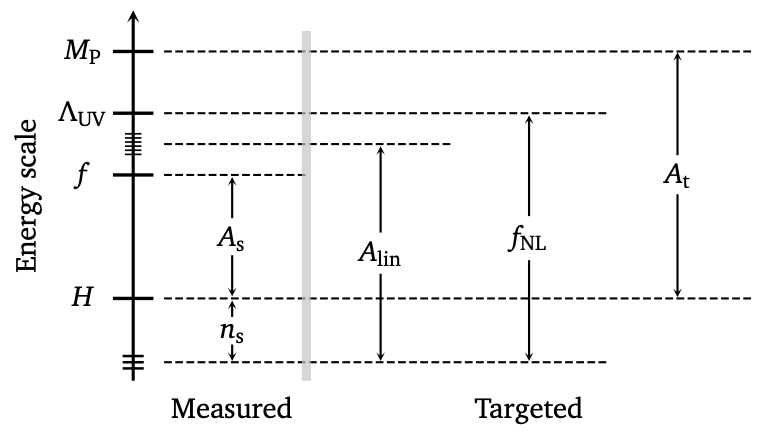}
    \caption{Cartoon of the various energy scales probed by studies of cosmic inflation.  The scalar amplitude $A_s$ and spectral index $n_s$ of the density fluctuations have already been well-measured via the temperature power spectrum of the CMB.  Small features of amplitude $A_\mathrm{lin}$ in this spectrum may indicate new BSM resonances in inflaton-inflaton scattering.  Primordial non-Gaussianities with amplitude $f_\mathrm{NL}$ measured via 3-point correlation functions can be used to discern inflaton scattering dynamics, e.g. soft scatters versus high-$p_t$ event topologies, at even higher energy scales $\Lambda_\mathrm{UV}$ and also distinguish between models with different particle content.  Finally, the tensor amplitude $A_t$ of the spectrum of gravitational waves emitted by the inflationary horizon can be measured via its imprint on the B-mode CMB polarization patterns and will provide an absolute determination of the Hubble scale $H$ of inflation relative to the Planck scale $M_\mathrm{P}$.  Interestingly, the high energy scales accessible by CMB measurements coincide with the scale of grand unification.  From Ref.~\cite{Achucarro:2022qrl}.}  
        \label{fig:inf_scales}
\end{figure}

During the inflationary epoch, cosmic acceleration is believed to be sourced by a non-vanishing vacuum energy density of a scalar inflaton field which, analogously to the Higgs field of the electroweak sector, is rolling to the minimum of its potential energy function in the midst of a cosmological phase transition.  At the end of this phase transition, the decay of the inflaton field ``reheats" the universe, releasing its potential energy to repopulate empty space with matter and radiation that become the photons, baryons, and dark matter that are observed today.  Quantum fluctuations of the inflaton field therefore produce a primordial spectrum of energy density fluctuations that source the large-scale structure of the spatial distribution of dark matter and ultimately the corresponding distribution of galaxies formed by the infall of baryons into the dark matter gravitational potential wells.  The spectrum of these fluctuations is measured statistically via the spatial two-point correlation function of fluctuations in the observed CMB temperature and in the LSS using angle and redshift-resolved measurements of the matter distribution.  A key prediction of a nearly scale-invariant power spectrum produced during the period of uniform exponential expansion has been verified by CMB data.  As described below, current and future experiments are now focused on determining the energy scale of inflation by measuring B-mode polarization patterns in the CMB, determining the dynamics and active degrees of freedom in the inflation sector by measuring 3-point correlation functions, and searching for resonances in the various power spectra due to new BSM scattering processes using the inflaton waves as the ultra-high-energy beam \cite{Achucarro:2022qrl}.  The various energy scales accessed by these various probes are summarized in figure~\ref{fig:inf_scales}.

Thus far, measurements of the CMB have provided the most powerful probes of inflation.  Polarization-sensitive experiments like CMB-S4 offer unique windows on the B-mode patterns, which reveal the absolute energy scale of inflation and will continue to provide precision measurements of the other key observables.  Upcoming cosmic surveys including DESI and an envisioned Spec-S5 project will provide much larger datasets, using precise redshift determinations to map the matter distribution in all 3 dimensions instead of being constrained to a single 2-dimensional slice at the surface of last scattering.  A new technique of line intensity mapping of the 3-d distributions of neutral hydrogen or other gas will possibly access the largest range of redshifts to provide the highest statistics measurements of the large scale structure.

\subsection{The Energy Scale of Inflation via CMB B-mode measurements}
\begin{figure}[t]
    \centering
    \includegraphics[width=0.9\linewidth]{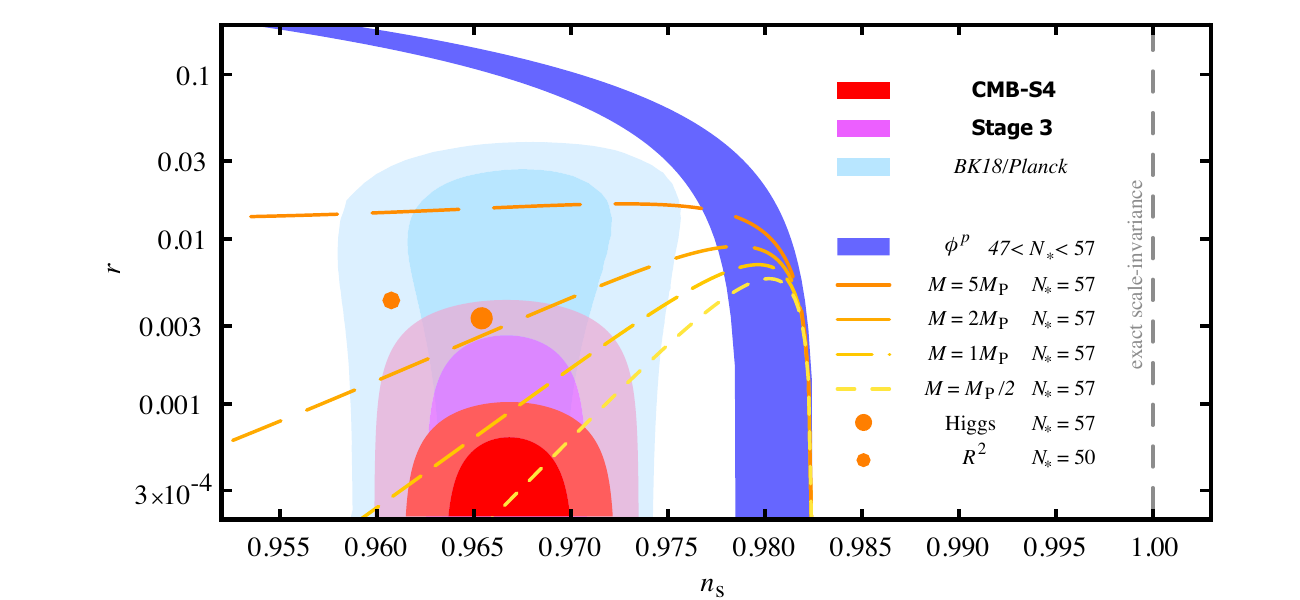}
    \caption{The tensor-to-scalar ratio $r$ plotted vs. the matter power spectral index $n_s$.  $r$ is observed from B-mode polarization patterns in the CMB sky, and is a measure of the absolute energy scale of cosmic inflation in the $10^{16}$~GeV range.  The light blue region represents current constraints, the purple region the projected constraints from ongoing experiments including the South Pole Observatory and the Simons Observatory, and the red region the targeted sensitivity of CMB-S4.  Dots and lines represent various models of inflation which will be tested.  Adapted from Ref.~\cite{Abazajian:2019eic}.
        \label{fig:cmb_bmodes}}
\end{figure}

In many of the simplest models of inflation, a stochastic background of primordial gravitational waves is emitted by quantum fluctuations at the inflationary horizon.  The amplitude of this spectrum directly provides the ratio of the scale of inflation to the Planck scale.  Propagation of these gravitational waves through the primordial plasma creates characteristic B-mode polarization patterns which constitute a primary science target for CMB experiments.  The sensitivity of CMB-S4 to the ratio $r$ of tensor power in B-modes to scalar power observed in the temperature fluctuation power spectrum is shown in Figure~\ref{fig:cmb_bmodes}, overlaid on predictions from some of the simplest models of the inflaton sector.  For example, a detection of $r \gtrsim 0.01$ by stage 3 experiments such as the South Pole Observatory or the Simons Observatory would provide evidence for an approximate shift symmetry in quantum gravity while a detection of $r \gtrsim 0.001$ by CMB-S4 would be expected in the simplest, natural models of high scale inflation.  A positive detection would identify a new fundamental energy scale of nature, possibly coincident with the putative scale of grand unification, and also provide the first evidence for the quantum nature of gravitational waves.  Conversely, a non-detection would point towards more complicated models of inflation, possibly occurring at lower energy scales.

The B-mode power of 10's of nK being probed by CMB-S4 is orders of magnitude below the scalar temperature anisotropy power, and so the signal is expected to be extremely faint. A key challenge is to accurately measure, model, and subtract polarized foregrounds due to thermal dust and synchotron emission within our Galaxy, as well as ``lensing B-modes" produced by weak gravitational lensing of the CMB.  Given the tremendous increase in volume and complexity of the upcoming datasets, continued development in the modeling and simulation of these foregrounds and of lensing, in computational analysis techniques, and in theoretical work will be necessary to fully realize the science potential of this project.

\begin{figure}[th!]
    \centering
    \includegraphics[width=0.75\linewidth]{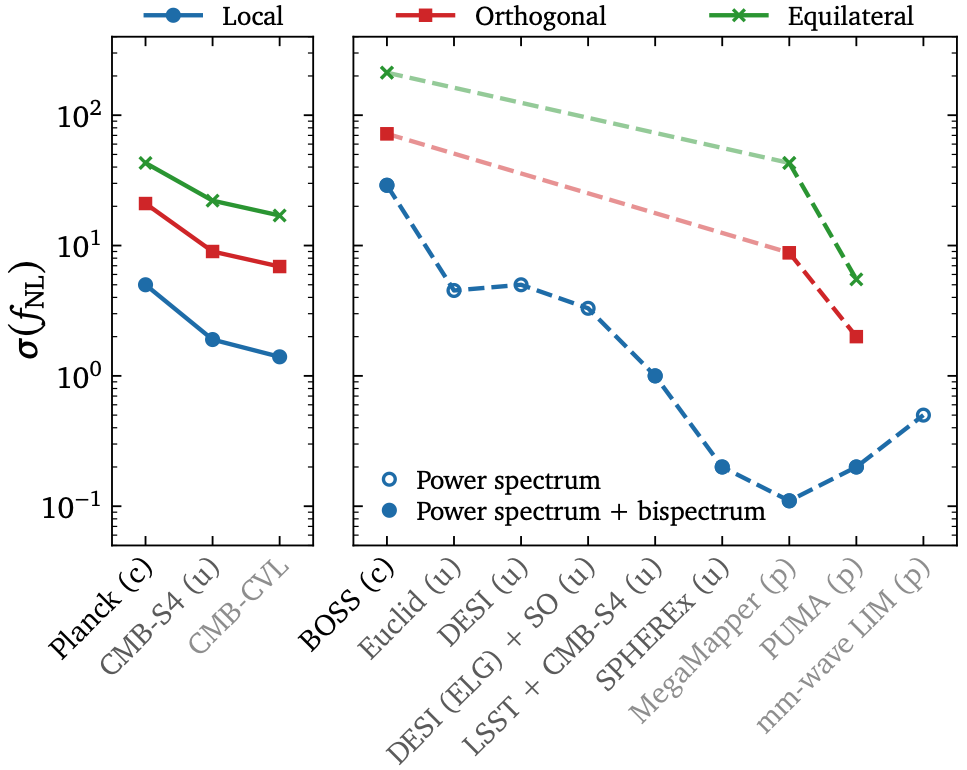}
    \caption{Resolution on the amplitude $f_\mathrm{NL}$ of the power bispectrum of the 3-point correlation function of matter density fluctuations for existing and planned future cosmic survey projects.  Three representative shapes (local, orthogonal, equilateral) are used for the 3-point function to measure scattering dynamics and field content in different event topologies. The sensitivity of CMB probes (left panel) to new dynamics in the inflaton sector should greatly improve with the advent of CMB-S4.  Planned future Stage V spectroscopic surveys (such as MegaMapper, as shown in the right panel) will provide larger data sets by incorporating redshift information to reconstruct the full, 3-dimensional matter distribution.  From Ref.~\cite{Achucarro:2022qrl}.}
    \label{fig:inf_fnl}
\end{figure}

\subsection{Primordial Non-Gaussianities:  Reconstructing the Physics of Inflation}
Current observations of the primordial density fluctuations are consistent with the Gaussian white noise statistics of a non-interacting gas of inflatons, freely propagating after being generated by quantum fluctuations at the horizon.   However, deviations from Gaussianity are necessarily present in even the simplest models of inflation.  For example, a generic inflaton potential energy function will contain higher-order terms describing inflaton self-interactions, and inflaton self scattering events would redistribute the spectral power.  Detecting primordial non-Gaussianities (PNG) in the inflaton gas as mirrored in the resulting density fluctuations would provide valuable information about the dynamics at play during the inflationary epoch.

It is helpful to think of the inflationary epoch as a collider experiment operated at extremely high energy densities.  Low-mass particles produced during inflation propagate long distances and are primarily responsible for the fluctuations in energy density produced after they decay and reheat the universe.  The scalar power spectrum measured in the CMB then characterizes the energy distribution and intensity of the particle beam.  To measure scattering and therefore interactions, it is easiest to probe higher-point statistics whose correlation functions encode the scattering dynamics as well as information about the different particle species participating in the interactions.  For example, just as the topology of hadronic jets produced in a collider experiment provide information about the underlying event, measurements of the 3-point correlation function of density fluctuations can provide statistical information about the fundamental interactions and microphysics within the inflaton sector as well as those of other BSM physics at energy scales accessible to the inflaton beam.  The 3-point functions measured on triangles formed at each distance scale probe scattering events at different center-of-mass energies, and the lengths of the triangle sides encode the kinematics of the events.  The collection of such events in the sky forms a high-statistics data set of independent scattering events at each energy.  Surveys which incorporate spectroscopic information can characterize a greater number of different modes on scales that evolve linearly, $N_{\rm modes}$, especially at higher redshifts where they encompass more volume. 

Primordial non-Gaussianity is typically parameterized by the strength of the interaction and by the momentum dependence or shape or topology of the signal. Depending on how strong correlations are between short and long distances, different information can be inferred about the dynamical processes responsible for the correlations. For example, self-interactions between inflaton perturbations generate correlations among modes of comparable wavelength and therefore the signal tends to peak around an equilateral triangle configuration. When there are strong correlations between short and long wavelength modes, a messenger particle is responsible for the long-range interaction. The resulting signal is called "local non-Gaussianity," and is a robust probe of new physics beyond the inflaton fluctuation and its self-interactions.  Said differently, equilateral non-Gaussianity probes interactions of the inflaton with itself and local non-Gaussianity probes light extra mediator fields (and also potentially can distinguish between single field and multi-field inflation models). While this is a useful general classification, there are many interesting scenarios in between these two extreme cases.

The primordial non-Gaussianity signal is generally characterized by an amplitude $f_{\rm NL}$; signatures of nonzero $f_{\rm NL}$ can manifest in both the observed power spectrum of density fluctuations as well as in their bispectrum — i.e., the Fourier transform of the 3-point function calculated for various triangle topologies.  Current and projected sensitivity to $f_\mathrm{NL}$ is shown in figure~\ref{fig:inf_fnl}.  The upcoming CMB-S4 experiment will provide much greater sensitivity to interesting models.  For example,
for local non-Gaussianity, models with models with $f_{\rm NL}^{\rm local} > 1$, which would produce signals detectable by CMB-S4, point to the existence of extra low-mass species active during or after inflation which are emitted colinearly in the primordial scattering events.  This measurement thus constitutes a powerful search for new BSM particles over a broad range of masses up to the beam energy given by the inflationary Hubble scale $H$.    For equilateral and orthogonal non-Gaussianity, models with $f_\mathrm{NL} > 1$ tend to favor scenarios with a strong breaking of boost symmetries of the inflationary background~(or small sound speed of the scalar perturbations).  

Future planned optical spectroscopic surveys will provide even greater sensitivity by incorporating redshift information to measure the full 3-dimensional matter distribution.  By measuring this distribution over the larger volumes available at higher redshifts, future datasets will increase the scattering event statistics by a factor of 10-100 relative to the $N_\mathrm{modes}$ achieved by DESI, an improvement factor measured by the "primordial figure-of-merit."   For example, the first factor of 10 would come from a Stage V Spectroscopic Facility targeting redshift $2 < z < 5$ galaxies using the LSST dataset.  The statistics could be improved by a factor of 100 using a relatively inexpensive line intensity mapping (LIM) technique to probe the same physics and by using 21 cm observations to map the large scale distributions of neutral hydrogen in the universe via 21 cm emissions.  
The long term goal is to explore the vast pristine volume before
the first stars ignited. 
LuSEE-Night pathfinder is focusing on the Dark Ages monopole and will make the first steps in understanding systematics to determine whether the full statistical power can be achieved.

\subsection{Spectral Features:  Uncovering New Dynamics in the Primordial Universe}
While the observed flat spectral index $n_s \simeq 0.97$ of the linear power spectrum of density fluctuations is consistent with a mild time dependence of the inflationary perturbations due to the small slope of slow-roll inflation models, more dramatic deviations from scale invariance can arise from (i) sharp or oscillatory features in the potential, (ii) power-law changes to the power spectrum on small scales, or (iii) new BSM resonances accessible at the energies of the inflaton waves.  Oscillatory features arise naturally from a variety of microscopic models of inflation. This periodic potential would then create a spectrum of harmonics on top of the predominant power law distribution.  

\begin{figure}[t]
    \centering
    \includegraphics[width=0.75\linewidth]{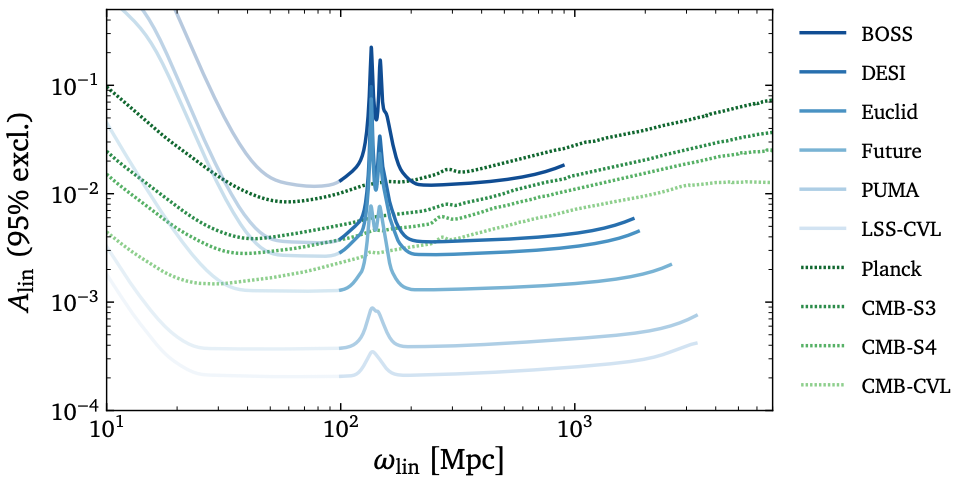}
    \caption{Current and projected sensitivity to small spectral features in the matter power spectrum for CMB and LSS probes.  Sensitivity limits on $A_{\rm lin}$, the fractional amplitude of the feature to the amplitude of the nearly scale-independent power spectrum, is plotted vs. the wavelength of the perturbation.  Searches for features in the matter power spectrum can also reveal new BSM dynamics at energy scales accessible at inflaton scattering energies. For example, new resonances at these ultra high energy scales may be revealed by bump-hunting searches.  More generally, spectral features which break scale invariance indicate new BSM dynamics.  Again, while CMB data currently provide the greatest sensitivity, a Stage V spectroscopic survey (exemplified by the ‘future’ limit depicted here) would provide more sensitive constraints at high $\omega_{\rm lin}$ by accessing 3-dimensional information over a large volume.  From Ref.~\cite{Achucarro:2022qrl}.
        \label{fig:inf_alin}}
\end{figure}

To date, cosmological data has constrained departures from a pure power law spectrum at the 1\% level; future CMB and LSS data will improve sensitivity by 1-2 orders of magnitude.  Figure~\ref{fig:inf_alin} shows the current and projected sensitivity to the amplitude of small spectral features for both CMB and LSS probes. Sensitivity to structure at smaller scales will utilize measurements of CMB spectral distortions and the stochastic gravitational wave background.  While scale-dependent features will also be targeted by 3-point function measurements of primordial non-Gaussianities, the linear 2-point function measurements allow a cleaner separation of primordial effects from late-time nonlinearities induced by gravitational clumping on smaller scales. 
Note that in Figure~\ref{fig:inf_alin}, the projected limits labeled ``Future'' correspond to what we expect from a Spec-S5 project. Also shown are several CMB projections, including CMB-S4. 

\subsection{CMB-S4}
CMB-S4 \cite{CMB-S4:2022ght} is a Stage IV cosmic microwave background project that plans to field multiple telescopes at the South Pole and in the Atacama desert, Chile. See the Snowmass 2021 White Paper~\cite{Chang:2022tzj} for a discussion on the broader experimental context. CMB-S4 has an enormously broad science case with key science goals including searching for primordial gravitational waves through the B-mode signal in the CMB as predicted from inflation ({\it detecting $r> 3\times 10^{-3}$ at $5\sigma$ or limiting $r\leq 10^{-3}$ at 95\% confidence if $r$ is very small}) and searching for the imprint of relic particles including neutrinos ({\it constraining $\Delta {N_{\rm eff}} \leq 0.06$ at 95\% confidence}).  CMB-S4 will also offer unique insights into dark energy and tests of gravity on large scales, find large samples of high-redshift galaxy clusters, elucidate the role of baryonic feedback on galaxy formation and evolution, open a window onto the transient Universe at millimeter wavelengths, and explore objects in the outer Solar System, among other investigations.

The current CMB-S4 instrument design calls for 500,000 polarization-sensitive bolometers that measure the sky at frequencies from 20--280 GHz.  The superconducting detectors will be dichroic and use the feedhorn-coupled orthomode transducer architecture. They will be read out using  time-domain multiplexed electronics and will be distributed between a set of telescopes at two sites: two 6-meter cross-Dragone reflecting telescopes in Chile, a 5-meter three-mirror-astigmatic reflecting telescope at the South Pole, and eighteen 0.5-meter refracting telescopes at the South Pole, grouped as triplets on six mounts, with each mount sharing a cryogenic system. This preliminary baseline design for CMB-S4 uses proven technology scaled up to much higher detector counts.  

The already-established integrated project office has addressed the technical challenges of the required scale-up with a detailed design and implementation plan that includes a full work-breakdown structure.  The project office also maintains a register of risks and a detailed cost estimation plan.  The preliminary, technology-limited project schedule contains nearly nine thousand milestones and catalogs the dependent relationships between them. The CMB-S4 project is moving forward and is ready to advance our compelling scientific program immediately.

\section{Dark Matter}
\label{sec:dm}

\begin{figure}[t]
\begin{center}
\includegraphics[width=\linewidth]{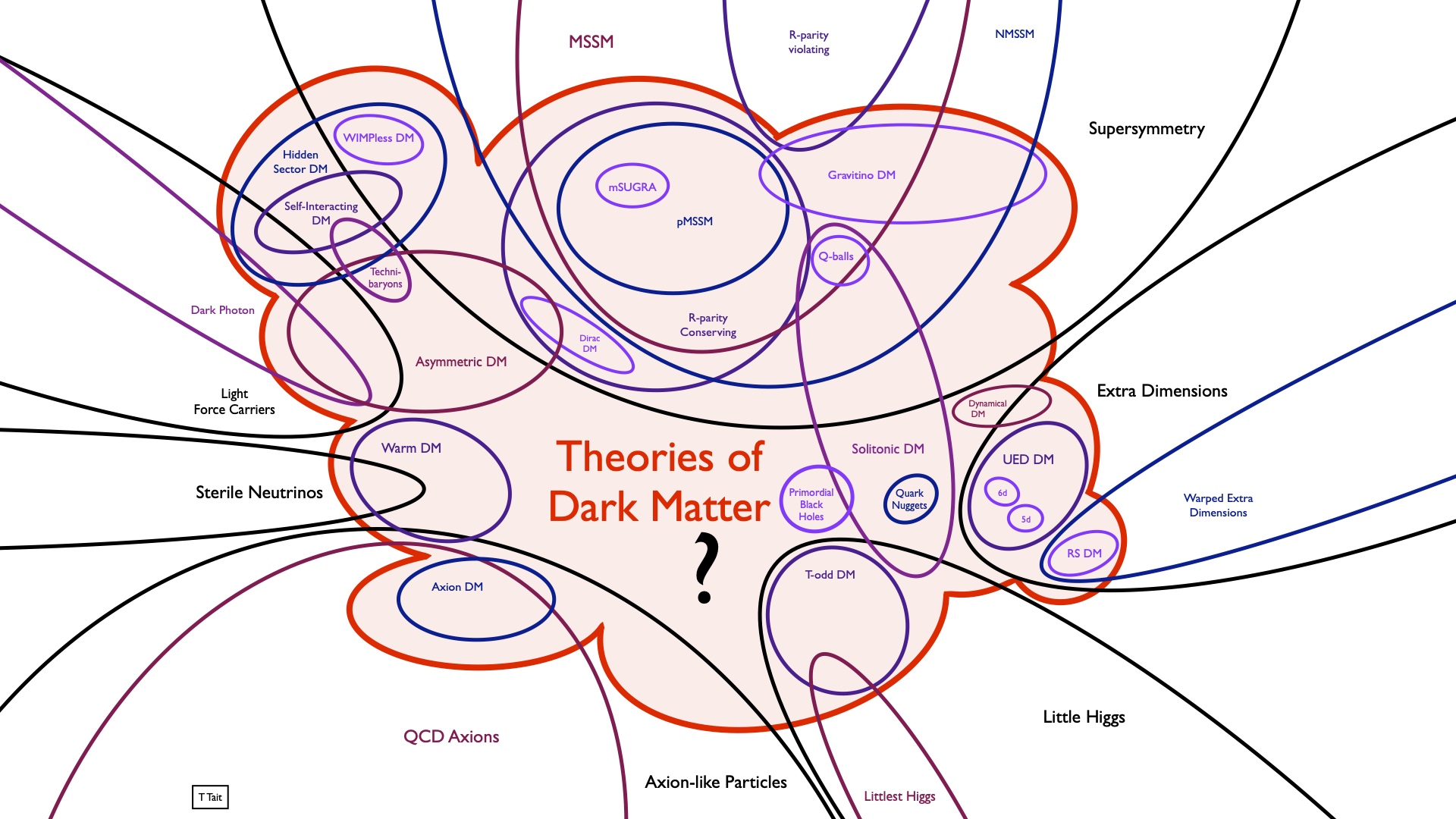}
\end{center}
\caption{Venn diagram of dark matter models, showing relationships between different ideas for the fundamental nature of dark matter.}
\label{fig:VennDM}
\end{figure}

Astronomical and cosmological observations of the gravitational influence of dark matter provide one of the strongest indications of new physics Beyond the Standard Model \cite{Bertone:2004pz}. 
Ascertaining the nature of these mysterious new particles, their interactions both with Standard Model particles and with themselves, and their cosmological origin is perhaps the grand challenge of this generation.  While the ambitious program to understand dark matter will require tools and techniques from across the HEP frontiers \cite{Bertone:2018krk,Boveia:2022syt}, the Cosmic Frontier is unique in that its experiments seek to detect and measure dark matter in its natural habitat -- the halo of our Galaxy, the halos of distant galaxies, and the large-scale structure of the Universe.  
While other frontiers may hope to discover new particles that could play the role of dark matter, only the Cosmic Frontier can  establish that a given discovery is, in fact, associated with the dark matter in the Universe.

Our understanding of the landscape of dark matter theories has evolved significantly in the past several years, as theoretical exploration has better defined the boundaries of what models are consistent with observations.  
As of Snowmass 2013 \cite{Feng:2014uja}, the classification of dark matter candidates was largely based on the particle physics features of the underlying models
(see Figure~\ref{fig:VennDM}).  Since then, focus has shifted toward exploring wide ranges of the possible phenomena in an effort to understand how well existing experimental searches cover the space of possibilities, and how new experimental opportunities provide sensitivity to regions of theory-space that are not captured by the current program \cite{Battaglieri:2017aum}.
There is great freedom to construct microphysical descriptions of dark matter, and a vast landscape of theoretical extensions of the Standard Model have been proposed.
These models range from very simple extensions of the Standard Model containing a single new particle to complex dark sectors containing multiple dark matter states, composite dark matter blobs, or even towers of dark particles that could constitute several different components of dark matter simultaneously \cite{Dienes:2022zbh}.

Cosmic observations currently delineate the allowed properties of dark matter. 
Dark matter must interact gravitationally, must be produced sufficiently non-relativistically that it clusters to form galaxies, must be sufficiently long-lived that it is present in the Universe today, and must not interact frequently with the Standard Model or it would produce signals that would have been observed. 
Going forward, cosmic probes offer unique opportunities to learn about the properties of dark matter via the influence that it exerts on ordinary matter \cite{Drlica-Wagner:2022lbd}. 
Measurements of the distribution of dark matter, including observables such as the matter power spectrum, the characteristics of dark matter halos (e.g., the mass spectrum, distribution, and density profiles), and the abundances of compact objects have become precise enough to place bounds on fundamental properties of dark matter such as particle mass and interaction strengths with itself and with the Standard Model.
Since the last Snowmass study, there have been a series of breakthroughs in modeling cosmic structure formation in novel dark matter scenarios \cite{Banerjee:2022qcb}.
These studies set the basis for quantifying astrophysical uncertainties and disentangling dark matter physics from baryon physics, a key step for extracting the Lagrangian parameters that describe a particular dark matter model from cosmic observations. 
We are at the threshold of definitively testing predictions of the cold dark matter paradigm on galactic and sub-galactic scales, and any observed deviation would revolutionize our understanding of the fundamental nature of dark matter.
In many cases, cosmic measurements of dark matter can be achieved by the same facilities that are being constructed to measure cosmic expansion assuming sufficient resources are provided \cite{Chakrabarti:2022cbu, Valluri:2022nrh, Mao:2022fyx, Dvorkin:2022bsc, DESI:2022lza}.  

The primordial cosmic density of dark matter, which can be measured precisely from CMB data \cite{Planck:2018vyg}, sets an important backdrop for any viable theory of dark matter.
Cosmic inflation typically erases any pre-existing density of dark matter, and so models must provide a physical `origin story' to explain its presence
in the Universe at the observed density.  
Two very plausible scenarios for how the dark matter could be produced include thermal freeze-out and misalignment production.
In freeze-out, the dark matter has substantial interactions with the Standard Model which causes it to be thermally populated in the primordial plasma until 
freezing out of thermal equilibrium when the Hubble expansion catches up to the rate at which the dark matter and Standard Model interconvert.
A dark matter mass and coupling of roughly weak scale sizes, assuming a standard cosmological history (e.g., no relevant additional physics Beyond the Standard Model
aside from the dark matter itself), results in approximately the correct relic abundance to match observations \cite{Lee:1977ua,Feng:2008ya}.  In misalignment production,
the bosonic dark matter field is initially displaced from the minimum of its potential.  Rolling to the minimum releases the initial vacuum energy as cold dark matter, whose
abundance depends on the physics of the dark matter potential as well as the initial displacement of the field.
While both of these mechanisms in principle connect the observed dark matter abundance to specific regions of parameter space, it is worth bearing in mind
that additional new physics beyond the dark matter itself can change the expectations for early cosmology, and thus the predicted
abundance as a function of the model parameters \cite{Gelmini:2006pq,Gelmini:2006pw,Hamdan:2017psw,Barman:2021ifu,Berger:2020maa,Dienes:2021woi,Howard:2021ohe}.

Experimental efforts over the past two decades have focused on experiments to search for
weakly interacting massive particles 
and QCD axions, both of which are strongly and independently motivated by other unsolved scientific mysteries.  In vanilla models of GeV-TeV mass dark matter associated with solving the electroweak gauge hierarchy problem \cite{Craig:2022uua}, WIMPs with electroweak couplings to Standard Model particles would be naturally populated at approximately the correct density by the freeze-out mechanism described above.   The QCD axion inevitably arises in PQ-type models in which the strong-CP problem (the otherwise inexplicable vanishing of the neutron electric dipole moment) is solved by promoting the strong CP-violating phase to a dynamical axion field \cite{Adams:2022pbo}.
The axion is produced via misalignment during the QCD phase transition, which causes the field roll to vanishing angle while releasing the original vacuum energy as axion dark matter.  Much progress has been made in experimental searches on both fronts.  Large second-generation WIMP detectors based on a variety of scattering targets, combined with indirect searches for high-energy annihilation products, have excluded $Z$-mediated couplings to Standard Model particles for WIMP masses up to $\sim$~TeV, and are now probing weaker couplings such as those mediated by Higgs boson exchange.  Concurrently, axion experiments have finally achieved sensitivity to the predicted QCD axion coupling strengths by deploying various microwave quantum sensing technologies and have begun to slowly scan the axion mass parameter space using resonant cavity detectors.  A cartoon encapsulating the current experimental situation is 
shown in the upper panel of Figure~\ref{fig:dm_now_and_future}.

While the WIMP and the QCD axion remain perhaps the most strongly motivated dark matter candidates, the breadth of dark matter searches has expanded enormously as new tools and sensing techniques have been identified that are able to probe large swaths of previously unexplored parameter space.  Many of the new techniques have been proposed by theorists identifying and studying models beyond WIMPs and QCD axions, which are compatible with all experimental and observational data and furthermore have their own
well-formed dark matter origin stories.

A particularly compelling class of new models are those of portal dark matter in which a hidden dark sector interacts weakly with the Standard Model sector through low dimension  Standard Model gauge singlet operators coupled to messenger particles.  
For example, in the case where the messenger is
a vector particle connected to the Standard Model via kinetic mixing with the hypercharge field strength portal, 
the result is a massive dark photon whose interactions are
predominantly proportional to electric charge scaled down by the kinetic mixing parameter $\varepsilon$.  Similarly, a dark Higgs can result from
a scalar or pseudo-scalar messenger whose potential contains mixing with the Standard Model Higgs mass portal. 
Such theories offer natural scenarios in which the dark matter may have much smaller interactions with the Standard Model, realizing the correct
freeze-out relic density for masses in the eV to GeV range, far below the weak scale.
Detecting the scattering of such low-mass dark matter requires much lower threshold calorimetric detectors sensitive to recoil energies in the range $\mu$eV -- keV.

A generic prediction of hidden-sector dark matter is that dark matter carries its own forces, which may produce novel signals in direct and indirect detection experiments. Such a force may also change cosmic structure formation, leading to signatures beyond the prevailing cold dark matter paradigm. For example, a dark force could operate at the most fundamental level with a range of $\mathcal{O}(10^{-12})~{\rm cm}$, but it would change the dark matter distribution within $\mathcal{O}(10^{22})~{\rm cm}$ in galactic halos, which could be detected in cosmic observations. In particular, dark matter self-interactions provide a compelling mechanism to produce diverse dark matter distributions ranging from tiny dwarf galaxies to huge galaxy clusters as inferred from observations~\cite{Tulin:2017ara,Adhikari:2022sbh}, which are a long-standing puzzle in cold dark matter~\cite{Bullock:2017xww}. In the next decade, Rubin LSST will measure the dark matter distribution at unprecedented small scales and hence it will provide a unique opportunity for probing interactions in the hidden sector~\cite{Drlica-Wagner:2019mwo,Drlica-Wagner:2022lbd}. Furthermore, detection of additional relativistic degrees of freedom by CMB-S4 would immediately imply the existence of a hidden sector~\cite{CMB-S4:2022ght}.  Eventually, 
CMB-S5 could directly measure the matter power spectrum on sufficiently small scales to distinguish between e.g. CDM, self-interacting DM, and fuzzy DM, independently from baryonic tracers.

The range of possible dark matter masses spans at least 50~orders of magnitude from ${\sim}10^{-22}$ to ${\sim}10^{28}$~eV for individual particles with the lower end constrained by the smallest dwarf galaxies and the upper end constrained by the Planck mass. While composite baryonic matter (e.g., MACHOs) has been ruled out by measurements of the CMB and Big Bang Nucleosynthesis, exotic macroscopic compact objects, including primordial black holes, are still allowed to comprise some fraction of the dark matter up to mass scales greater than a Solar mass (${\sim}10^{57}$~GeV). 
 
For direct detection experiments, where the dark matter interacts directly with the detector, different detection techniques are needed to match the energy scale and overall detector size to the expected energy/momentum transfer and event rates for the traversing dark matter for each regime of dark matter mass.
As such, the following broad classification of types of dark matter helps to distinguish between classes of experimental techniques that can be deployed as part of the``search wide" strategy to achieve comprehensive coverage of dark matter parameter space.
\begin{itemize}
    \item $10^{-22}$ eV$\lsim m_\chi\lsim 1$ eV: ``Ultralight" Dark Matter 
    \item 1 eV$\lsim m_\chi\lsim 1$ GeV: ``Light" particle Dark Matter 
    \item 1 GeV$\lsim m_\chi\lsim 100$ TeV: ``Heavy" particle Dark Matter 
    \item $m_\chi\gsim$ 100 TeV: ``Ultra-Heavy" Dark Matter (UHDM)
\end{itemize}
In the following sections, the status and prospects for the detection of dark matter in each category will be discussed.

\subsection{Waves of ultralight bosonic dark matter}

Provided that gravitational clumping of a single dark matter species is responsible for the formation of all galaxies and galactic substructures, the $\sim$kpc sizes of the dark matter halos hosting the smallest observed dwarf galaxies imply that the mode volume is $(\Delta x)^3 = 1/(\Delta p)^3 \gtrsim (\mathrm{kpc})^3$.  Estimating the momentum dispersion from the escape velocity from this local gravitational potential well, $\Delta p = m v_\mathrm{escape}$, a lower bound $m \gtrsim 10^{-22}$~eV can be derived.  Dark matter with smaller masses than this ``fuzzy dark matter" limit cannot be spatially localized within the volume of the smallest observed galaxies.  
In addition, for dark matter mass $m\leq 100$~eV, the mode occupation number must exceed unity in order for there to be enough dark matter to account for the local gravitational well that formed the dwarf galaxy.  One may therefore infer from these cosmic probes that dark matter in the mass range $10^{-22}$~eV -- $10^2$~eV must be bosonic since the Pauli exclusion principle would prevent sufficient fermionic dark matter from fitting inside these small galaxies.  At these low masses, the local Galactic dark matter takes the form of a gas of classical sine waves, each with large mode occupation number.    

\begin{figure}[th]
\begin{center}
\includegraphics[width=0.8\linewidth]{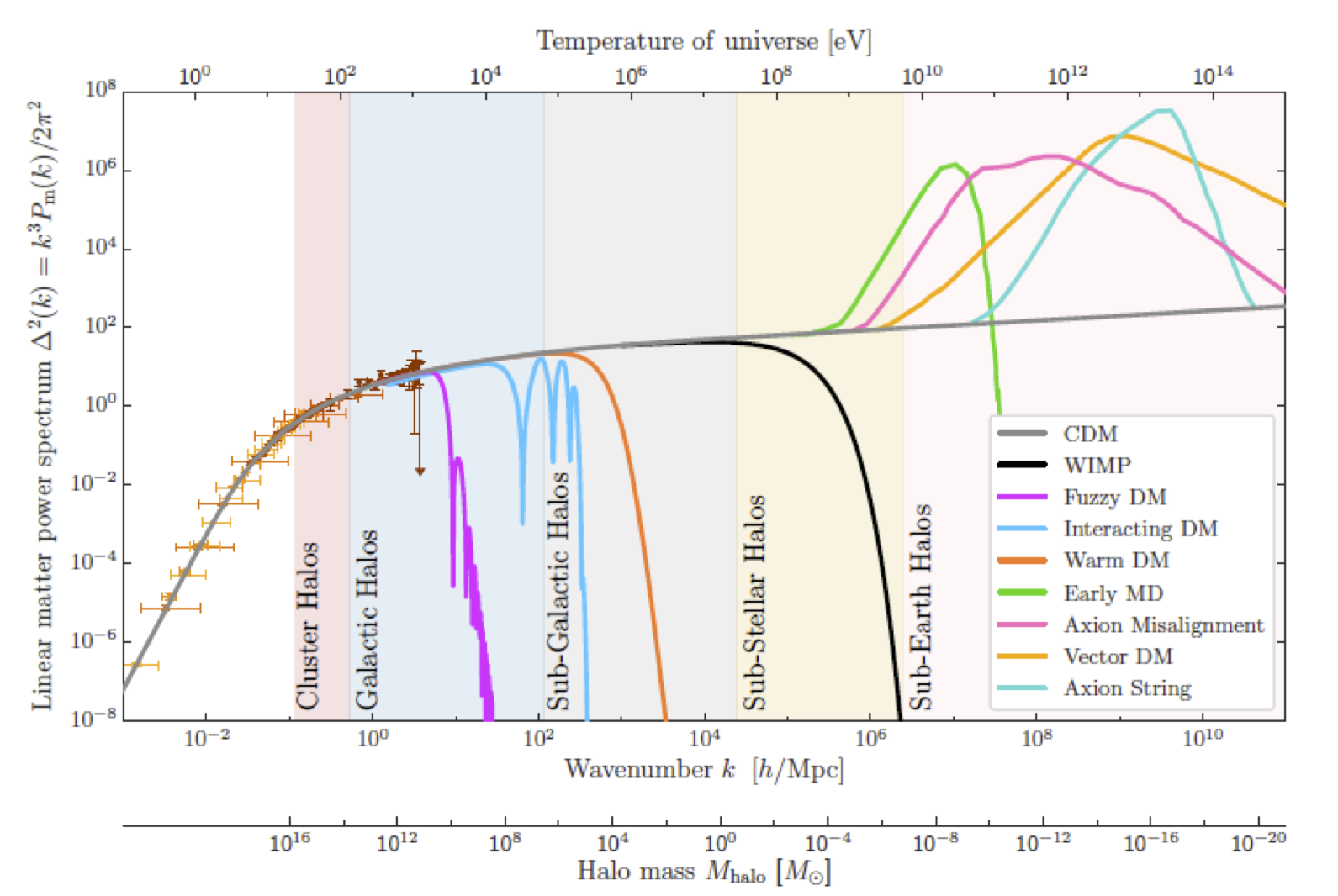}
\caption{Measurements of the distribution of dark matter provide information about its fundamental nature. The shape of the linear matter power spectrum (colored lines) and the properties of the dark matter halos that form at each scale (shaded regions) are sensitive probes of the dark matter mass and interactions. From Ref.~\cite{Bechtol:2022koa}.
\label{fig:dm_halo_size}}
\end{center}
\end{figure}

In the near future, cosmic surveys will gain sensitivity to even smaller dark matter halos through measurements of dwarf galaxies, stellar streams, strong lensing, and the Lyman-$\alpha$ forest \cite{Bechtol:2022koa}.
While the canonical collisionless cold dark matter model predicts that the mass spectrum of dark matter halos extends far below the mass at which halos are known to form luminous galaxies (Figure~\ref{fig:dm_halo_size}), ultra-light bosonic dark matter that has a mass close to the fuzzy dark matter limit ($\sim 10^{-22}$~eV) or fermionic dark matter that is close to the Pauli exclusion limit (${\sim} 100$~eV) would suppress the formation and density profiles of these small halos.
Thus, cosmic surveys such as Rubin LSST have an opportunity to discover signatures of these models through a measured absence of small halos, or further constrain the fundamental mass scale of dark matter if small dark matter halos are observed.
These are just two examples of dark matter models where fundamental dark matter properties alter the distribution of dark matter in the universe in testable ways. 
Similar arguments apply for a large range of specific dark matter particle models, as shown in Figure~\ref{fig:dm_halo_size}.

\begin{figure}[t]
\begin{center}
\includegraphics[width=0.8\linewidth]{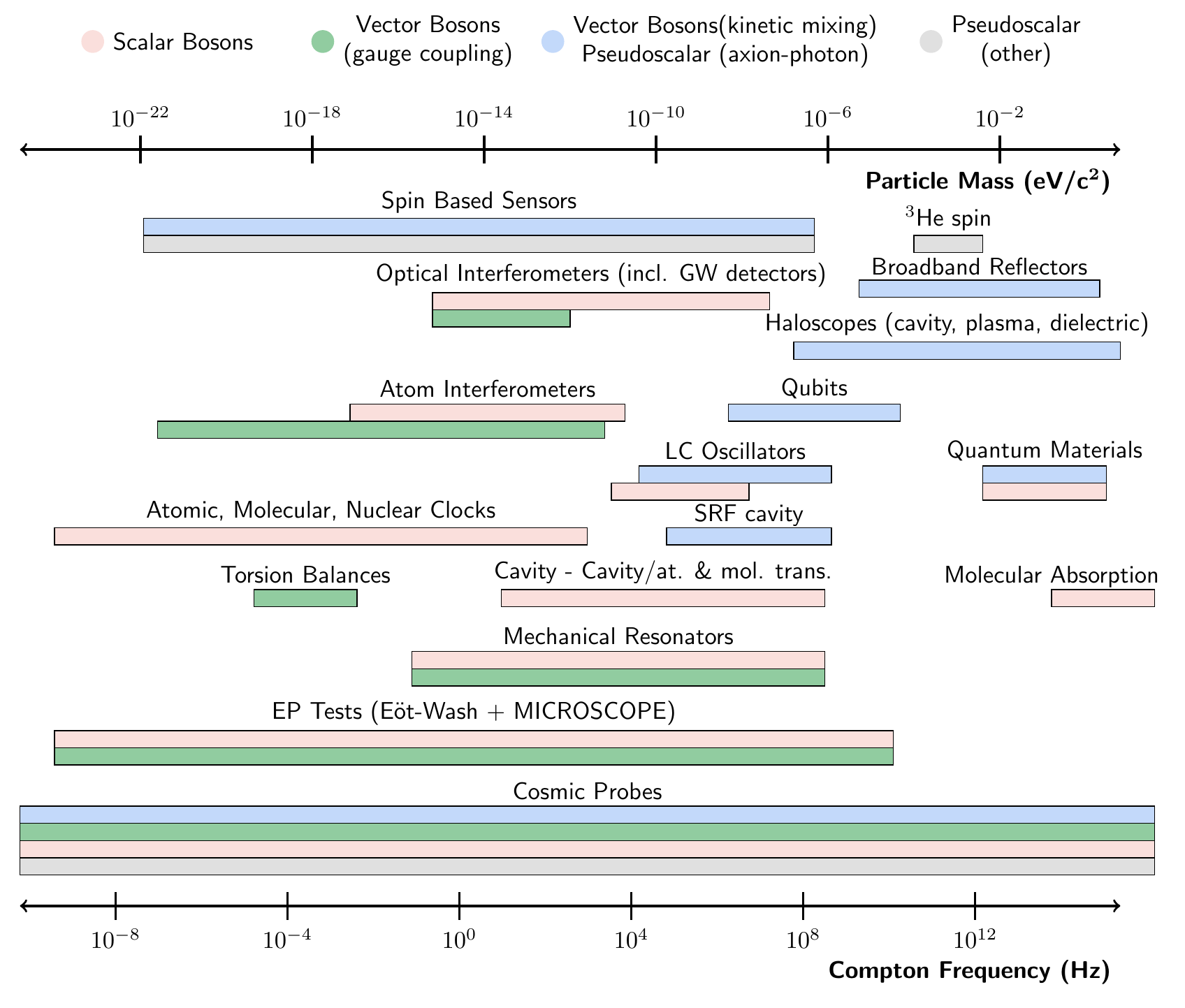}
\caption{Waves of ultra-light bosonic dark matter can be probed through a range of experimental and observational approaches including cavity experiments, quantum sensing technologies, AMO techniques, and cosmic probes.
Assembled from the CF2 and CF3 reports~\cite{Jaeckel:2022kwg,Drlica-Wagner:2022lbd}.}
\label{fig:dm_wave_techniques}
\end{center}
\end{figure}

Specialized terrestrial searches can look for modulating signals due to the oscillation of the dark matter sine wave on experimentally accessible time scales, as shown in figure~\ref{fig:dm_wave_techniques}.  For example, the lowest mass ($m\sim 10^{-22}$~eV) wave  may produce a detectable time-dependent classical force on test masses over an oscillation period of approximately one year.   Another class of theories couples scalar moduli dark matter (which are ubiquitous in string theory compactifications) to the sizes of the Standard Model parameters such that the oscillation of the dark matter wave mimics a modulation of the fundamental constants. Techniques from the field of AMO physics can immediately be brought to bear on this new detection challenge, where instead of a slow drift of the fundamental constants, one optimizes the apparatus and analysis to search for narrow line, single frequency perturbations.  Tools of choice include atomic clocks, atom interferometers, torsion balances, nuclear magnetic resonance, electric dipole moment experiments, and gravitational wave interferometers.  For example, atomic clocks have now reached fraction frequency precision of $10^{-18}$ and when deployed in space missions can search for perturbations from scalar dark matter; nuclear clocks under development will have even greater reach.

While much of this work remains in the domain of AMO physics, in some cases, capabilities of the HEP labs can greatly enhance the state of the art for these techniques beyond university-scale experiments in the process of adapting them to the dark matter challenge.  For example, the MAGIS pathfinder experiment will utilize an access shaft of the NuMI beamline at Fermilab to achieve a world record 100~m vertical baseline for atom interferometry and probe dark matter interactions with period equal to the free-fall time.  An upgraded km-scale interferometer is envisioned for the Sanford lab.  Accelerator lab expertise in large scale experimental deployments of beam pipes, vacuum systems, etc. may also prove useful in construction and operations of future large optical interferometers.  

\subsubsection{A high priority target -- the QCD axion ($10^{-12}$~eV to $10^{-2}$~eV)}

\begin{figure}[th!]
\begin{center}
\includegraphics[width=0.8\linewidth]{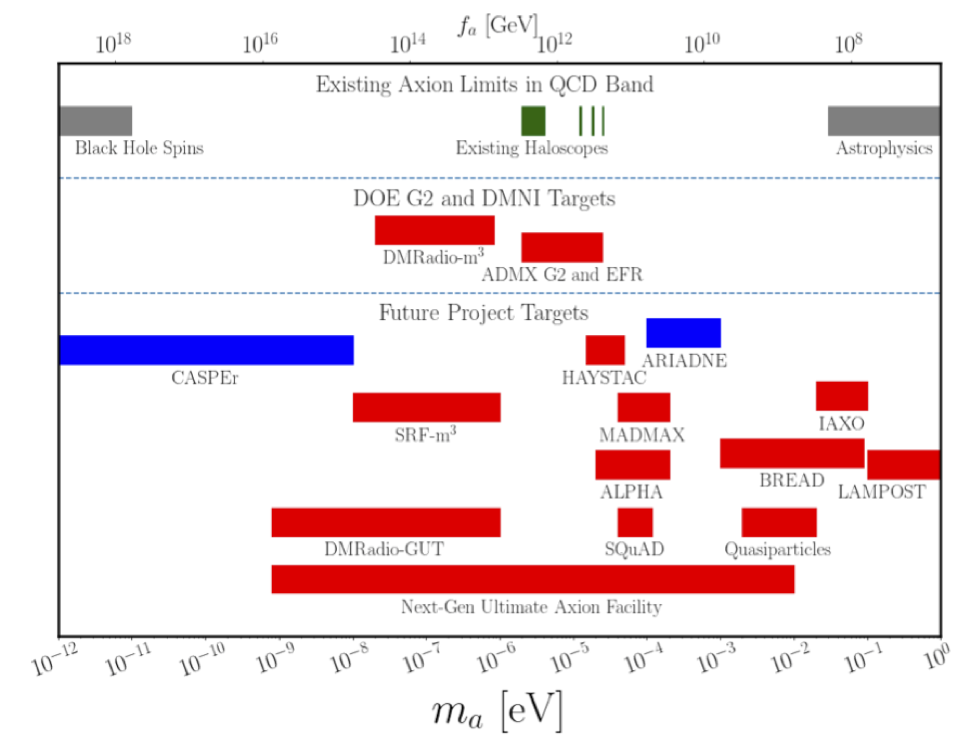}
\caption{A high priority target is the QCD axion which solves the strong CP problem as well as the origin of the dark matter.  The QCD axion model makes testable predictions for the interaction strengths as a function of mass, providing useful benchmarks.  This plot shows a suite of ongoing and future experiments which will test the QCD model by providing broad coverage of axion mass regions at the predicted coupling strengths to photons (red) and gluons (blue).  From the CF2 report~\cite{Jaeckel:2022kwg}.}
\label{fig:dm_axion_coverage}
\end{center}
\end{figure}

Perhaps the most significant technological development in the field of terrestrial dark matter searches since the last Snowmass is the demonstration by ADMX and HAYSTAC of experimental sensitivity to the invisible QCD axion.  The large improvement in signal/noise ratio achieved by these experiments was made possible by the deployment of new quantum sensing technologies including dilution refrigerators, quantum-limited amplifiers, and squeezed state receivers.  The ability to finally begin testing the QCD axion dark matter model has renewed both experimental and theoretical interest in the field of ultralight dark matter detection with several new techniques proposed to cover nearly the entire range of allowed QCD axion dark matter masses.  These experiments would concurrently provide sensitivity to large areas of model-independent parameter space for more general axion-like particles and dark photons.

The QCD axion emerges from the Peccei-Quinn (PQ) solution to the strong CP problem -- the 70-year-old mystery of why the neutron has a vanishing electric dipole moment despite being made up of a Fermi-sized collection of charged quarks.  The PQ model begins with a standard wine-bottle phase transition in which a global U(1) symmetry is broken at some high energy scale $f_a$ which results in a massive Higgs mode and a massless Goldstone mode called the axion.  This axion field is coupled to the gluons of QCD and thus identified with a dynamical CP-violating angle.  Later, during the QCD phase transition at energy $\Lambda_\mathrm{QCD}$, a gluon instanton condensate slightly tilts the wine-bottle potential, causing the axion vacuum expectation value to roll to its minimum potential energy.  At this true minimum, the CP-violating angle vanishes and thus zeroes out the neutron electric dipole moment.  The axion gains a tiny mass $m_a=\Lambda_\mathrm{QCD}^2/f_a$ via this see-saw mechanism between the QCD scale and the PQ scale.  Furthermore, during this vacuum relaxation process, the potential energy associated with the original CP-violating value of the misaligned vacuum axion field is released as ultracold dark matter.

The QCD axion search program is simplified in that the coupling to SM fields is proportional to the axion mass, both being proportional to $1/f_a$ as a result of the see-saw mechanism.  A couple of benchmark models (KSVZ and DFSZ) with order unity coupling coefficients determine a well-defined target region which cuts a diagonal band in a more general coupling vs.\ mass parameter space.  As shown in figure~\ref{fig:dm_axion_coverage}, coverage of axion parameter space can therefore be represented via a single parameter, the axion mass, with a lower bound of $\sim 10^{-12}$~eV coming from the requirement that $f_a<M_\mathrm{Planck}$ for the validity of the effective field theory. 
Cosmic probes of astrophysics in extreme environments further constrain the allowed QCD axion mass range to lie between $\sim 10^{-12}$--$10^{-2}$~eV \cite{Baryakhtar:2022hbu}.
For example, stronger axion-nucleon couplings predicted at larger axion masses would have caused supernova 1987a to cool more quickly than observed and would lead to anomalous cooling of neutron stars \cite{Carenza:2019pxu,Buschmann:2021juv}, while stronger axion-photon or axion-electron couplings would be inconsistent with stellar evolution as observed in globular clusters \cite{Ayala:2014pea,Dolan:2022kul,Capozzi:2020cbu}.
Detections of black holes with non-vanishing spins also provide constraints on axions with masses near $10^{-12}$~eV, since axions would populate a gravitationally bound state around black holes extracting energy and angular momentum through superradiance \cite{Brito:2015oca}.
Future observations of extreme astrophysical environments, along with improved theoretical understanding of the Standard Model physics at play in these environments, promise to further improve our sensitivity to axion physics.

The mass parameter space can further be split into pre-inflationary and post-inflationary axion production mechanisms.  The mass determines the characteristic cosmological time scale at which the initial axion potential energy in the tilted potential can be released as dark matter.  Because the photon and baryon densities are redshifting away, the dark matter production time determines its relative proportion of the overall cosmological energy density.  At a mass of around $10^{-5}$~eV, the release of an initial potential energy density of order $\Lambda_\mathrm{QCD}^4$ would produce the observed dark matter energy fraction.  At higher masses, the energy would be released too early while the baryon density is still too high, and so the dark matter would be underproduced.  However, the potential energy could be bound to networks of topological defects which would also prevent early release.  At lower masses, the dark matter is released too late and would be overproduced.  A solution to this overproduction problem is to posit that the initial PQ phase transition happened prior to cosmic inflation.  In this case, the earth could live in an inflated patch of the universe which had a downward statistical fluctuation in initial axionic potential energy density with correspondingly less energy released as dark matter.  

\subsubsection{Joint probes of axion and cosmic inflation parameter space}

The interplay of axion production histories with cosmic inflation enables complementary probes of the joint parameter space using both axion experiments and inflation experiments.  For example, if a low mass axion $m_a \leq 10^{-5}$~eV is discovered in a direct detection experiment, this implies that the PQ symmetry is broken at a high energy scale $f_a = \Lambda_\mathrm{QCD}^2/m_a \geq 10^{12}$~GeV.  The requirement that this symmetry breaking happened prior to cosmic inflation then forces the energy scale of inflation to be quite low, $E_I \leq f_a$.  Furthermore, even for very low axion masses for which $f_a \rightarrow M_\mathrm{Planck}$, radiation of massless axions from the inflationary horizon would produce isocurvature perturbations in the CMB which are strongly constrained by the Planck experiment.  In general, the detection of a low mass axion requires that $E_I \leq 10^{14}$~GeV to avoid overproducing axion isocurvature, but then concurrently, the Hawking radiation in gravitational waves from the horizon would also produce too little tensor mode power to be observed in the CMB B-mode spectrum.  CMB experiments should then instead focus on improving precision on isocurvature measurements to study this low energy scale inflation.  Conversely, if inflation occurs at a high energy scale $E_I \geq 10^{16}$~GeV, the higher amplitude gravitational waves may first be measured in CMB experiments via the B-mode polarization spectrum, and low mass axions would be excluded due to overproducing isocurvature.  The axion direct detection program should then focus on higher mass axions in which the PQ symmetry is broken after inflation so that no massless isocurvature modes can be produced. 

\subsubsection{Wave dark matter status and strategies for detection}

In the U.S., the DOE-supported ADMX-G2 resonant cavity experiment has excluded the QCD axion in the axion frequency range 645~MHz - 1.1~GHz (2.66-4.2~$\mu$eV) at even the more pessimistic DFSZ axion-photon coupling strength, and will continue to scan upward to 2~GHz using nearly quantum limited preamplifiers in their radio receiver.  The NSF-supported HAYSTAC experiment has deployed a squeezed state quantum receiver and demonstrated KSVZ coupling sensitivity in a narrow band near 4.1~GHz (17~$\mu$eV).   

Two new small-scale U.S. axion experiments are being developed within the DOE's DMNI program for small projects.  The first is ADMX-Extended Frequency Range (ADMX-EFR) which will increase the frequency scan rate by deploying and simultaneously operating multiple cavities within a single magnet bore, targeting higher frequencies 2-4~GHz.  The other is DMRadio-m$^3$ which will search over a wide range of lower frequencies and lower axion masses $<250$~MHz (1~$\mu$eV) using a toroidal antenna incorporated into a lumped element LC circuit to couple to the axion waves.  In each case, the total signal power will scale with the volume of the magnetized region which serves as the detector target for axion-photon conversion.  Decisions on new starts within the DMNI program are expected to happen before the P5 report.

Meanwhile, private foundation-funded R\&D continues on the CASPEr-Electric and CASPEr-Wind nuclear magnetic resonance experiments which aim to measure the dark matter-induced modulation of the nucleon electric dipole moment and the axion-nucleon spin coupling respectively.  The AC effects on the spin energy levels can be resonantly detected by tuning the spin precession frequency to match the axion wave frequency and observing the resulting tilt and precession of the spin ensemble polarization vector.  Both experiments are currently targeting lower frequencies/masses below 1~MHz and have a longer term goal of reaching sensitivity to the tiny signals predicted for the QCD axion.  This R\&D is critical as only by observing the axion in multiple, predicted detection channels can it be distinguished from more general dark matter models and proven to solve the strong-CP problem by inducing the defining nucleon electric dipole moment.

Various quantum sensing technologies are being transferred to the dark matter axion search.  For example, the low frequency experiments are limited by thermal noise.  In this case, single quantum mixers used can be used to transduce signal photons to higher frequency ranges where quantum limited amplification is available.  The low level of quantum noise relative to thermal noise then allows the resonant search to occur at constant ratio of signal to thermal noise over many Lorentzian linewidths to recover some of the speed advantage of a broadband search -- the same strategy as was used in the Weber bar technique originally used to search for gravitational waves.  Another strategy is to utilize an energized superconducting RF cavity (SRF) and use the AC magnetic field instead of a DC magnetic field to upconvert the low frequency axion wave into an RF signal photon.

At signal frequencies higher than those being probed by ADMX, the experimental challenge is that the size of the resonant cavities must shrink with increasing frequency to match the smaller wavelength of the signal photons, and so the signal power shrinks according to this rapidly decreasing target volume.  While ADMX-EFR will pack many cavities into a single magnet to try to recover the lost volume, other ideas like ALPHA, the Axion Longitudinal Plasma Haloscope will try to create a large volume metallic metamaterial whose plasma frequency can be tuned into resonance with the dark matter wave.  Alternatively, instead of attempting to maintain or increase the signal level, another strategy is to drastically reduce noise by using photon counting detectors which can have arbitrarily low dark count rates. The resulting noise from background count fluctuations can be far below the zero-point noise that is incurred by the currently employed frequency-resolved power readout techniques.  New, ultra-low noise single microwave photon detection techniques based on superconducting qubits and on Rydberg atoms are being transferred from neighboring fields of quantum computing and AMO.  The Superconducting Qubit Advantage for Dark Matter (SQuAD) prototype has demonstrated a factor of nearly 40 noise reduction relative to zero-point noise which, combined with high-Q dielectric cavities could enable QCD axion searches up to 20~GHz.  R\&D is also underway with the Rydberg Atoms at Yale (RAY) experiment to perform single photon counting at even higher microwave frequencies.

Finally, broadband techniques are also being developed which use large impedance mismatches at metallic or dielectric plates immersed in magnetic field to create axion to photon transition radiation.  This radiation can be focused, even by the plate geometry itself, onto low noise single photon detectors.  In Germany, a large R\&D effort is underway to develop the MADMAX experiment which seeks to deploy a set of aligned plates to create multiple scattering targets within a large, high-field dipole magnet bore and read out signal power coherently using HEMT amplification.  In the U.S., smaller-scale experiments such as BREAD and LAMPPOST are performing R\&D funded by DOE-OHEP's QuantISED program to achieve high sensitivity to smaller signal levels via ultra low noise photon counting detectors such as quantum capacitance detectors, MKIDs, and SNSPDs.  This new quantum photon counting technology covering 10~GHz - 100 THz may find use in various other areas of HEP outside of dark matter.  

As many of these direct detection technologies are nearing maturity and are already or will be shovel-ready in the next decade, there is a key opportunity to cover most of the remaining $\sim$10 orders of magnitude in the QCD axion mass parameter space while concurrently exploring much of dark photon parameter space.  The detector and sensor component of most of the experimental techniques falls firmly in the small experiment category and so a portfolio of small experiments could be strategically devised to divide and conquer the remaining axion parameter space.  However, a critical path and long lead time item for most experiments is the large bore, high field magnet needed to induce the interaction between axion waves and signal photons.  While resonant cavity experiments are viable in the frequency range 1-30 GHz and have been operated with commercial research solenoids or MRI magnets, the frequency scan speed has been in practice limited by the availability of magnet space.  For example, if ADMX could operate two magnets instead of one, then nearly identical radios deployed in each magnet could be tasked to scan through independent frequency ranges to cover twice as much mass range in the same amount of time.  Unlike most other techniques, these frequency scanning experiments enjoy linear scaling of science return with project cost which is usually dominated by magnet cost.

While the resonant experiments are constrained in size to the wavelength scale of the signal, lumped element experiments targeting lower frequencies and  broadband antenna targeting higher frequencies each rely on using larger, customized magnets to intercept enough of the dark matter flow to achieve sensitivity to the QCD axion.  A serious effort to comprehensively cover the many decades in allowed axion mass parameter space in finite operations time will thus require a significant investment in magnet facilities both in large, custom magnets for signal-limited experiments and a fleet of smaller commercial magnets for the resonant cavity experiments.  Given that a Mu2e or CMS-scale solenoid may take a decade to construct, and even smaller superconducting magnets take several years to fabricate, a possible path forward is for the magnets to be planned, constructed, and operated as user facilities in which multiple experiments could be concurrently or sequentially deployed.  A HEP magnet facility would focus on large bore magnets for longer term tenants, in contrast to the current portfolio of small bore, ultra high-field magnets that provided for shorter term condensed matter experiments by the NSF's National High Magnetic Field Facility. 

\begin{figure}[t]
\begin{center}
\includegraphics[width=0.7\linewidth]{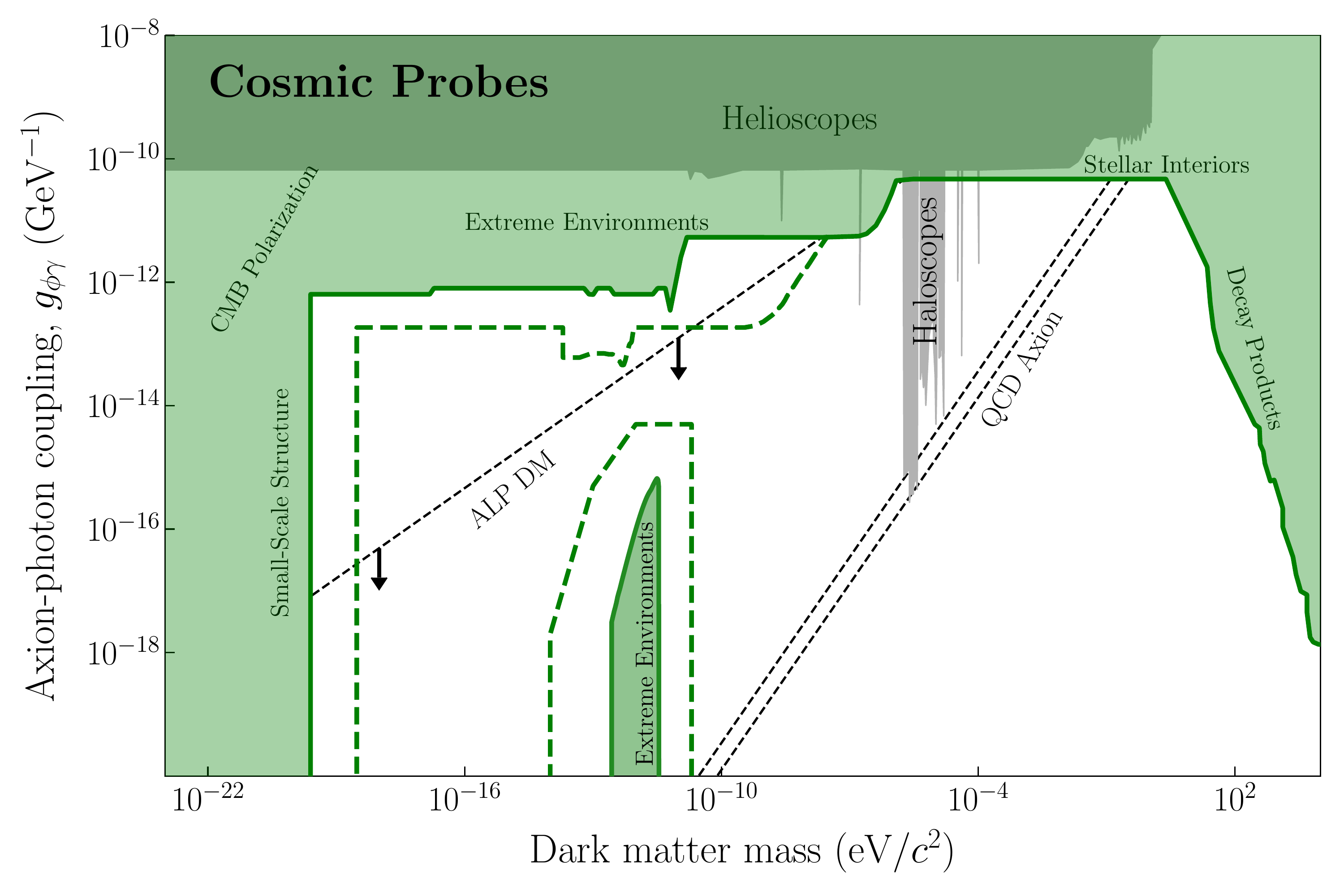}
\caption{Cosmic probes of extreme astrophysical environments combined with measurements of dark matter halos and other cosmological observations set strong constraints on the parameter space of axion-like particles (green regions). Projected improvements in sensitivity coming from future facilities and observations are indicated with a dashed green lines. Cosmic probes are sensitive well-motivated regions for the QCD axion and axion-like particles, and they are highly complementary to other experimental searches with helioscopes and haloscopes (gray regions). From the CF3 report~\cite{Drlica-Wagner:2022lbd}.}
\label{fig:dm_cf3_alp}
\end{center}
\end{figure}

Important constraints also arise from dark matter measurements with cosmic surveys (e.g., Rubin LSST, CMB-S4, Spec-S5) and observations of extreme astrophysical environments, which combined with advances in theory and simulations will provide sensitivity to orders of magnitude of parameter space of axions and ALPS~\ref{fig:dm_cf3_alp}.  Fifth force searches for spin dependent forces mediated by the axion such as ARIADNE and other torsion pendulum experiments may also determine a characteristic range for the BSM Yukawa interaction and thus focus the targeted mass range for direct detection experiments.  Similarly, solar searches including CAST and the proposed upgrade IAXO may determine the mass of axions escaping the hot core of the sun and provide an experimental target for the dark matter experiments.

Gravitational waves have been recently shown to be an excellent new  cosmic probes of ultralight bosonic matter.  For example, if the PQ phase transition is sufficiently strongly first order, then a stochastic background of gravitational waves could be created by the boiling of the vacuum during this phase transition.  Measuring the spectrum of this stochastic background with LIGO or with future GW observatories would determine the energy scale of the PQ transition and then via the axion model, uniquely determine the mass of the axion dark matter.  The axion direct detection program could then be focused on a narrow mass window rather than needing multiple experiments with different technologies to cover each of the $\sim$10 decades the currently allowed mass window.  

Searches for narrow line gravitational wave signals also constrain ultralight bosons of any type due to the prediction of the formation of gravitational atoms around spinning black holes.  Just as in the hydrogen atom, particles bound to a black hole are confined to orbits of quantized angular momentum.  Since bosons do not have a Pauli exclusion principle, these orbits become classically occupied with boson condensates which are created from the gravitational superradiance effect when the boson Compton wavelength is comparable to the black hole size.  Transitions of bosons between different orbitals then results in narrow line gravitational wave radiation.  A range of current and planned GW observatories can conduct model-independent searches for low-mass bosons from $10^{-21}$--$10^{-11}$~eV.

\subsection{Light particle dark matter (1~eV to 1~GeV)}

Above the $\sim$eV scale, the dark matter typically manifests as individual quanta.  A priority target is the sterile neutrino, whose existence is suggested by the need to incorporate neutrino masses in the Standard Model, and which typically becomes sufficiently long-lived to play the role of dark matter at masses below around 10 keV.  
If sufficiently mixed with the active neutrinos, sterile neutrinos with masses below a few keV would thermalize as warm dark matter, and would suppress structure formation due to free-streaming to a degree that is inconsistent with cosmic observations.  
Evading this cosmological bound requires a mechanism to produce the dark matter in a very cold state, and appropriately weak coupling to the Standard Model such that it does not kinetically thermalize.

\begin{figure}[t]
\begin{center}
\includegraphics[width=0.7\linewidth]{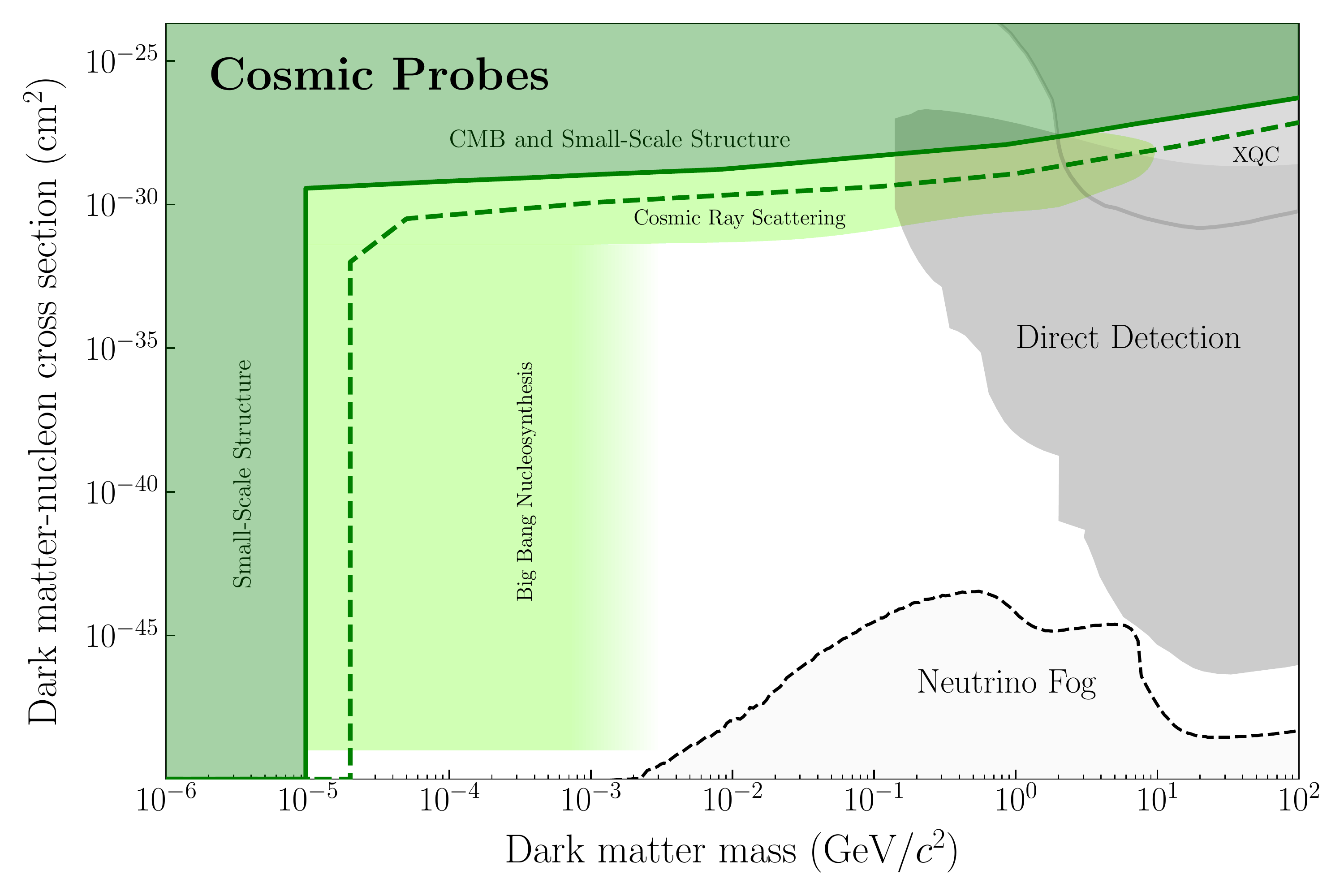}
\caption{Cosmic probes of the matter power spectrum, dark matter halos, Big Bang nucleosynthesis, and cosmic ray upscattering set strong constraints on the minimum thermal dark matter particle mass and spin-independent dark matter–nucleon scattering cross section (green regions). Projected improvements in sensitivity coming from future facilities and observations are indicated with a dashed green lines. These constraints are highly complementary to constraints from direct detection experiments (gray regions). The neutrino fog for xenon direct detection experiments is shown with dashed black line.  From the CF3 report~\cite{Drlica-Wagner:2022lbd}.}
\label{fig:dm_cf3_idm}
\end{center}
\end{figure}

The universe itself provides an exceptional calorimeter in the form of the primordial baryon--photon plasma prior to recombination.
Scattering between light dark matter and nucleons and/or electrons would transfer heat and momentum between the Standard Model and dark matter. This would have the effect of heating the dark matter and resulting in  a suppression of the matter power spectrum similar to that seen in warm dark matter, which would be observable in the CMB, Lyman-$\alpha$ forest, and dwarf galaxies \cite{Gluscevic:2017ywp,Boddy:2018wzy,Nadler:2019zrb,DES:2020fxi,Rogers:2021byl}.
These cosmic probes provide access to the full range of light dark matter masses at large scattering cross sections that are inaccessible to terrestrial experiments due to shielding by the Earth's atmosphere. 
Expected improvements will come from current and near-future cosmic survey experiments like DESI, Rubin LSST, CMB-S4, and Spec-S5 \cite{Chakrabarti:2022cbu,Drlica-Wagner:2022lbd}.
Cosmic probes are highly complementary to direct detection searches, and as illustrated in Figure~\ref{fig:dm_cf3_idm} will provide sensitivity to lighter dark matter particles and higher cross sections than are accessible to direct searches.

For sub-GeV-mass dark matter, direct detection technologies generally rely on individual dark matter particles scattering or being absorbed on various target media in detectors with low energy and momentum threshold.  
Additional handles to distinguish the dark matter scattering signal from backgrounds are provided by the direction of the incoming dark matter, which lead to both annual and daily modulations in the signal rate.
In scattering processes, the signal energy is at most the initial dark matter kinetic energy and the momentum transfer must be less than twice the initial dark matter momentum.  For typical Galactic dark matter velocities of $v=10^{-3} c$, $E_\mathrm{max} \approx 10^{-6} M_\mathrm{DM}$, and $p_\mathrm{max}\approx 10^{-3} M_\mathrm{DM}$, detectors with energy thresholds in the meV--keV range are needed to search for scattering of dark matter in the mass range keV--GeV.   At these lower dark matter masses, kinematic matching for efficient billiard-ball scattering requires using lower mass nuclei, electrons, or collective modes such as phonons or magnons as scattering targets.  

Noble liquid detectors have achieved signal thresholds of $\sim 10$~eV when detecting electrons in the ionization channel, demonstrating the lowest background rate per unit target mass in this signal energy range.  Various ways of doping the target medium are being explored in order to increase scintillation and ionization yield and to improve kinematic response to lower mass particles with low-A nuclei.  Odd-A nuclei also provide sensitivity to spin-dependent interactions.

Semiconductor-based detectors have a threshold near the typical band gap energies of 1~eV in the electron scattering channel, providing sensitivity to dark matter masses near 1~MeV, approximately an order of magnitude below that of the noble liquid detectors.  A recent significant development has been the demonstration of sub-electron charge resolution using skipper CCDs which perform repeated, non-destructive measurements of the charge in individual pixels to average away the readout noise.  The ongoing SENSEI and DAMIC-M experiments are planning a factor of 10--100 scale-up in silicon CCD target mass in a jointly proposed experiment Oscura, part of the the DOE's DMNI program.  Meanwhile, semiconductor crystal detectors such as SuperCDMS HVeV and EDELWEISS are also reaching single electron sensitivity using voltage bias across the semiconductor diode to increase the single electron signal energy.
The semiconductor GaAs is a bright scintillator, and single-photon-sensitive GaAs-based detectors are being developed as part of the TESSERACT program funded through DMNI, with instrumental and nuclear recoil background rejection achieved through coincident photon and phonon detection.

To further lower thresholds for the electron scattering channel, novel target materials with collective electronic excitations with meV scale bandgaps are being investigated.  For example, superconducting nanowire single photon detectors can also be used to detect energy deposited from dark matter scattering events with trigger thresholds near 250 meV for localized energy deposits which quench the bias supercurrent.  SPICE, also part of the TESSERACT program, is investigating low threshold dark matter detection via its electric couplings to the oscillating dipoles exhibited by optical phonon modes in polar crystals.  The 10-100~meV scale bandgap of optical phonons provides both a low detection threshold and a large phase velocity for good kinematic matching to the relatively fast 300~km/s speeds of the galactic dark matter.   Novel semiconductor and semimetal compounds with meV-scale electronic bandgaps are also being developed.   A key challenge is scaling up these 2-d materials to sufficiently large target masses to provide deep coverage of dark matter parameter space.  Collective plasmon modes including those in new heavy fermion materials may also be better kinematically matched to low mass dark matter scattering.  Directional detection of low mass dark matter may also be enabled by using bulk target materials with intrinsically anisotropic band gaps, resulting in large daily modulation in signal rates. 

In the phonon channel for dark matter-nucleon scattering, the semiconductor and crystalline detectors such as CRESST, EDELWEISS, SuperCDMS, and MINER are targeting eV or sub-eV resolution by reducing the readout noise in transition edge sensors.   Superfluid helium is also being investigated as part of the TESSERACT program, and is a promising material for detecting dark matter-induced nuclear recoils, combining a light, pure target with intrinsic background rejection via multiple signal channels and multi-pixel coincident readout.
Beyond the NTDs and TESs traditionally used to read out microcalorimeters, new sensor technologies being explored include kinetic inductance detectors, magnetic microcalorimeters, SNSPDs, and superconducting qubits, with some of this next-generation technology being funded through various DOE quantum programs.  While development of ultra-low-threshold quantum sensors may justifiably be funded through dedicated quantum programs, the deployment of these sensors and the studies of dark matter detector response including both measurements and simulations, and the characterization of new low energy backgrounds are uniquely HEP activities that could best be carried out in small, pathfinder experiments.  As in the case of the axion searches described above, the DOE DMNI program currently provides funding to develop a limited number of proposed techniques.  Given the variety of new sensing technologies, an opportunity exists to cover broad swaths of unexplored sub-GeV dark matter parameter space  by expanding the portfolio of small experiments.

 In addition to sensors designed to measure excitations of internal bulk phonon modes, a complementary approach is to monitor the center-of-mass motion of a composite object \cite{Carney:2020xol}. Optical or electrical readout of such mechanical sensors, in either the classical or quantum regime, offers a novel approach to detecting particle dark matter on a wide range of mass scales, including both heavy \cite{Monteiro:2020wcb} and light \cite{Afek:2021vjy} candidates. These sensors are naturally directionally dependent because they monitor changes to three-momentum of the mechanical system, and can reach energy thresholds well below the eV scale \cite{Afek:2021vjy}.
 
 \begin{figure}[th!]
\begin{center}
\includegraphics[width=0.8\linewidth]{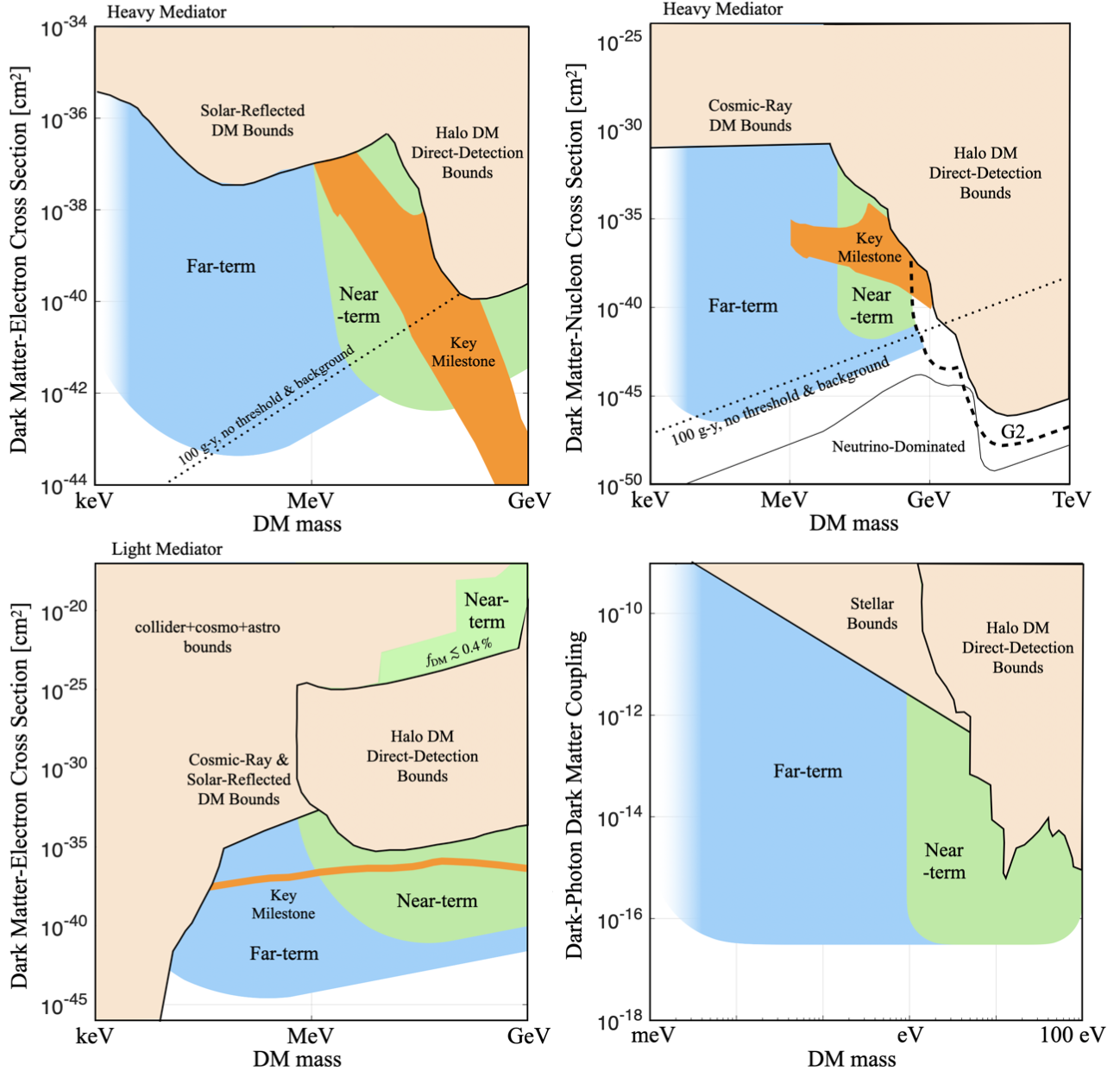}
\caption{Predicted near-term and far-term sensitivities to dark matter with mass in the range 1~eV--1~GeV through different types of mediators
connecting to the Standard Model.  Key milestones indicate representative target regions in which the dark matter is thermally produced for each scenario.  Near-term projections are based on demonstrated sensitivities to electron recoil and nuclear recoil in the 1~eV and 10~eV range respectively.  As the maximum energy deposit in scattering is the initial kinetic energy of the dark matter of order $10^{-6} M_{\rm DM}$, far term projects with lower energy thresholds will be anabled with ongoing R\&D in novel target materials with band gap below the 1~eV scale of semiconductors and of chemistry.  Examples include optical phonons in polar materials with band gap 10-100~meV and superconductors and other quantum materials with sub-meV gap.  Concurrently, demonstration of low threshold sensor technology, including those based on quantum sensors will be required. From Ref.~\cite{Essig:2022dfa}.
\label{fig:dm_portal_sensitivity}}
\end{center}
\end{figure}

Beyond the technical capabilities of any particular detector, 
the distribution of deposited energies is also sensitive to the microphysics, being largely determined by nature of the mediator connecting the dark matter to
the Standard Model.  Heavy mediators manifest as non-renormalizable interactions which favor larger energy transfer (up to the kinematic
limit), whereas light mediators enhance lower transferred energy.  Figure~\ref{fig:dm_portal_sensitivity}, shows four cases of mediators,
corresponding to heavy mediators with preferred coupling to electrons or nuclei, and light mediators coupled to electrons or in the form of dark photons.
In each panel, existing bounds are shaded in tan,
potential advances in the probed parameter space of dark matter mass versus coupling/cross section 
in the near term ($\sim 5$~years) based on existing research 
investments are shaded green, whereas those achievable in the far term (longer time scales) are shaded blue.  Each shaded region represents the combined footprint of several experiments.

Low-mass particle dark matter can be detected through its decay or annihilation into Standard Model messengers.  Below a GeV, the available Standard Model channels include photons,
neutrinos, electrons, muons, and the lightest hadrons.  Because of poorly understood
astrophysical backgrounds and confounding factors such as Galactic magnetic fields (which scramble the incident direction of charged particles), photons with energies $\gtrsim$~keV are typically considered among the most promising messengers.

\begin{figure}[t!]
\begin{center}
\includegraphics[width=0.8\linewidth]{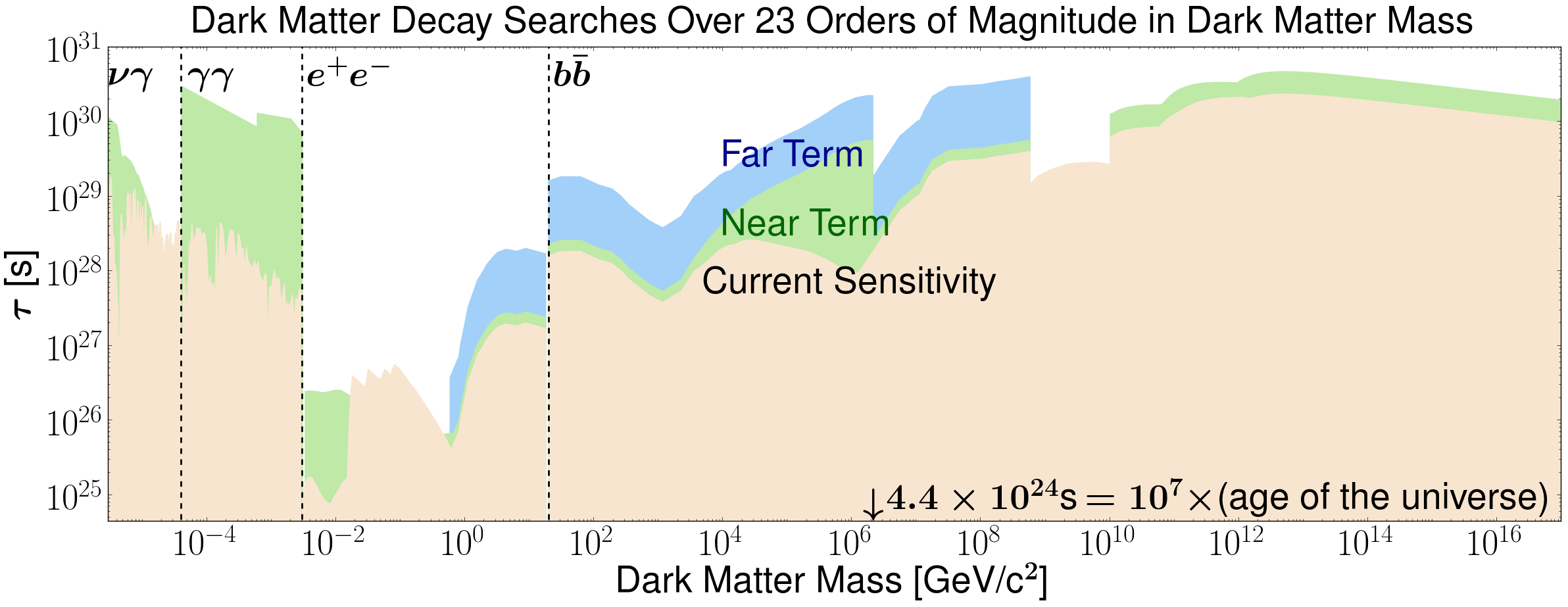}
\caption{Constraints on the lifetime of dark matter based on null searches for produced $X$-rays, gamma rays, and neutrinos, as well as planned improvements in the near term (green) and far term (blue).
The sharp boundaries are artifacts of analysis, rather than instrumental thresholds.  From the CF1 report~\cite{Cooley:2022ufh}.\label{fig:dm_decay}}  
\end{center}
\end{figure}

Indirect searches are uniquely powerful for probing the dark matter lifetime.  A compilation of constraints based on the searches for X-ray, gamma-ray, and neutrino signals across
many decades of mass is shown in Figure~\ref{fig:dm_decay}.  It is striking that such constraints require the dark matter to live many orders of magnitude longer than the current age
of the Universe.

In thermal freeze-out scenarios, indirect detection probes the same interactions that fix the dark matter abundance in the early universe. In the sub-GeV mass range, current limits already generically rule out the simple thermal freeze-out scenario for $s$-wave annihilation, unless the dark matter annihilation products are almost exclusively neutrinos, significantly constraining the space of viable dark matter candidates that were once in thermal contact with the Standard Model (e.g.,~\cite{Cirelli:2020bpc}). In particular, constraints from measurements of the CMB \cite{Planck:2018vyg} have a key role in establishing the viable portals by which dark sectors could communicate with the Standard Model.

At the low end of this energy range, X-ray telescopes have placed stringent constraints on sterile neutrinos, complementary to cosmic probes of warm dark matter~\cite{CF1WP5}. Upcoming X-ray telescopes (XRISM~\cite{Tashiro2020,XRISM2020}, Micro-X~\cite{2020JLTP..199.1072A}, Athena~\cite{Nandra2013}, HEX-P~\cite{Madsen2019HEX}, Lynx~\cite{Zhong:2020wre}) will further improve constraints on decaying/annihilating light dark matter in general, and sterile neutrinos in particular; some will have sufficient energy resolution to seek to resolve DM-sourced spectral lines.

At higher energies, there is currently a sensitivity gap in the MeV--GeV gamma-ray band.  The last major experiment was NASA's Imaging Compton Telescope (COMPTEL)\cite{schonfelder1993instrument}, which operated from 1991--2000.  Several proposed future experiments aim to address this gap: the Compton Spectrometer and Imager (COSI) \cite{Tomsick:2021wed} has a planned launch date in 2025, and will survey the gamma-ray sky at energies of 0.2--5 MeV; 
AMEGO-X~\cite{Fleischhack:2021mhc} has a planned satellite launch date in the late 2020s and will probe energies from 25 keV-–1 GeV using a silicon pixel tracker, CsI calorimeter, and plastic anti-coincidence detector;
and e-ASTROGAM~\cite{e-ASTROGAM:2017pxr} would target the 0.3 MeV--3 GeV energy range using a similar approach. 
Other proposals focused on the MeV band include SMILE~\cite{takada2021first,tanimori2020mev,hamaguchi2019space,takada2020smile}, GRAMS~\cite{Aramaki:2019bpi}, and GammaTPC~\cite{Aramaki:2022zpw}, which rely on gaseous or liquid time projection chambers (TPCs), 
and GECCO~\cite{moiseevsnowmass2021} (based on a novel CdZnTe imaging calorimeter with a deployable coded aperture mask).  

Investment in such experiments would enable both data-driven studies of backgrounds relevant for lower and higher energy indirect searches, as well as providing greater sensitivity to decaying and/or annihilating light dark matter in the MeV--GeV mass range.  They have the potential for sufficient sensitivity to probe thermal freeze-out even when the dominant annihilation is $p$-wave~\cite{CF1WP5, Coogan:2021sjs} (which is suppressed at non-relativistic velocities and thus challenging to observe via indirect detection), further complementing terrestrial experiments focusing on this mass range.

\subsection{Classic WIMPs}

Dark matter masses between $\sim$~GeV and hundreds of TeV populate the classic WIMP
parameter space, where theories aimed at explaining the electroweak hierarchy typically reside.  Such theories have been the subject of extensive theoretical
exploration \cite{Roszkowski:2017nbc,Servant:2002aq,Hubisz:2004ft,Cirelli:2005uq,Cahill-Rowley:2014boa,GarciaGarcia:2015fol}, driven both by the connection to electroweak physics
as well as the fact that roughly electroweak-sized couplings naturally lead to a relic abundance close to the one required by cosmological observations through freeze-out.
As a result, searches for WIMPs remain extremely well-motivated, with many interesting models inhabiting viable and testable parameter space \cite{Leane:2018kjk,Arakawa:2021vih}.

For these masses, the most efficient scattering is with nuclei, which can be described as an expansion in the dark matter velocity by terms in an
effective field theory \cite{Goodman:2010ku,Fitzpatrick:2012ix,Gresham:2014vja,Hill:2014yxa}
Two classes of interactions dominate in the non-relativistic limit: spin-dependent (SD) scattering, which requires a nuclear target with non-zero spin, and spin-independent (SI) scattering, which is coherently enhanced in many models for large-$A$ targets.  Detectors employing heavy nuclei to target SI scattering have seen rapid advance in their size and control of backgrounds, achieving many decades of improved sensitivity over the past twenty years.  Such experiments are now within an order of magnitude of sensitivity to the expected background neutrinos produced in the atmosphere, Sun and supernovae.  This `neutrino fog' is interesting to measure in its own right, but will require new technologies and analysis techniques to be distinguished from a dark matter signal, such as
the expected annual modulation of the DM rate~\cite{Drukier:1986tm} and directional detectors (see {\it e.g.} the review found in~\cite{Vahsen:2021gnb}).  Directional DM detectors would further open up the tantalizing potential to map out the local DM structure of the Galaxy.

\begin{figure}[t!]
    \centering
    \includegraphics[width=0.9\textwidth]{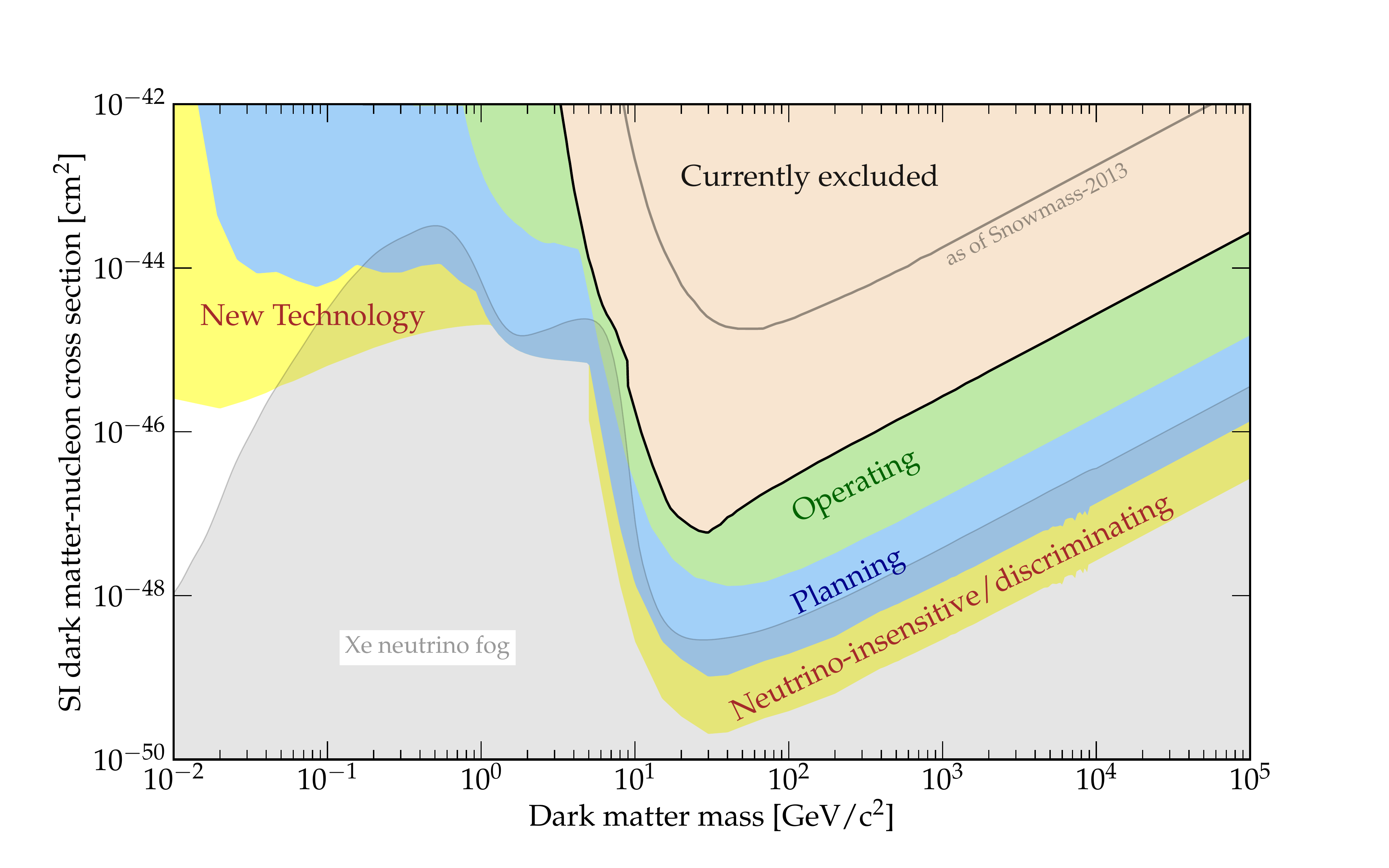}
   \caption{Combined Spin-independent dark-matter nucleon scattering cross section space. 
   Current 90\% c.l.~constraints are shaded beige,
   while the reach of currently operating experiments are shown in green (LZ, XENONnT, PandaX-4T, SuperCDMS SNOLAB, SBC).
   Future experiments are shown in blue (SuperCDMS, DarkSide-20k, DarkSide-LowMass, SBC, XLZD, ARGO) 
   and yellow (Snowball and Planned$\times$ 5). The neutrino fog for a xenon target is shaded light grey. 
   From Ref.~\cite{Akerib:2022ort}.}
    \label{fig:limitplot_SI}
\end{figure}

Currently, the most stringent SI bounds for $\gtrsim$ 10 GeV dark matter masses are from liquid xenon (LXe) and liquid argon (LAr) detectors employing time projection chamber (TPC) technology. Charge and phonon detectors are able to reach lower DM masses and are expected to reach sensitivity for dark matter masses between 0.5--5~GeV to within a decade of the neutrino fog, limited by cosmogenically activated isotopes including $^3$H and $^{32}$Si, $^{210}$Pb, and by dark counts. Isotopically purified LAr detectors~\cite{GlobalArgonDarkMatter:2022xgs} or phase change detectors such as bubble chambers~\cite{Alfonso-Pita:2022akn} can provide powerful background-discriminating technology and the combination of scalability, low threshold, and background discrimination \emph{at low threshold} could allow noble-liquid bubble chambers to explore the neutrino fog in the 1--10~GeV WIMP mass range. 

\begin{figure}[th!]
    \centering
\includegraphics[width=0.9\textwidth]{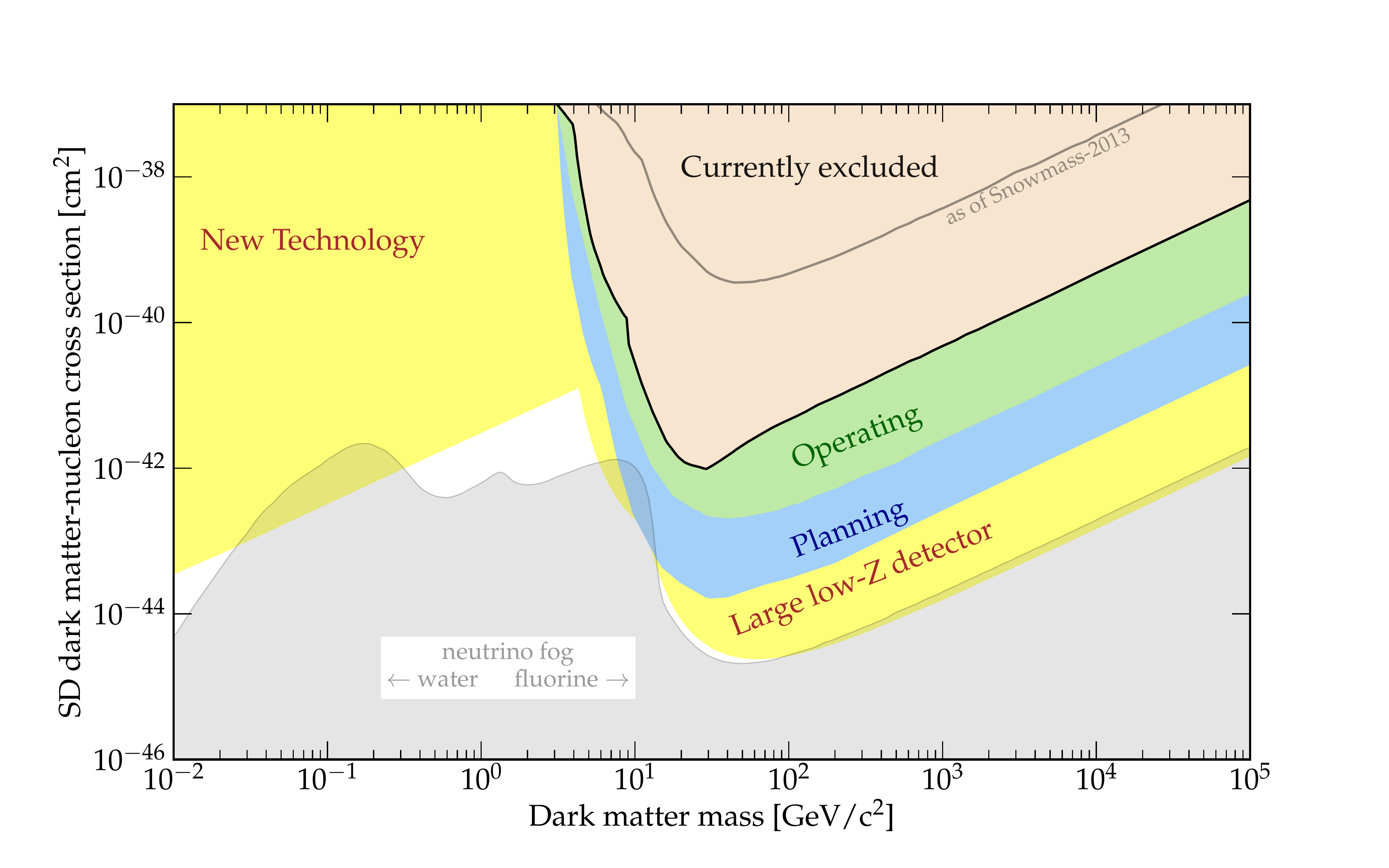}
   \caption{Combined Spin-dependent dark-matter nucleon scattering cross section space for scattering with neutrons or protons.
   Current 90\% c.l.~constraints are shaded beige, whereas the reach of currently operating experiments are shown in green (LZ, XENONnT).  
   Future experiments are shown in blue (PICO-500, XLZD) 
   and yellow (Snowball, PICO-100 ton). The neutrino fog for a water or fluorine target is shaded light grey. 
   From Ref.~\cite{Akerib:2022ort}.}
    \label{fig:limitplot_SD}
\end{figure}

Figure~\ref{fig:limitplot_SI} shows the current, operating, and future projected $90\%$ CL constraints for WIMP DM interacting spin-independently~\cite{Akerib:2022ort}.  
The beige regions show the current best exclusion for each mass, combining previously reported limits~\cite{XENON:2020gfr,SuperCDMS:2018gro,Adhikari:2018ljm,CRESST:2019jnq,Bernabei:2018yyw,DarkSide:2018bpj,DEAP:2019yzn,EDELWEISS:2016nzl,LUX:2016ggv,NEWS-G:2017pxg,PandaX-II:2017hlx,PICO:2016kso,PICO:2017tgi,XENON:2019zpr,PandaX-4T:2021bab,LZ:2022ufs}, as collected by Ref~\cite{OHare:2021utq}.  A measure of the
incredible progress of the past decade is provided by the comparison with the grey line from Snowmass 2013~\cite{Cushman:2013zza}.  The projected
reach of currently operating experiments is shaded in green, whereas proposed upgrades are shown in blue
(including SuperCDMS, DarkSide-LowMass, SBC, XLZD, and ARGO) and the reach of proposals based on new technologies are shaded in
yellow (Snowball and Planned$\times$ 5). The region corresponding to the neutrino fog for a xenon target is shaded light grey.

Similarly, Figure~\ref{fig:limitplot_SD} shows the current, operating, and future projected $90\%$ CL constraints for WIMP DM interacting spin-dependently~\cite{Akerib:2022ort}.  
Shaded in beige is the union of the currently excluded parameter space, led by LXe TPCs and freon-based bubble chambers~\cite{Behnke:2012ys,Kim:2018wcl,Archambault:2012pm,Amole:2016pye,Amole:2017dex,CRESST:2019jnq,Archambault:2012pm,CDMS-II:2009ktb,LUX:2017ree,PandaX-II:2018woa,XENON100:2016sjq,Aprile:2019jmx,XENON:2019rxp}, as collected by Ref~\cite{OHare:2021utq}.  The expected reach of currently operating experiments is shaded green,
and future proposed experiments are shaded blue (including PICO-500 and XLZD) and yellow (Snowball and PICO-100 ton). 
The region corresponding to the neutrino fog for a water or fluorine target is shaded light grey. 

Taken together, Figures~\ref{fig:limitplot_SI} and \ref{fig:limitplot_SD} illustrate the exciting prospects for direct searches in the WIMP regime, with a suite of experiments based on
different technologies capable of probing WIMP dark matter all the way down and into the neutrino fog.

Indirect searches for dark matter play a very important role in covering the parameter space.  At lower masses, as discussed above, measurements of the CMB
have a key role in constraining the viable portals to communicate with the SM.  At the highest masses, indirect detection achieves sensitivity
to WIMP parameter space that is inaccessible to direct or collider searches, and in general provides a broadly model-agnostic probe of thermal freeze-out scenarios.  The most effective messengers of WIMP annihilation are gamma rays (which point back to
their origin, giving an additional analysis handle) and energetic anti-matter signals in cosmic rays. Especially for higher DM masses, where abundant energy is available to produce the full range of SM particles, indirect signals are generically expected to be multi-messenger and multi-scale~\cite{CF1WP8}.

\begin{figure}[t!]
\begin{center}
\includegraphics[width=0.6\linewidth]{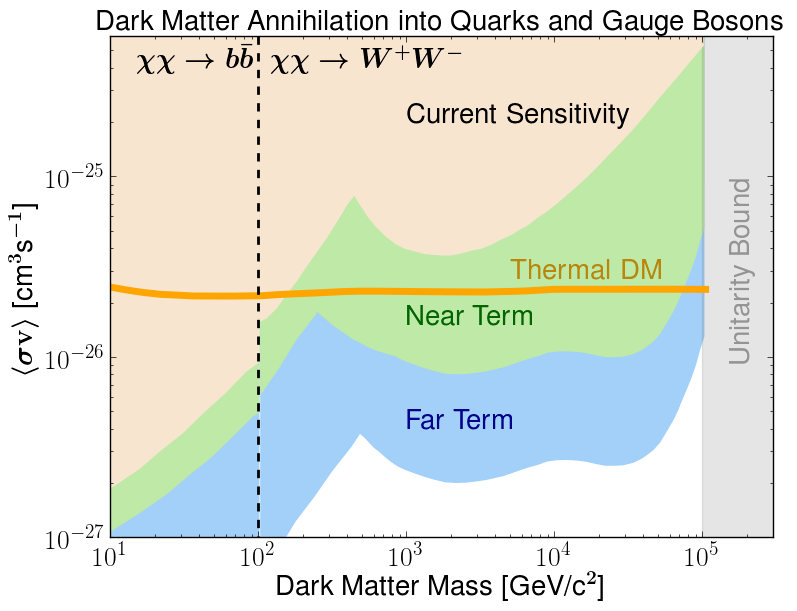}
\caption{Limits on WIMP annihilations into pairs of bottom quarks (for masses below $\sim 100$~GeV and $W$ bosons (for larger masses) based on null searches
by $gamma$-ray observatories.  The beige regions indicate the current limits for each mass, whereas the green shaded region indicates near future gains based on planned
missions, and the blue shading indicates the reach that would be enabled by long term investments in ground- and space-based observatories.  From the CF1 report~\cite{Cooley:2022ufh}.
\label{fig:dm_annihilation}}
\end{center}
\end{figure}

The rate of annihilation depends on the underlying microphysics, which determines both the rate into each specific annihilation channel as well as its dependence on
the relative velocity between the annihilating dark matter particles.  Dark matter making up the halos of galactic structures is typically highly non-relativistic ($v \sim 10^{-4} - 10^{-3}$),
and thus $s$-wave annihilations (for which $\langle \sigma v \rangle$ is $v$-independent) typically dominate over annihilations at higher partial wave.  Additional challenges involve
determining the dark matter distribution along the line of sight of an indirect search, which strongly impacts the rate of annihilation, since $\langle \sigma v \rangle \propto \rho_\chi^2$.
This systematic uncertainty benefits strongly from advances in simulation of galaxy formation.  Modeling of astrophysical backgrounds with sufficient precision as to be able
to distinguish a subtle signal of dark matter annihilation or decay from more prosaic astrophysical processes is also a key ingredient (see Refs.~\cite{Cooley:2022ufh,CF1WP6,CF1WP7} for further discussion); this is already manifest in a number of puzzling excesses in indirect searches~\cite{CF1WP6}.

Gamma rays may be detected both by space- and ground-based  telescopes.  The space-based Fermi-LAT is optimized to reconstruct gamma rays 
from 1 -- 100 GeV and is effectively able to observe the entire sky as it orbits, 
whereas ground-based air and water Cherenkov telescopes are effective at energies above $\sim 100$ GeV. Most current Cherenkov telescopes are located in the Northern Hemisphere, but Southern Hemisphere locations are advantageous for observing both the Galactic Center and the central dense region of the Milky Way's dark matter halo.  For $s$-wave annihilation,
Fermi-LAT has already strongly constrained
the thermal freeze-out cross section for masses up to $\sim 100$~GeV based on observations of the Milky Way's dwarf spheroidal galaxies (e.g.~\cite{Fermi-LAT:2016uux,Ando:2020yyk}), which are
(as measured by stellar dynamics) dark matter rich and baryon poor, limiting the expected sources of background.  These limits are expected to improve in the near future, as surveys
such as Vera Rubin discover (about a factor of two) more dwarf spheroidal galaxies~\cite{Hargis:2014kaa}.

In the near future,
the SWGO~\cite{Abreu:2019ahw} (water Cherenkov) and CTA~\cite{2019scta.book.....C} (air Cherenkov) telescopes, successors using similar technology to the successful HAWC and VERITAS Cherenkov telescopes but with larger installations and Southern Hemisphere sites,
have the potential to probe the thermal freeze-out scenario up to 10s of TeV masses 
(depending on the annihilation channel)~\cite{CTA:2020qlo, Viana:2019ucn}, approaching the 100 TeV scale where we can begin to set unitarity-based limits on the capacity for freeze-out to generate the correct relic abundance. 
Further in the future, APT~\cite{buckleyastro2020} is a concept for a space-based successor instrument to the Fermi-LAT (with a demonstrator suborbital mission scheduled for 2025), which aims 
to improve sensitivity at lower masses
by an order of magnitude. 
Figure~\ref{fig:dm_annihilation} shows the limits on WIMP annihilations into pairs of bottom quarks (for masses below $\sim 100$~GeV) and $W$ bosons (for larger masses) 
based on null searches by gamma-ray observatories.  The beige regions indicate the current limits for each mass, whereas the green shaded region shows improvements expected
in the near future based on planned Cherenkov observatories and including new populations of dwarf spheroidal galaxies expected to be discovered by Rubin LSST.  The blue shading
indicates parameter space that could be probed by longer term investment into future large Cherenkov arrays and new space-based missions.  The orange band indicates the benchmark cross
section corresponding to the correct relic abundance from freeze-out. The gray ``unitarity bound'' region corresponds to the general mass range (the exact bound is model-dependent and depends on assumptions about long-range forces, compositeness of the dark matter, etc.) in which obtaining the correct dark matter abundance via freeze-out becomes inconsistent with unitarity in the early universe, under standard assumptions for the cosmological history.

Cosmic-ray experiments (primarily AMS-02) currently set competitive constraints on heavy annihilating or decaying DM, with systematic uncertainties independent from gamma-ray probes. 
In the near future, GAPS~\cite{Aramaki2015} will provide the first dedicated search for low-energy anti-deuterons, which are expected to have only tiny astrophysical backgrounds and could serve as a very clean discovery channel, and HELIX~\cite{f2f8e111be344c99b12978ac18b696ce} will provide new constraints on cosmic-ray propagation. 
Far-future proposals ALADInO~\cite{Battiston:2021org,instruments6020019} \& AMS-100~\cite{2019NIMPA.94462561S} involving superconducting magnetic spectrometers seek greatly improved sensitivity to anti-deuterons and anti-helium; GRAMS~\cite{Aramaki:2019bpi} would employ a liquid-argon TPC to search for both gamma rays and charged cosmic rays, and ADHD~\cite{CF1WP5} is a proposal to pursue a novel delayed-annihilation signal from anti-nuclei in a helium detector. 
Accelerator experiments including NA61/SHINE~\cite{Aduszkiewicz:2017sei}, ALICE~\cite{ALICE:2020zhb}, LHCb~\cite{Aaij:2018svt}, and AMBER~\cite{Denisov:2018unj} can also provide complementary measurements to constrain cosmic-ray production and propagation, helping to reduce systematic uncertainties in both signals and backgrounds.

\subsection{Ultra-Heavy Dark Matter}

Measurements of baryon acoustic oscillations and the CMB set some of the strongest limits on the total baryonic fraction of matter in the universe.  The sound waves of the primordial plasma have compressions due to gravitational attraction and the restoring force due to plasma pressure causes the following rarefactions.  The higher amplitude in the odd-numbered power spectrum peaks indicate that most of the gravitating matter does not exhibit pressure when compressed and is hence not electromagnetically charged like protons or electrons.  So even if dark matter took the form of non-luminous, cold baseballs (a model that is difficult to constrain observationally), these baseballs would still have to be made of some non-standard exotic particles.  

\begin{figure}[t!]
\begin{center}
\includegraphics[width=\linewidth]{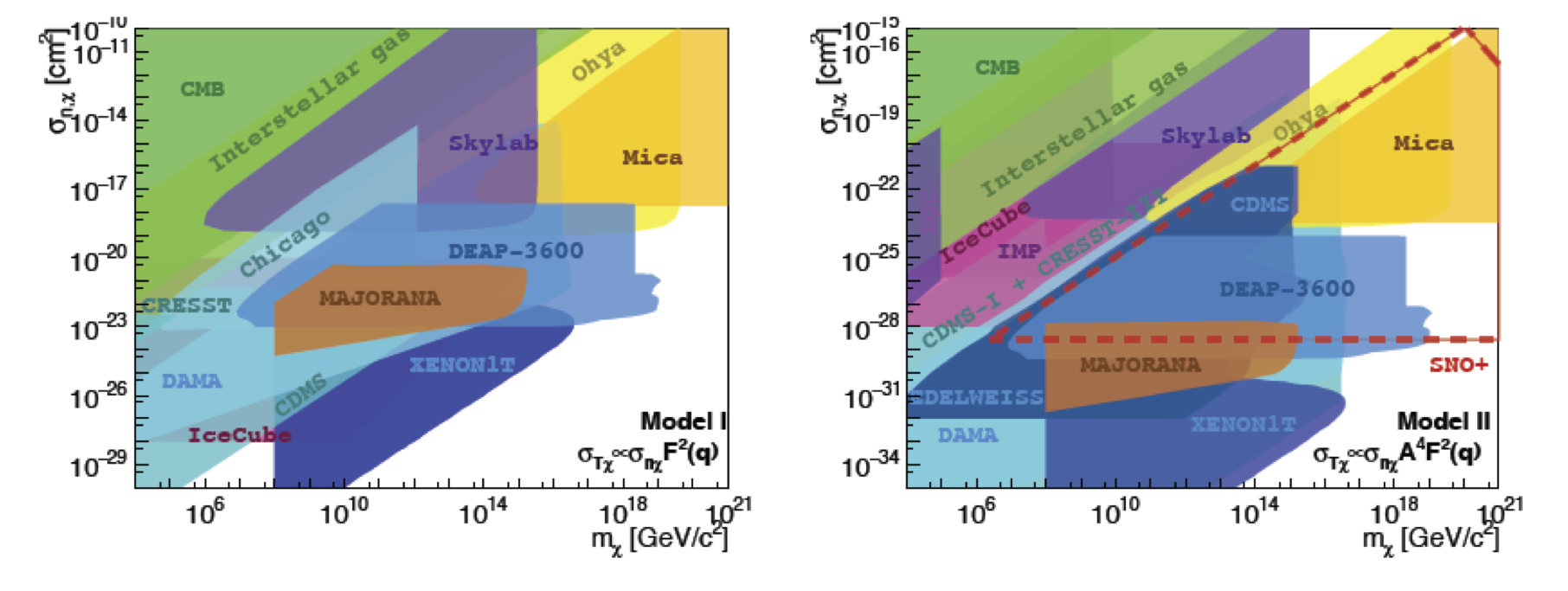}
\caption{Ultra-heavy dark matter can be studied through a range of terrestrial experiments, indirect detection experiments, and cosmic probes. The sensitivity of direct detection experiments is primarily limited by the cross-sectional area of the detectors due to the fact that the kinematics of nuclear recoil asymptotes to fixed recoil energy for large dark matter masses.  From Ref.~\cite{Carney:2022gse}.\label{fig:dm_heavy_sensitivity}}
\end{center}
\end{figure}

For masses beyond the unitarity limit, it becomes a challenge to produce dark matter via freeze-out.  One interesting set of models produces dark matter in this mass range as
composite blobs of more fundamental particles, held together by new dark forces.  
Cosmic observation of the CMB and interstellar gas set robust constraints at large cross sections, while observations of extreme astrophysical environments can also constrain the presence of UHDM.
Both the searches for WIMPs described in the previous subsection as well as
more specialized searches have sensitivity to this UHDM regime, depending on the strength and nature of its interactions with the SM (see Ref.~\cite{CF1WP8} for a more in-depth discussion).  For strongly interacting blobs,
a typical event in a terrestrial experiment could involve multiple scattering, necessitating new analysis techniques and implying that the area of the detector 
is the most relevant factor rather than its volume.  For indirect searches, high energy neutrinos and cosmic rays become especially important at high masses. 
Cascades of secondary particles from high-energy primaries often lead to observable indirect signatures at much lower energies than the DM mass, such that the 
gamma-ray telescopes described above have sensitivity to DM masses much higher than their target energy ranges. 
This situation represents an ongoing theoretical challenge, requiring new techniques for accurate predictions to take advantage of
the opportunity to observe rich and complementary multi-wavelength and multi-messenger signals~\cite{CF1WP8}. For DM masses of order GUT-scale, however, the secondary photons  {\it en route} to Earth travel unscathed. Accordingly, AugerPrime and next-generation cosmic ray observatories anchor unique DM indirect detection experiments which are free of astrophysical background: a clear detection
of an extreme energy photon would be a momentous discovery~\cite{Anchordoqui:2021crl}. Fig.~\ref{fig:dm_heavy_sensitivity} shows the current and projected experimental regions of ultra-heavy parameter space accessible to direct searches. 
 
 \subsubsection{Dark Matter Beyond the Planck Scale}
 
Dark matter at the Planck mass (${\sim}10^{-5}$~g) and above is very difficult to detect directly due to its extremely low flux.  
Detection techniques rely on scattering to be mediated via long range forces so that effects can be seen in sparsely instrumented detectors with large collecting area.  For example, if dark matter had a long range Yukawa force $10^3$ times greater than gravity, then interactions of 10~kg mass dark matter with the 10~kg scale mirrors of the LIGO gravitational wave observatory could potentially be detected as the clumps traversed within the near field of the 4~km long interferometers~\cite{Hall:2016usm}.  Proposed larger observatories such as the 40~km long Cosmic Explorer or laser ranging between astroids would provide a larger collection area~\cite{Baum:2022duc}. 

 \begin{figure}[t!]
\begin{center}
\includegraphics[width=0.7\linewidth]{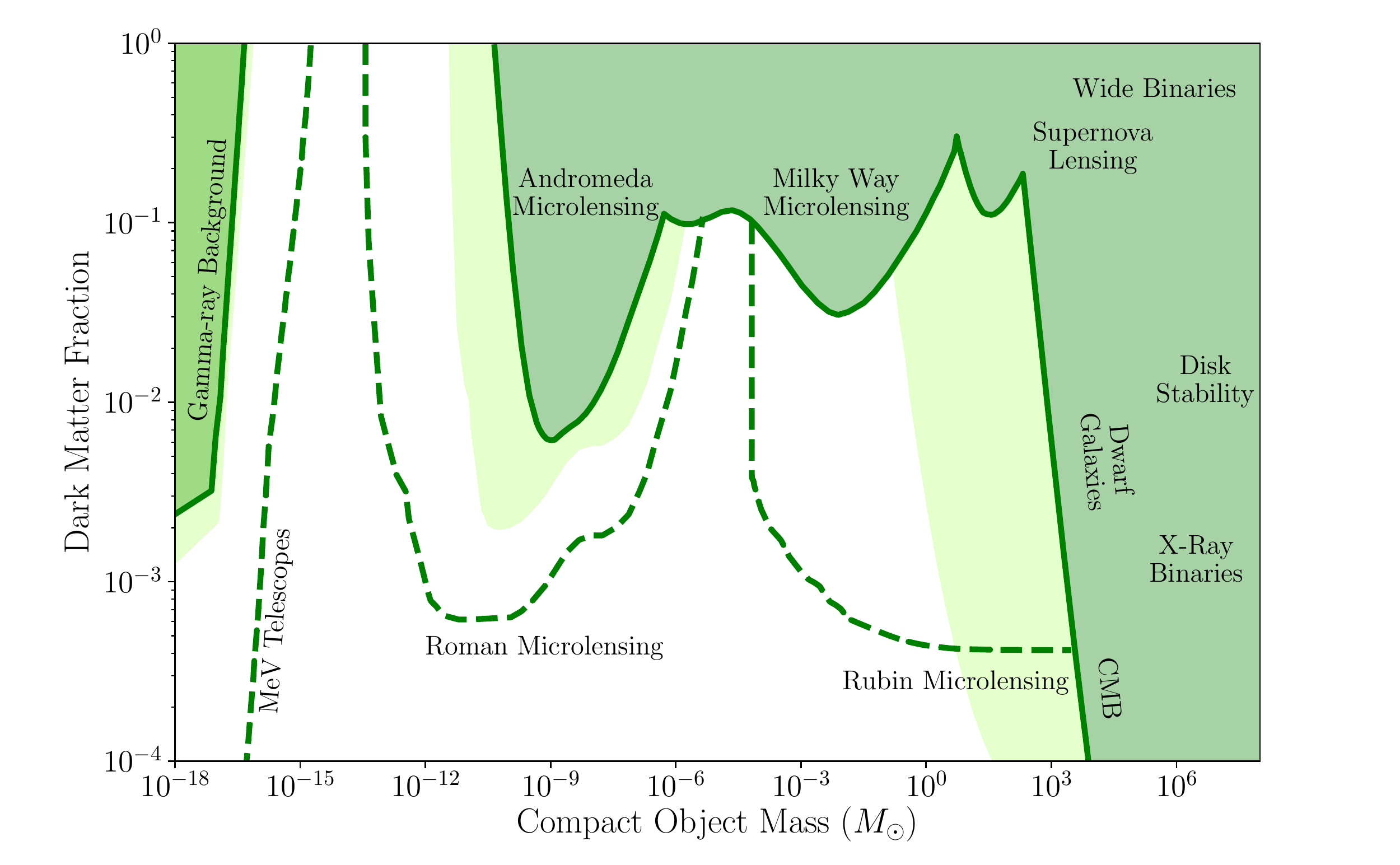}
\caption{Some fraction of the dark matter could take the form of macroscopic, compact objects such as primordial black holes. Cosmic observations provide sensitivity to primordial black holes over a wide range of masses (green regions). Projected improvements in sensitivity coming from future microlensing and gamma-ray searches are indicated with dashed green lines.  From the CF3 report~\cite{Drlica-Wagner:2022lbd}. \label{fig:dm_pbh}}
\end{center}
\end{figure}

Heavier compact constituents of dark matter including primordial black holes (PBHs), can be detected through cosmic probes (Figure~\ref{fig:dm_pbh}). 
PBHs with masses of ${\sim}10^{-18}\,M_\odot$ (${\sim} 10^{15}$~g) should be evaporating today, creating short bursts of Hawking radiation in the GeV-TeV range that could be observed with space-based and ground-based gamma-ray telescopes, as well as neutrino observatories.  
For higher masses, gravitational micro-lensing provides the most sensitive probe of the abundances of PBHs.
Near-future observatories such as Rubin LSST have the potential to radically increase the sensitivity of micro-lensing searches if scheduled optimally \cite{Bird:2022wvk}.
Sub-solar mass PBHs may be detected through microlensing with Rubin LSST and other observatories or through mergers with next-generation gravitational wave detectors.
High redshift searches with optical or gravitational wave probes may also reveal the existence of primordial black holes.  

\subsection{Delve Deep, Search Wide}

The parameter space of dark matter models is vast, yet experiments which will provide significant inroads into this space are relatively inexpensive.  The Cosmic Frontier plans to explore dark matter parameter space over the next decade through targeted searches to make deep progress on high-priority science targets, and an expanded portfolio of cosmological and astrophysical observations, and small pathfinder experiments implementing new detector technologies to search wide and provide broad coverage of the range of possibilities.   This strategy will cover large regions of model-space, and could easily lead to a transformational discovery within the next decade (see the lower panel of Figure~\ref{fig:dm_now_and_future}).

The WIMP and the QCD axion remain highly motivated and clear experimental targets.  Tremendous progress has been made in the past two decades in liquid noble detectors and a multi-national collaborative project would continue this momentum to push sensitivity down to the solar neutrino background.  Concurrently, indirect searches with proposed TeV gamma-ray observatories would conclusively test the self-annihilation cross section which defines the WIMP model, up to the scale where unitarity bounds become relevant.  In the axion field, a portfolio of complementary techniques including NMR, lumped element resonators, cavity resonators, quantum sensors, and novel scattering targets would cover most of the remaining parameter space of the QCD models that solve the strong-CP problem. Axion-like particles are also well-motivated with a parameter space that can be targeted over the next ten years and beyond.  Concurrently, a portfolio of new technologies including CCDs, novel materials as scattering targets, and ultra-low threshold readouts 
including low heat capacity transition edge sensors, kinetic inductance detectors, and Josephson junction-based sensors, is being assembled to test a broad range of portal dark matter models. Pathfinder experiments using AMO techniques including atomic clocks, atom interferometers, and opto-mechanical oscillators are also being developed.  Cosmic and indirect probes will complement and inform all of these searches, while broadly exploring parameter space and phenomena that are currently inaccessible to terrestrial experiments. With the exception of the medium-scale multi-national WIMP experiments, all of the other terrestrial dark matter experiments would be best characterized as ``small."  Similarly, modest investments in theory, computation, and cross-disciplinary analysis infrastructure would enable access to dark matter science in the larger cosmic surveys program.

All combined, a coordinated U.S.\ experimental program to delve deep and search wide for dark matter might reasonably be expected to cost as much as a single large experiment.  Such a comprehensive and coordinated program would be very reasonably justified by its broad science reach in addition to its broader impacts in technology development, contributing to the quantum ecosystem, and training the next generation of experimental hardware specialists.

Investment in new experimental techniques for both low-mass portal dark matter, axions, and more general bosonic waves has been slow to ramp up as most DOE/NSF dark matter funding has been invested in the larger dark matter experiments as per the recommendations of the previous P5 report.  Identifying this gap in the U.S. dark matter strategy, private foundations including the Simons Foundation, the Heising-Simons Foundation, and the Moore Foundation have begun to invest in university programs to fund dark matter instrumentation R\&D including pathfinder experiments at the \$10M level.  Some of the technology development and instrumentation work has also been enabled by leveraging various quantum programs, given that ultraweak dark matter interactions are a natural target for new quantum sensor technology.  Cross-disciplinary engagement with neighboring scientific communities including atomic, molecular, and optical physics, condensed matter physics, and quantum information science have been facilitated by DOE-OHEP's QuantISED program and by the National Quantum Initiative Science Research Centers.  This collaborative research enables the direct transfer of novel detection concepts as well as mature detector technologies from neighboring fields to HEP applications as well as the direct engagement of hardware experts who are able offer fresh perspectives on the detection challenges.  

A positive development has been two DOE Basic Research Needs workshops, the first creating the DMNI program to fund small dark matter experiments of a scale smaller than the medium-size LZ and SuperCDMS experiments, and the second aimed at identifying instrumentation R\&D needs including quantum sensing for fundamental physics.  The DMNI program has funded the technical design studies of a limited portfolio of small experiments as recommended in the previous P5 report.  These include the axion experiments ADMX-Extended-Frequency-Range and DM-Radio, the 1~eV threshold Oscura silicon CCD experiment, the sub-eV threshold TESSERACT program that utilizes novel condensed matter targets, as well as the fixed target experiments LDMX and CCM that probe portal dark matter models via light dark particle production.  While these experimental concepts have not yet proceeded to project status, the DMNI program provides a prototype for implementing a broad and coordinated dark matter program with a portfolio of complementary experiments to target different regions of unprobed parameter space.  Continuation of DMNI or similar programs targeting smaller pathfinder experiments is therefore a key element of the `delve deep, search wide' strategy for discovering the nature of dark matter.

Similar support could significantly strengthen cosmic and indirect probes of dark matter. These approaches have the potential to yield large scientific dividends by searching wide ranges of dark matter parameter space. For example, Rubin LSST has enormous potential to discover new physics beyond the prevailing cold dark matter paradigm~\cite{Drlica-Wagner:2019mwo, Mao:2022fyx}. Rubin LSST can measure the distribution of dark matter on unprecedentedly small scales, thereby probing microscopic properties of dark matter, including thermal particle mass, self-interactions, interactions with radiation, and quantum wave features. Microlensing measurements will directly probe primordial black holes as a component of dark matter. The planned CMB-S4~\cite{CMB-S4:2022ght} and future Spec-S5~\cite{DESI:2022lza} and CMB-S5 projects will extend access to rich dark matter particle physics. For instance, detection of additional relativistic degrees of freedom by CMB-S4 would imply the existence of a dark sector. 
However, building the infrastructure to perform dark matter analyses with these experimental facilities requires dedicated support for experimental collaborations, theoretical research, and large numerical simulations (which are crucial to distinguish novel dark matter physics from baryonic astrophysics). 
Rubin LSST and other cosmology facilities should be explicitly identified as dark matter facilities to enable what promises to be an exciting decade of dark matter research.


\section{Neutrinos}
\label{sec:nus}

Cosmic probes can provide crucial information about the neutrino sector and exhibit a high degree of complementarity to terrestrial probes \cite{Abazajian:2022ofy}.  Cosmology is sensitive to the number and masses of the neutrinos through their
impact on the evolution of the Universe, providing information that is currently inaccessible through other means.

\subsection{Cosmic Measurements of Neutrino Masses}
The current cosmological limits on the sum of the neutrino masses is $\sum m_\nu < 0.12$~eV as obtained using cosmic microwave background, baryon acoustic oscillations, supernova Ia, and large scale structure measurements.  Additional information on impact of cosmological neutrinos on the growth of structure both will be obtained from high redshift galaxy surveys. CMB-S4 will reach sensitivity $\sigma(\sum m_\nu) < 0.02$~eV and if the data indicate a mass sum $<$ 0.1~eV, this would rule out the inverted neutrino mass hierarchy which predicts higher mass sum, providing key synergy with long baseline neutrino experiments and important input to searches for neutrino-less double beta decay attempting to unravel whether neutrino masses are fundamentally Majorana or Dirac in nature.   A Stage V Spectroscopic Facility would reach similar constraints from an independent experiment at $2 <z<5$. 
Further gains in accuracy on $\sigma(m_{\nu})$ beyond what CMB-S4 and Spec-S5 each will achieve will be limited by the currrent $\tau$ prior, unless that parameter is constrained better. Proposals to achieve such improvements are described in the CF4 report \cite{Annis:2022xgg}. 
Higher statistics observations of the matter distribution at high redshifts could be measured by line intensity mapping surveys (in 21-cm and mm-wave) and other probes, such as future high resolution CMB imaging, and would reach resolution $\sigma(\sum m_\nu) < 0.01$~eV.  

\subsection{New Opportunity: High-energy neutrinos}
The discovery of TeV-PeV astrophysical neutrinos opens up unique opportunities to probe the neutrino sector at energy scales not accessible with laboratory neutrino beams. High-energy neutrinos from IceCube were used to discover the Glashow resonance and measure the high-energy neutrino-nucleon cross section and inelasticity distribution. Future statistics will allow further tests of the Standard Model neutrino physics between 1~TeV and 10~PeV. In addition, ultrahigh energy neutrinos (UHE; $\gtrsim$ 100 PeV neutrinos) have been long-predicted but remain undetected. They provide a path to probe weak-scale physics at center-of-mass energies above 50 TeV. Next-generation UHE neutrino experiments have the potential to detect these neutrinos and push the forefront of neutrino physics~\cite{Ackermann:2022rqc}. 

\begin{figure}[t]
  \centering
  \includegraphics[width=0.7\textwidth]{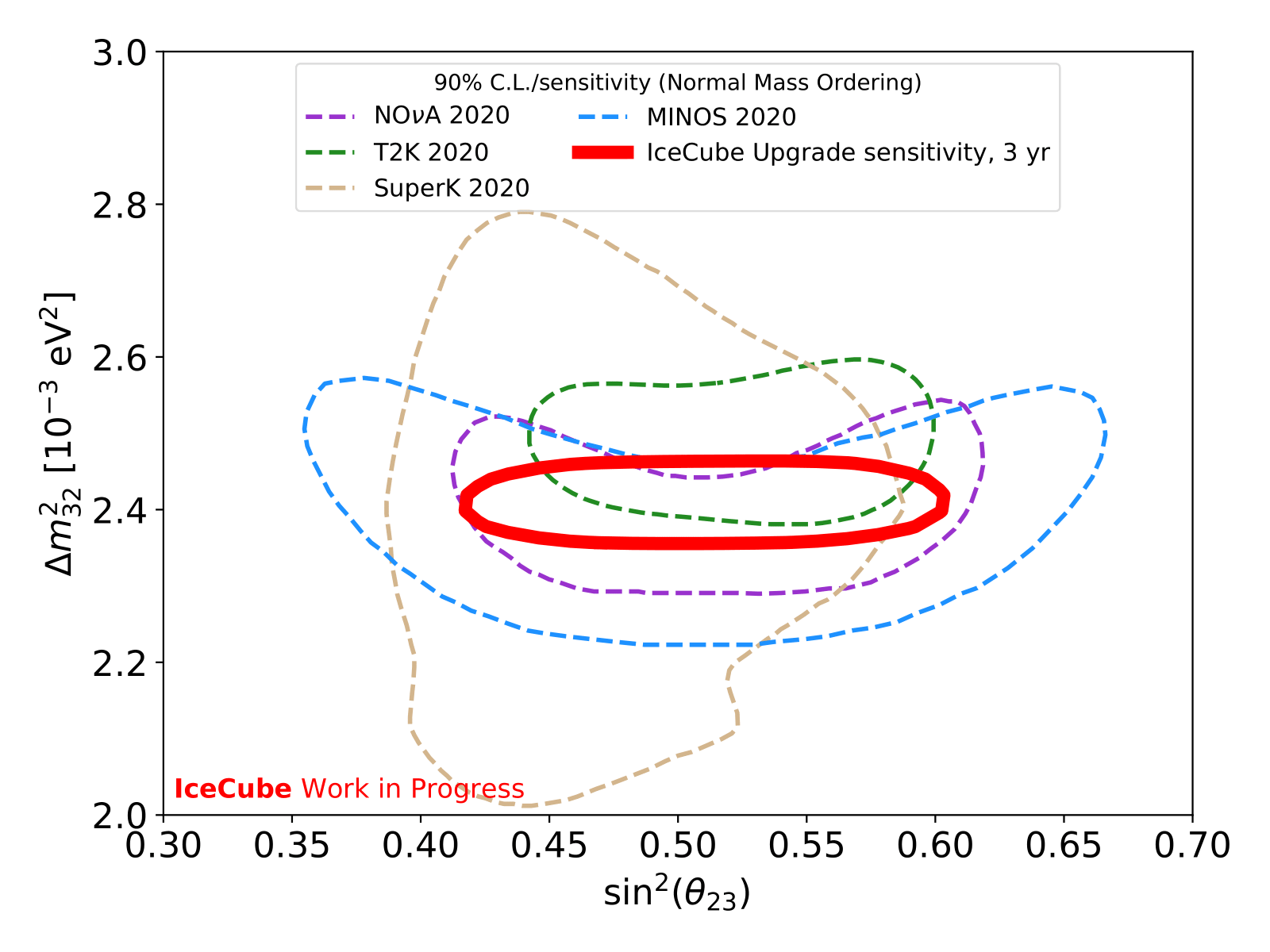}
  \caption{
  Measurements of $\sin^2 (\theta_{23})$ and $\Delta m^2_{32}$ with the IceCube Upgrade (inner fiducial volume) comparing with long-baseline neutrino oscillation facilities and other Cherenkov detectors.  From the CF7 report~\cite{Adhikari:2022sve}.  
  }
  \label{fig:oscillation}
\end{figure}

Observations of neutrino flavors and neutrino-antineutrino ratios may probe BSM physics~\cite{Arguelles:2022xxa}. The search for BSM neutrino interactions with other neutrinos or with dark matter may shed light on the UV theory of neutrinos and the origins of their low mass.  Such new interactions could be observed for example via spectral distortions in the  cosmic neutrino flux in the PeV-EeV range as they scatter on the cosmic neutrino background. Deviations from equal admixtures of the 3 neutrino species might occur at various energy thresholds when new interactions turn on.  Neutrino production measurements from CERN's Forward Physics Facility will provide important information to calibrate the atmospheric neutrino background produced by cosmic rays scattering on nuclei in the atmosphere and thus allow cleaner detection of cosmic neutrinos~\cite{Anchordoqui:2021ghd,Feng:2022inv}. Finally, at GeV energies, the upcoming IceCube Upgrade and other future cosmic neutrino experiments will provide neutrino oscillation sensitivity complementary to long baseline experiments (see figure~\ref{fig:oscillation}).


\section{Exploring the Unknown: New Particles, New Fields, New Principles of Nature}
\label{sec:unknown}

\subsection{Dark Radiation}

Dark radiation can be observed in various cosmological epochs while it is still relativistic, before its kinetic energy redshifts away.  
Within the Standard Model (SM), neutrinos form an important component of dark radiation,
and constraints from their impact on the cosmic evolution at the epochs of Big Bang Nucleosynthesis and the CMB
provide powerful constraints on the number of neutrino species, typically
reported as $N_\mathrm{eff}$, the ``effective number of neutrinos".
In addition, determinations of the dark radiation content of the Universe provide unique opportunities to search for particles produced in
the early universe, even when such particles are extremely weakly interacting with the SM.  A few
well-motivated examples include light particles invoked by models that aim to explain the physics of a dark sector, address the strong CP problem, solve the weak hierarchy problem, account for short baseline neutrino anomalies, and/or models of warm inflation.

\begin{figure}[th!]
    \centering
    \includegraphics[width=0.8\linewidth]{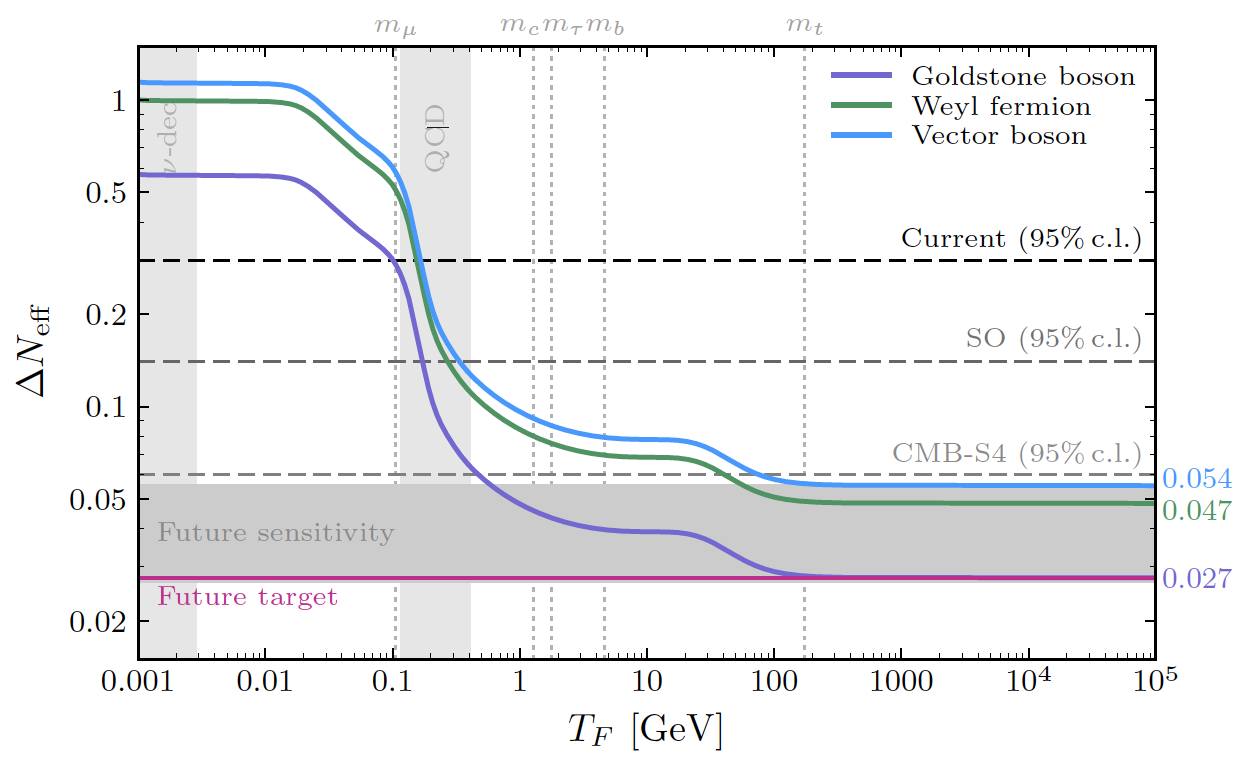}
    \caption{The presence of additional relativistic BSM particles beyond the three known neutrino species impacts the power spectrum of acoustic oscillations in the primordial plasma.  Measurements of the CMB can be used to discern the presence or absence of these exotic degrees of freedom, parameterized as $N_\mathrm{eff}$, the effective number of relativistic species active during this epoch.  Some representative models are shown in this plot along with the current constraints on additional particles and projections for the Simons Observatory and CMB-S4.  From Ref.~\cite{Dvorkin:2022jyg}.}
    \label{fig:cs_neff}
\end{figure}

Figure~\ref{fig:cs_neff} shows the current measurement of $\Delta N_\mathrm{eff}$,
and future prospects from operation of the Simons Observatory and CMB-S4, via temperature and polarization measurements on small angular scales over a large fraction of the sky.  The projections indicate the contribution to $\Delta N_\mathrm{eff}$ from a single species of one of three different types of relic particles (Goldstone or vector bosons and Weyl fermions) for a particle which was initially in chemical equilibrium with the SM plasma, but whose interactions decoupled at freeze-out temperature $T_F$. Current observations constrain $\Delta N_\mathrm{eff} < 0.3$ (at 95\% c.l.), which probes individual particles decoupling during or after the QCD phase transition (at $\sim$100MeV scale). The next generation of CF experiments is poised to reach very exciting levels of sensitivity to $\Delta Neff$. CMB-S4 will achieve $\Delta N_\mathrm{eff} < 0.06 (95\%)$ which would be sensitive to new particles with spin decoupling at 100 GeV and real scalars at 1 GeV, just prior to the QCD phase transition. The later is particularly important for axion-like particles coupling to heavy fermions, where CMB-S4 would be the most sensitive experimental or observational probe by orders of magnitude. 
CMB-S4, by reaching a precision of  $\Delta N_\mathrm{eff} \lesssim 0.06$, will thus be able to discover or rule out the existence of such weakly-interacting particles which have frozen out back to the temperature of the beginning of the QCD phase
transition ($\sim 0.5$~GeV).  Moreover, a Stage V Spectroscopic Facility would reach similar constraints from an independent experiment at $2 <z<5$ which could be combined to achieve even higher precision.  Building off CMB-S4, future experiments reaching a precision of 0.027, as could be achieved e.g. by a CMB-S5 facility and future LIM surveys, would be sensitive to such particles freezing out all the way back to the electroweak scale.

Relativistic dark particles can also be produced late in cosmological time from a variety of sources including dark matter decay, dark sector phase transitions, and cosmic string decays~\cite{Dror_2021} and could be detected with various wave dark matter detector technologies.  For example, quintessence models promote the dark energy from a cosmological constant to a dynamically evolving field which is slowly rolling in a scalar potential, triggering a new epoch of cosmic inflation.  A natural question is then whether this quintessence field is slowly rolling in a flat potential, or if it is rolling in a steeper potential but slowed down by dynamical friction.  In this warm dark energy scenario, the ongoing interactions during the slow roll in the present day universe would populate the universe with dark radiation in the form of relativistic, low mass particles with temperature and energy density typically characterized by the milli-eV scale (10~K) of the dark energy density~\cite{Berghaus:2020ekh}.  This late injection of entropy is not constrained by bounds on CMB distortion at earlier times.  Searches for this 10~K bath of exotic particles with repurposed, low-threshold dark matter detectors would nicely complement the cosmic surveys studies of the evolution of cosmic acceleration.  

\subsection{The Highest Energy Particles}

Cosmic particles above 100~TeV
provide a unique window into fundamental particle physics at energy scales beyond those reachable by terrestrial accelerators.  In addition to providing indirect probes of dark matter in various annihilation or decay channels and the measurements of the neutrino sector described elsewhere in this report, observatories of cosmic rays, neutrinos, gamma rays, and gravitational waves together constitute a rich, multi-messenger program to
explore the unknown through the study of the highest energy particle astrophysical phenomena in the universe.  

The increased precision of the measurement of ultra high energy (UHE; $>10^{18}$ eV) cosmic rays represents a unique opportunity to study particle physics at the very edge of the Energy Frontier via natural accelerators~\cite{Coleman:2022abf}.
In the area of hadronic shower physics, a long standing puzzle is the mysterious excess of muons in UHE cosmic ray air showers beyond what is expected from extrapolations of hadronization models calibrated with LHC and other collider data~\cite{PierreAuger:2016nfk}.  Ongoing upgrades of the Pierre Auger Observatory and IceCube(-Gen2) with the planned enhancement of its surface array will provide greater resolution of the electromagnetic and muonic flux in air showers as well as measurements of the depth of shower maximum which provides complementary information on the initial particle species.  The proposed Forward Physics Facility at the LHC~\cite{Anchordoqui:2021ghd,Feng:2022inv} would also provide muon and electron neutrino shower data which could be used to calibrate kaon production in high energy hadronic showers.  Higher than expected kaon production rates would enhance the muon production by reducing the production of neutral pions which siphon energy into the electromagnetic portion of the shower.  Whether the puzzle is resolved by a better understanding of the fragmentation process or potentially by BSM physics at high center-of-mass energies, these data will better inform the modeling and design of future high energy colliders.  
Next-generation UHECR experiments, such as the Global Cosmic-Ray Observatory (GCOS), the Probe of Extreme Multi-Messenger Astrophysics (POEMMA), and  IceCube-Gen2 with its surface array, will allow for even more precise measurements of particle physics at energies well-beyond the reach of human-made accelerators.  The high boosts and large, cosmological propagation scales of UHE cosmic particles also allow for powerful tests of Lorentz and CPT symmetries by searches for spectral distortions in the flux of UHE cosmic rays, very-high-energy gamma-rays ($0.1-100$~TeV photons), UHE gamma-rays ($>100$~TeV photons), and cosmic neutrinos.  

High energy gamma rays and neutrinos additionally offer probes of new physics including the decays of super-heavy relics left behind from the Big Bang, cosmic strings, and axion-photon conversion in large-scale magnetic fields. Next-generation gamma-ray telescopes such as the Southern Wide-field Gamma-ray Observatory (SWGO), the Cherenkov Telescope Array (CTA), and the All-sky Medium-Energy Gamma-ray Observatory (AMEGO) will open up the access to new sky regions and energy ranges, and advance fundamental physics studies with significantly improved sensitivities. Next-generation UHE neutrino ($>100$~PeV neutrino) observatories led by the U.S. community such as the IceCube-Gen2, The Radio Neutrino Observatory in Greenland (RNO-G), POEMMA, Beam forming Elevated Array for COsmic Neutrinos (BEACON), the Payload for Ultrahigh Energy Observations (PUEO), and the Extreme Neutrino Observatory (Trinity), have the potential to discover UHE neutrinos and use them to probe fundamental physics. 

The concept of multi-messenger astronomy is finally being realized with co-detection of gamma rays and gravitational waves in a binary neutron star merger and the co-detection of neutrinos and gamma rays in a blazar flare~\cite{Engel:2022yig}.  Studies of these violently energetic astrophysical events may reveal new physics including the nature of quark matter in neutron stars, the possible accumulation or scattering of dark matter on these objects, and the production of exotic particles in large astrophysical large magnetic fields.  Various tests of Einstein gravity may lead to discoveries of relevance to cosmology including modifications to the understanding of cosmic acceleration and the distribution of matter. As remarked in the dark matter section of this report, while a 20-year strategic plan is being developed for most cosmic particle channels across broad ranges in energy, an ongoing concern is the lack of coverage for MeV-GeV gamma rays beyond that provided by the aging Fermi telescope.

\subsection{Probes of Fundamental Physics with Gravitational Waves}

One of the most significant scientific developments of the past decade is the emergence of gravitational wave detection as a powerful new tool to study fundamental physics.  In addition to the extensive menu of gravitational physics that will be studied by current and future gravitational wave observatories, the new GW probes will also provide unique and independent information in a number of areas of interest to the high energy physics community.  For example, neutron star binary mergers will probe the neutron star equation of state and may also be sensitive to accumulations of various types of dark matter within the neutron stars.  Binary mergers provide a new standard candle for measurements of cosmic acceleration with systematics independent of those of the currently used optical probes.  Alternatively, the neutron star merger's gravitational wave luminosity distance may be combined with redshift information from the host galaxy to provide new measurements of the Hubble parameter at large redshifts.  

Gravitational wave observatories will also provide unique probes of early universe phase transitions which may have happened at times prior to big bang nucleosynthesis -- a period about which not much information is currently available.  These phase transitions may produce a stochastic background of gravitational waves from boiling of the vacuum during strongly first order phase transitions or from the oscillations and decay of cosmic strings and other topological defects which may form during these phase transitions.  The resulting gravitational wave power spectrum will provide information about these new fundamental energy scales which, once discovered will provide concrete new information for BSM model building.  As a concrete example, with its lower frequency signal response, LISA will be able to search for a stochastic gravitational wave background and provide evidence for an extended Higgs sector causing the a first-order electroweak phase transition.  Cross-correlating the gravitational wave background with the matter power spectrum may also reveal the dynamics of BSM early universe physics.

\begin{figure}[th!]
    \centering
    \includegraphics[width=0.8\linewidth]{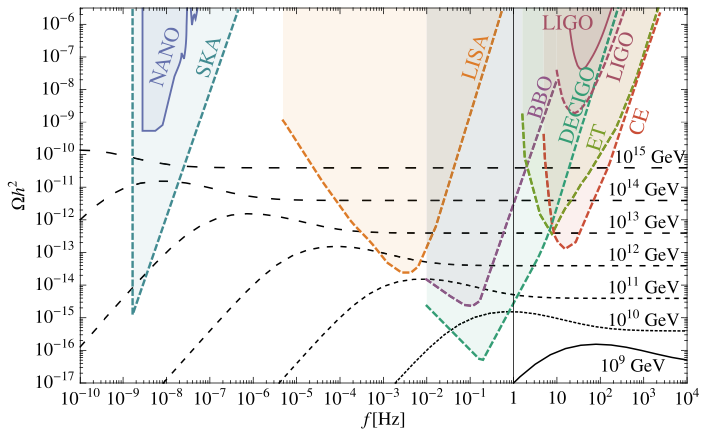}
    \caption{Future gravitational wave observatories will be sensitive to the stochastic spectrum of GWs created by cosmic string decay or by boiling of the vacuum in first order phase transitions.  This plot shows an example of the high energy symmetry breaking scales that can be probed in a particular model of thermal leptogenesis.  In other models, the predicted spectrum may be peaked at certain frequencies, in which case observatories optimized for specific frequency ranges may have greater sensitivity.  From reference~\cite{Dror:2019syi}.}
    \label{fig:unk_murayama}
\end{figure}

The signal will be visible even through the cosmic microwave background which obscures photon probes of earlier cosmological epochs.  Even phase transitions which might happen in secluded dark sectors which have no BSM couplings to standard model particles can still be probed due to the universal coupling of gravity to all matter and energy.  As example, figure~\ref{fig:unk_murayama} shows how a broad program of new gravitational observatories targeting different frequency ranges can probe deeply into the predicted power spectrum of phase transitions at energy scales from $10^{10}-10^{15}$~GeV for various models of thermal leptogenesis.  The observatories range from pulsar timing arrays at the lowest frequencies to space-based experiments at sub-Hz frequencies to next generation ground-based interferometers at the highest frequencies.

New physics of early universe phase transitions may even be observed with the current generation of gravitational wave observatories.  As a case study, LIGO may already be sensitive to the stochastic power spectrum from Peccei-Quinn phase transition in QCD axion models if a particularly violent phase transition occurs at relatively low energies around $10^8$~GeV where the red-shifted frequency spectrum of the GW emission becomes matched to the LIGO band.  If the energy scale of the phase transition can be determined in this way, then the axion model gives a firm prediction for the mass of the dark matter axion and will narrow the search window for the direct detection experiments from 10 decades in mass down to a single decade.  A confirmation of a dark matter signal would then provide corroborating evidence for the axion solution to the strong-CP problem.  Alternatively, if the dark matter signal is discovered first in the post-inflationary scenario with mass between $10^{-5}-10^{-2}$~eV, then this provides a firm target for the sensitivity and frequency range needed for future gravitational wave observatories to observe the phase transition.  Similar strategies may be employed to search for new fundamental energy scales and test see-saw models in other contexts.

The collection of currently operating gravitational wave observatories include the LIGO facilities in Livingston, Louisiana and Hanford, Washington, along with the Virgo facility in Italy and the recently constructed, underground KAGRA observatory in Japan.  LIGO-India will also be online at the end of the decade, and the space-based LISA observatory is planned for the next decade.  Pulsar timing arrays operating at lower frequencies include NANOGrav in North America, the Parkes Pulsar Timing Array in Australia, the European Pulsar Timing Array, and the Indian Pulsar Timing Array, while plans are being made for a future Square Kilometer Array.  Near term plans for the U.S. gravitational wave community include the LIGO Voyager upgrade in which the existing fused silica mirrors may be replaced with crystalline silicon mirrors which can be more easily thermalized to avoid geometric distortions from heating, and the Cosmic Explorer proposal which would increase the interferometer arm length and hence antenna size from 4~km to 40~km.  The AMO community is also developing long baseline atom interferometry as a potential new technique targeting lower frequency gravitational waves as well as oscillatory forces from dark matter in the 1~Hz band.  The MAGIS-100 pathfinder experiment will utilize a 100~m vertical beamline access shaft at Fermilab with possible future expansion to a 2~km drop at SURF and eventually a space mission to avoid terrestrial Newtonian noise.

Participation of the HEP community may be key to the success of these ambitious projects and to ensure that optimizations for HEP science can be integrated into the designs, data pipelines, and operations plans.  As an example, searches for stochastic gravitational wave power can be performed within a narrow band to reject noise as opposed to measurements of time domain waveforms which intrinsically require broader bandwidth template searches.  Similarly, searches for kg-mass dark matter passing within an antenna length of the interferometers will need dedicated data pipelines and analyses.  More direct engagement by national laboratories may also be critical for the construction of future large scale GWO projects.


\section{Conclusion} 
\label{sec:conclusion}

As we have described in this report, the coming decade will provide us with tremendous opportunities to make new breakthroughs in our knowledge of fundamental physics through Cosmic Frontier experiments. With a scientific scope directly encompassing four of the five 2014 P5 science drivers of the field (dark matter, dark energy and cosmic acceleration, neutrinos, and exploring the unknown), much of the discovery potential of the HEP community in the twenty-first century is driven by accomplishments in the Cosmic Frontier. The comprehensive plan envisioned by our community will ensure that this potential is fully realized. Taking full advantage of the scientific opportunities will require us to simultaneously pursue the strategies of Aiming High, Delving Deep, and Searching Wide. These strategies complement and reinforce each other: by developing powerful experiments that are able to simultaneously address many key science questions, pursuing definitive constraints on the most promising scenarios for Beyond the Standard Model physics, and exploring new regions of parameter space where the truth may lie, we can ensure that we will make new discoveries and approach ever closer to a full understanding of the fundamental physics of our Universe.



\bibliographystyle{JHEP}
\bibliography{Cosmic/myreferences} 


\end{document}